%% file: CGMB.tex
\documentclass[11pt,a4paper]{article}

\usepackage{amsmath,amssymb,mathrsfs}
\usepackage{subfigure}
\usepackage{url}
\usepackage{xcolor}
\usepackage{graphicx}
\usepackage{hyperref}
\usepackage{siunitx}
\usepackage{cite}

\allowdisplaybreaks[4]

\title{
Gravitational Waves as a Big Bang Thermometer}

\begin{document}

\renewcommand{\theequation}{\thesection.\arabic{equation}}


\hfill{} DESY 20-187

\hfill{} TUM-HEP-1293-20

\vspace{1.0 truecm}

\begin{center}

{\textbf{\LARGE
Gravitational Waves as a Big Bang Thermometer}
}

\bigskip

\vspace{0.5 truecm}

{\bf Andreas Ringwald$^1$, Jan Sch\"utte-Engel$^{2,3,4}$ and Carlos Tamarit$^5$} \\[5mm]

\begin{tabular}{lc}
&\!\! {$^1$ \em Deutsches Elektronen-Synchrotron DESY, Notkestra\ss e 85,}\\ 
&{\em D-22607 Hamburg, Germany}\\[.4em]
&\!\! {$^2$ \em 
Department of Physics, Universit\"at Hamburg, Luruper Chaussee 149,}\\[.4em]
&{\em D-22761 Hamburg, Germany}\\[.4em]
&\!\! {$^3$ \em Department of Physics, University of Illinois at Urbana-Champaign,}\\
&{\em Urbana, IL 61801, U.S.A.}\\[.4em]
&\!\! {$^4$ \em Illinois Center for Advanced Studies of the Universe,}\\
&{\em  University of Illinois at Urbana-Champaign, Urbana, IL 61801, U.S.A.}\\[.4em]
&\!\! {$^5$ \em Physik-Department T70, Technische Universit\"at M\"unchen,}\\
&{\em James-Franck-Stra\ss e, D-85748 Garching, Germany}\\[.4em]
\end{tabular}

\vspace{1.0 truecm}

{\bf Abstract}
\end{center}

\begin{quote}
There is a guaranteed background of stochastic gravitational waves produced 
in the 
thermal plasma in the early universe. Its energy density per 
logarithmic  frequency interval scales with the  
maximum temperature $T_{\rm max}$ which the primordial plasma attained at the beginning of the standard hot big bang era.
It peaks in the microwave range, at around $80\,{\rm GHz}\,[106.75/g_{*s}(T_{\rm max})]^{1/3}$, where 
$g_{*s}(T_{\rm max})$ is  the effective number of entropy degrees of freedom in the primordial plasma at $T_{\rm max}$.
We present a state-of-the-art prediction of this Cosmic Gravitational Microwave Background (CGMB) 
for general models, and carry out calculations for the case of the Standard Model (SM) as well as for several of its extensions. On the side of minimal extensions we consider 
 the Neutrino Minimal SM ($\nu$MSM) and the SM - Axion - Seesaw - Higgs portal inflation model (SMASH), which provide a complete and consistent cosmological history including inflation. 
As an example of a non-minimal extension of the SM we consider the Minimal Supersymmetric Standard Model (MSSM). Furthermore, 
we discuss the current upper limits and the prospects to detect the CGMB 
in laboratory experiments and thus measure the maximum temperature and the effective number of degrees of freedom 
at the beginning of the hot big bang. 
\end{quote}

\thispagestyle{empty}

{
\vfill\flushleft
\noindent\rule{6 truecm}{0.4pt}\\
{\small  E-mail addresses: \tt andreas.ringwald@desy.de, jan.schuette-engel@desy.de, carlos.tamarit@tum.de}
}

\newpage

\tableofcontents

\renewcommand{\thepage}{\arabic{page}}
\renewcommand{\thefootnote}{\arabic{footnote}}
\setcounter{footnote}{0}

\input{introduction}

\input{gw_background_from_plasma}

\input{observational_constraints_on_CGMB}

\input{laboratory_searches_for_CGMB}

\input{discussion_outlook}

\input{acknowledgments}

\appendix

\input{appendix_complete_leading_order}

\input{appendix_eff_number_dog}

\input{appendix_upper_bound_tmax_inflation}

\input{appendix_3d_eff_gw_emw}

\input{bibliography}
\end{document}

%% file: introduction.tex
\section{Introduction}
\label{sec:introduction}
\setcounter{equation}{0}

Standard hot big bang cosmology provides a successful description of the evolution of the universe back to at least a 
fraction of a second after its birth, when the primordial plasma was radiation-dominated, with temperatures around a few MeV.
It nicely explains 
the Hubble expansion, the cosmic microwave background (CMB) radiation, and the abundance of light elements. 
But it does not predict the maximum temperature, $T_{\rm max}$, which the thermal plasma had at the beginning 
of the radiation-dominated era. It must be larger than a few MeV~\cite{Kawasaki:1999na,Kawasaki:2000en,Giudice:2000ex,Hannestad:2004px,Hasegawa:2019jsa}, but it could be arbitrarily high, although there are  arguments that the maximum temperature 
is bounded from above by the Planck scale, 
$T_{\rm max}\lesssim M_P\equiv 1/{\sqrt{8\pi G}}\simeq 2.435\times 10^{18}$\,GeV~\cite{Sakharov:1966fva}. 
At temperatures higher than that quantum gravity effects become very important and we simply do not know what happens in that regime. Nevertheless, it is natural to assume that for $T_{\rm max}>M_P$ the gravitons would reach thermal equilibrium and acquire a blackbody spectrum which would decouple at $T_{\rm dec}\approx M_P$~\cite{Kolb:1990vq}. After decoupling, the blackbody spectrum would simply redshift with the expansion of the universe, ending up with an effective temperature around $0.9\,{\rm K}\,[106.75/g_{*s}(T_{\rm dec})]^{1/3}$,
where $g_{*s}(T_{\rm dec})$ is the effective number of entropy degrees of freedom at decoupling.

For $T_{\rm max}<M_P$ gravitons are not expected to  thermalize, as the Planck-suppressed gravitational interaction rates will remain below the expansion rate of the universe. 
Nevertheless, out-of-equilibrium gravitational excitations can still be produced from the thermal plasma, and remarkably $T_{\rm max}$ can be probed by gravitational waves (GWs) and bounded by corresponding 
limits~\cite{Ghiglieri:2015nfa,Ghiglieri:2020mhm} (see also Ref.~\cite{Hu:2020wul}). 
In fact, any plasma in thermal equilibrium emits GWs produced by 
physical processes ranging from macroscopic hydrodynamic fluctuations to microscopic particle 
collisions. 
The magnitude and spectral shape of the corresponding stochastic GW background that 
is produced  
during the thermal history of the 
universe -- from the beginning of the thermal radiation dominated epoch after the big bang, at a temperature 
$T_{\rm max}$,  until the electroweak crossover, at a 
temperature $T_{\rm ewco} \simeq 160$\,GeV -- has been calculated in Refs.~\cite{Ghiglieri:2015nfa,Ghiglieri:2020mhm}. As the thermal emission
always peaks at energies of the order of the temperature, and as the frequency of the emitted waves redshifts in correlation with the temperature, the 
 spectral shape of the ensuing gravitational wave background resembles a bit the blackbody spectrum of photons and neutrinos, its power 
peaking today in the same microwave domain -- that is, for frequencies around $\sim100\,$GHz -- as the ones for 
photons and neutrinos. We dub it therefore the Cosmic Gravitational Microwave Background (CGMB), similar to the Cosmic Microwave 
Background (CMB).

Even though at small frequencies, in the sub-10\,kHz range, where all the 
ongoing and near-future planned GW detectors operate, 
this stochastic background is many orders of magnitude below the 
observable level and tiny compared with that from astrophysical and other, more speculative, non-equilibrium sources,  
the total energy density carried by the microwave part of the spectrum near the peak frequency is 
non-negligible if the production continues for a long time, that is if $T_{\rm max}\gg T_{\rm ewco}$. 
This is due to the fact that, 
although the thermal rate of production is Planck suppressed, peak emissions at different times add up constructively because of the correlated redshifting of frequency and temperature leads to an approximate linear relation between the total energy emitted in gravitational waves around the peak frequency and the temperature $T_{\rm max}$.
Observing this part directly sets an ambitious but worthwhile 
 goal for future generations of GW detectors, allowing to probe properties of the primordial thermal plasma at the beginning of the hot big bang era, such 
as its maximum temperature and, as we will see, its effective number of degrees of freedom.

This paper is organized as follows. In Section \ref{sec:GWSM} we determine -- based on the work of 
Refs.~\cite{Ghiglieri:2015nfa,Ghiglieri:2020mhm} -- the frequency spectrum of the CGMB in a general theory, and subsequently focus on the Standard Model (SM) and several of its extensions. As minimal extensions we choose the Neutrino Minimal Standard Model ($\nu$MSM)~\cite{Asaka:2005an,Asaka:2005pn} and the SM - Axion - Seesaw - Higgs portal inflation model (SMASH)~\cite{Ballesteros:2016euj,Ballesteros:2016xej}. Both  explain neutrino masses and mixing, the non-baryonic dark matter (DM) abundance, the baryon asymmetry of the universe (BAU), and eventually also solve
 the horizon and flatness problems of the standard hot big bang cosmology. As an example of nonminimal extension of the SM we focus on the Minimal Supersymmetric Standard Model (MSSM) \cite{Fayet:1977yc,Farrar:1978xj,Baer:2006rs},  which is motivated by the Higgs naturalness problem, gauge coupling unification and dark matter.  In Section \ref{sec:observational_constraints}
we confront the CGMB predictions with upper limits on the total energy density of any extra relativistic radiation field 
at the time of big bang nucleosynthesis (BBN) or of decoupling of the CMB photons.  
In Section \ref{sec:laboratory_searches} we compare the predictions with current limits from direct laboratory searches for 
GWs and we discuss laboratory experiments which may ultimately probe sub-Planck-scale values of 
$T_{\rm max}$. Finally, we summarize our findings and give an outlook for further investigations in Section \ref{sec:summary}.

%% file: gw_background_from_plasma.tex
\section{GW background from primordial thermal plasma}
\label{sec:GWSM}
\setcounter{equation}{0}

In this section we exploit the results from Refs.~\cite{Ghiglieri:2015nfa,Ghiglieri:2020mhm} concerning the CGMB produced in the primordial thermal plasma at sub-Planckian temperatures, $T_{\rm max}<M_P$.  While the former references focused mainly on the SM case, we provide when possible expressions generalized to an arbitrary theory with gauge fields, real scalars and Weyl fermions. 
The fields are treated as massless, which is  a good approximation for temperatures much above the masses of particles in the vacuum. For temperatures below the mass threshold of a given particle, the former decouples from thermal plasma and one can work in an effective theory in which the heavy particle has been integrated out. Therefore, the general results given below can be applied at different temperature ranges when using the appropriate effective theories for the light excitations.

In the following we will start  with the general formulae for the production rate of gravitational waves from the primordial plasma and derive expressions for the current energy fraction of gravitational waves per logarithmic frequency interval. Next, we will focus on the predicted spectrum for the SM, to be followed by calculations for  three different theories Beyond the SM (BSM):  the $\nu$MSM,  SMASH, and the MSSM.

\subsection{Production rate of GWs from a general thermal plasma}
\label{sec:prod_gen_plasma}

Here we  revise the state-of-the art results for the rate of emission of gravitational waves from a thermal plasma in a generic quantum field theory coupled to gravity. We draw from the results of refs.~\cite{Ghiglieri:2015nfa,Ghiglieri:2020mhm}. While the explicit expressions in the former references were tailored for variations of the SM with different numbers of Higgs doublets, generations and colors, we will rewrite the results in a way that facilitates the application to arbitrary quantum field theories with gauge fields, real scalars and Weyl fermions.
For our general theory we consider $n=1,\dots\,{\cal N}_g$ gauge groups with coupling constants $g_n\equiv \sqrt{4\pi\alpha_n}$ and Lie algebras of dimension $N_n$ spanned by generators $T^a_n, a=1,\dots,N_n$. We further assume real scalar fields $\phi_i$, $i=1,\dots,N_s$ and Weyl spinors $\psi_\alpha$, $\alpha=1,\dots,N_f$. The real scalars transform under each group $n$ under a direct sum of irreducible representations $r_{n,\hat{\i}}$, which can include several copies of the same representation. For each irreducible representation of each gauge group we consider the Dynkin index $T_{n,\hat{\i}}$ defined from the identity ${\rm Tr}_{r_{n,\hat{\i}}} T_n^a T_n^b = \delta^{ab} T_{n,\hat{\i}}$. 
Analogously, we define fermion representations $r_{n,\hat{\alpha}}$ with Dynkin indices $T_{n,\hat{\alpha}}$. 
The Dynkin indices of the adjoint representations of the gauge fields themselves will be denoted as $T_{n,{\rm Ad}}$. 

Regarding the interactions of the fields, it turns out that scalar quartic couplings do not contribute to gravitational wave production at leading order, and thus we will focus on gauge and Yukawa interactions. For the latter we use the convention 
\begin{align}
 {\cal L}\supset -\sum_{i,\alpha,\beta} y^i_{\alpha\beta} \phi_i \psi_\alpha \psi_\beta+c.c..
\end{align}

With the representations of the matter fields defined as above, one may recover the Debye thermal masses of the gauge fields in the plasma from the following expression:
\begin{align}
 m^2_n(T)= g_n^2(T) T^2\left(\frac{1}{3}\,T_{n,\rm Ad}+\frac{1}{6}\sum_{\hat{\i}}  T_{n,\hat{\i}}+\frac{1}{6}\sum_{\hat{\alpha}}  T_{n,\hat{\alpha}}\right) \equiv T^2\hat{m}^2_n(T).
\end{align}
In the equation above we included a temperature dependence of the couplings $g_n(T)$, arising from choosing a renormalization scale proportional to the temperature. This is expected to provide optimal accuracy for the computations of thermal effects, as they involve excitations whose typical momenta are of the order of $T$. This choice of renormalization scale implies that  the dimensionless quantity $\hat{m}^2_n(T)$ inherits a logarithmic temperature dependence which has been explicitly indicated. 

Within the gauge interactions, we will assume that  hypercharge is the weakest. This has an impact in the production of gravitational waves with low frequencies, which as will be seen later is related to the plasma's shear viscosity \cite{Ghiglieri:2015nfa}, which is known to be dominated by the effect of the weakest gauge interaction \cite{Arnold:2000dr}.
Due to this, following the former reference it is convenient to define  a number $N_{\rm species}$  given by one half of the sum over the hypercharge Dynkin indices of the real scalar and Weyl fermion representations. Analogously one can define $N_{\rm leptons}$ as one-half of the squared hypercharges of the Weyl fermions that  interact with no other SM gauge group than hypercharge. Assigning $k=1$ to the hypercharge group, one has
\begin{align}\label{eq:Nspeciesleptons}\begin{aligned}
 N_{\rm species}=&\,\frac{1}{2}\sum_{\hat{\i}} T_{1,\hat{\i}}+\frac{1}{2}\sum_{\hat\alpha}T_{1,\hat{\alpha}}\,,
\\
 N_{\rm leptons}=&\,\frac{1}{2}\hskip-.9cm\sum_{\hskip.9cm\hat\alpha: T_{n,\hat{\alpha}}=0,n>1}\hskip-.9cmT_{1,\hat{\alpha}}\,.
\end{aligned}\end{align}
We note that for the expression for $N_{\rm leptons}$ it was assumed that the only fields that interact exclusively with hypercharge are fermions. We expect that scalar fields with similar properties will also contribute to $N_{\rm leptons}$; however, the estimates of transport coefficients in Ref.~\cite{Arnold:2000dr} did not account for such fields. For the MSSM studies in Section~\ref{sec:GW_MSSM} we will assume a contribution to $N_{\rm leptons}$ coming from the supersymmetric partners of the right-handed leptons, obtained by adding to $N_{\rm leptons}$ in Eq.~\eqref{eq:Nspeciesleptons} the analogous sum over representations of real scalar fields.

In a homogeneous and isotropic universe, with scale factor $a$ and Hubble expansion rate $H=\dot a/a$,  the energy density 
$\rho_{\rm CGMB}$ carried by  the CGMB, which was generated 
in a thermal plasma with temperature $T$,  evolves in cosmic time $t$ as
\begin{equation}
 \left(\partial_t + 4 H(t)\right) \rho_{\rm CGMB}(t) = \frac{4\, T^4}{M_P^2}   
\int \! \frac{{\rm d}^3\mathbf{k}}{(2\pi)^3}\, \hat\eta\left(T,\frac{k}{T} \right) 
 \,
\label{evol1}.
\end{equation}\
The former equation assumes a very small energy density of gravitational waves, so that one can neglect the backreaction contribution from gravitational excitations annihilating  or decaying back into the plasma. As emphasized in the introduction, this is expected to be a good approximation for temperatures below the Planck scale.
For momenta lower than the temperature, the dimensionless source term $\hat{\eta}\left(T,\frac{k}{T} \right)$
can be understood from long-range hydrodynamic fluctuations,  while for momenta comparable or greater than the temperature,  $\hat{\eta}\left(T,\frac{k}{T} \right)$ is dominated by contributions from quasi-particle excitations in the plasma~\cite{Ghiglieri:2015nfa,Ghiglieri:2020mhm}. While the hydrodynamic contribution is known to leading-log order in the gauge couplings,  recently the quasi-particle contribution has been estimated to full leading order in the gauge and Yukawa couplings \cite{Ghiglieri:2020mhm}. The results, for temperatures above the electroweak crossover, can be written as:
\begin{eqnarray}\label{eq:etafull}
\hat\eta\left(T,\frac{k}{T} \equiv \hat{k}\right)  
&\simeq & 
 \left\{ 
  \begin{array}{ll}
    \displaystyle \frac{\bar\eta}{g_1(T)^4 \ln(5 / \hat{m}_{1})}, & \;\quad  \hat{k} \lesssim \alpha_1^2\,,\\
    \displaystyle\hat{\eta}_{\rm HTL}(T,\hat k)+\sum_{n=1}^{{\cal N}_g}g_n(T)^2 N_n\left(\frac{1}{2}\,T_{n,\rm Ad}  \,\eta_{gg}(\hat k )\right.\\
   \displaystyle\left. +\sum_{\hat{\i}}   T_{n,\hat{\i}}\,\eta_{sg}(\hat k )
   +\frac{1}{2}\sum_{\hat \alpha}   T_{n,\hat{\alpha}}\,\eta_{fg}(\hat k )\right) &  \;\quad \hat{k} \gtrsim {\rm max}\,\{\hat m_n\}.\\
   +\displaystyle \frac{1}{4}\sum_{i\alpha\beta}|y^i_{\alpha\beta}(T)|^2\,\eta_{sf}(\hat k ),
  \end{array}
  \right. 
\end{eqnarray}
In the equations above we defined a dimensionless momentum $\hat{k}=k/T$ and introduced a coefficient $\bar\eta$ and functions $\hat{\eta}_{\rm HTL}(T,\hat k)$, $\eta_{gg}(\hat k )$, $\eta_{sg}(\hat k )$, $\eta_{fg}(\hat k )$,  $\eta_{sf}(\hat k )$ which will be described next.

First, the hydrodynamic contribution for $k\lesssim\alpha_1^2 T$ coincides with the shear-viscosity of the plasma divided by $T^3$ \cite{Ghiglieri:2015nfa},
\begin{align}
 \hat{\eta}=\frac{\eta^{\rm shear}}{T^3}, \quad {\rm for\ }\ \hat{k}\lesssim\alpha_1^2.
\end{align}
The shear viscosity is inversely proportional to a scattering cross section and therefore large
for a plasma in which there are some weakly interacting particles. Under our assumption that hypercharge is the weakest gauge force, right-handed leptons (or additional fields only charged under $U(1)_Y$)
are the most weakly interacting degrees of freedom, changing their
momenta only through reactions mediated by hypercharge gauge fields above the electroweak crossover. The results of Ref.~\cite{Arnold:2000dr} give then the following value for the coefficient $\bar\eta$ in Eq.~\eqref{eq:etafull}:
\begin{align}
 \bar\eta=\zeta(5)^2\left(\frac{5}{2}\right)^3 \left(\frac{12}{\pi}\right)^5 \left(\frac{N_{\rm leptons}}{9 \pi^2 + 224 N_{\rm species}}\right),
\end{align}
where $\zeta$ is Riemann's zeta function. 

For $k\gtrsim{\rm max}\,\{\alpha_n^2\} T$, GWs are dominantly  produced via  
microscopic particle scatterings, despite the fact that their rates are suppressed by the 
coupling strengths responsible for the interactions and a 
Boltzmann factor $e^{-k/T}$, which takes into account that 
the energy carried away by the graviton must be extracted from thermal
fluctuations. These contributions  were first estimated at leading-log accuracy in Ref.~\cite{Ghiglieri:2015nfa} and then calculated at full leading order in Ref.~\cite{Ghiglieri:2020mhm}. In the latter calculation, the production rate is obtained from the imaginary part of the two-point correlator of the stress-energy momentum tensor at two-loop order.  Some of the loop integrals involved turn out to be infrared divergent when treating the fields as massless. A resummation of the effects of the thermal masses resolves the singularities, and  Ref.~\cite{Ghiglieri:2020mhm} implemented this in the following manner: a contribution containing the divergence was added and subtracted; the subtracted part was used to define infrared-finite loop integrals, while the added piece was rendered finite by  implementing the resummation of the thermal masses. In this way, the thermal resummation is only performed in the region of phase space near the singularities, but the procedure guarantees full leading-order accuracy. The function $\hat{\eta}_{\rm HTL}(T,\hat k)$  in Eq.~\eqref{eq:etafull} corresponds to the regulated divergence (with ``HTL'' alluding to the hard thermal loop resummation \cite{Braaten:1989mz}) and is given by
\begin{align}\label{eq:etaHTL}
 \hat{\eta}_{\rm HTL}(T,\hat k)=\frac{\hat k}{16 \pi(e^{\hat{k}} - 1)} \sum_n N_n \hat{m}^2_n(T) \log\left(1+4\frac{\hat{k}^2}{\hat{m}^2_n(T)}\right).
\end{align}
This is to be contrasted with the initial leading-log estimate of $\hat{
\eta}$, $\hat{\eta}_{\rm LL}$, that was computed in Ref.~\cite{Ghiglieri:2015nfa}:
\begin{align}\label{eq:etaLL}
 \hat{\eta}_{\rm LL}(T,\hat k)=\frac{\hat k}{8 \pi(e^{\hat{k}} - 1)} \sum_n N_n \hat{m}^2_n(T) \log\left(\frac{5}{\hat{m}^2_n(T)}\right).
\end{align}
Finally, the  remaining functions in Eq.~\eqref{eq:etafull} correspond to the infrared-finite two-loop integrals mentioned before, for diagrams involving  only gauge fields ($\eta_{gg}(\hat k )$), scalars and gauge fields ($\eta_{sg}(\hat k )$), fermions and gauge fields ($\eta_{fg}(\hat k )$), and scalars and fermions  ($\eta_{sf}(\hat k )$). The infrared subtraction has to be performed in the $gg,sg$ and $fg$ sectors. Explicit expressions for the loop functions are given in Appendix~\ref{app:complete_leading_order} (see Eqs.~\eqref{eq:etasIJKs},\eqref{eq:IJKs}). From the latter equations and from Eqs.~\eqref{eq:etafull}, \eqref{eq:etaHTL} it is clear that, as advertised earlier, all contributions to the production rate are suppressed by powers of the gauge and Yukawa couplings, as well as Boltzmann factors.


\subsection{Current stochastic GW background from a primordial thermal plasma}
\label{sec:current_stoch_backgd}

In this section we relate the thermal production rate discussed above with the stochastic background of gravitational waves in the present universe. At every value of the temperature, the Boltzmann suppression factor ensures a peak emission for momenta of the order of the temperature. The expansion of the universe redshifts temperature and momenta by approximately the same amount, and as a consequence of this the peak emission at a given temperature overlaps with the redshifted peak emissions of the previous history of the universe. Thus, the energy density of gravitational waves at the peak frequency is sensitive to the entire history of the hot primordial plasma, so that the weak Planck-suppressed production rates can be partly compensated. Similarly to the case of the CMB, the peak of the thermal spectrum lies currently in the microwave region -- simply because the current CMB temperature $T_0 \approx 2.73$ K is associated with frequencies in the 100 GHz regime -- leading to the CGMB.

Given that
$(\partial_t + 3 H(t))s = 0$, where $s$ is the entropy density, the factor $4H(t)$ in Eq. \eqref{evol1} 
can be taken care of by
normalizing $\rho_{\rm CGMB}$ by $s^{4/3}$. Subsequently, 
the equation can be integrated from the time $t_{\rm hbb}$, when the hot big bang era starts with a temperature $T_{\rm max}(<M_P)$,  to the 
time $t_{\rm ewco}$, when the electroweak crossover takes place, by assuming that at 
$t_{\rm hBB}$ there were no (thermally produced) 
gravitational waves present:  
\begin{eqnarray}
 \frac{\rho_{\rm CGMB}(t_{\rm ewco})}{s^{4/3}(t_{\rm ewco})}
 &\simeq & \frac{4}{M_P^2}
 \int_{t_{\rm hbb}}^{t_{\rm ewco}}
  {\rm d}t \,\frac{T^4}{s^{4/3}(T)} 
\int \! \frac{{\rm d}^3\mathbf{k}}{(2\pi)^3}\,\hat\eta\left(T,\frac{k}{T} \right)
\label{evol2}  \\
&= & 
 \frac{12\sqrt{10}}{\pi M_P} \,
\left( \frac{45}{2\pi^2}\right)^{4/3}
\int_{T_{\rm ewco}}^{T_{\rm max}} \! \frac{{\rm d}T }{T^3}
\,
\frac{g_{*c}(T)}{\left[g_{*\rho}(T)\right]^{1/2}\,\left[g_{*s}(T)\right]^{7/3}}\,
\int \! \frac{{\rm d}^3\mathbf{k}}{(2\pi)^3}\,\hat\eta\left(T,\frac{k}{T} \right) 
 \;. 
\nonumber
\end{eqnarray}
Here we have used that time and temperature are related as \cite{Laine:2015kra}
\begin{equation}
 \frac{{\rm d}T}{{\rm d}t} = - \frac{\pi}{\sqrt{90}} 
\left[g_{*\rho}(T)\right]^{1/2}\,
\frac{g_{*s}(T)}{g_{*c}(T)}\,
\frac{T^3}{M_P} 
 \;,
\end{equation} 
where
\begin{equation}
g_{*\rho}(T)=\frac{\rho(T)}{\frac{\pi^2}{30}T^4}, \hspace{3ex}
g_{*s}(T)=\frac{s(T)}{\frac{2\pi^2}{45}T^3}, \hspace{3ex}
g_{*c}(T)=\frac{c(T)}{\frac{2\pi^2}{15}T^3},
\end{equation}
are the effective degrees of freedom of the energy density, $\rho(T)$, the entropy density, $s(T)$, and 
the heat capacity,  $c(T)$, respectively.  The above result can be written in the form 
\begin{equation}
 {\rho_{\rm CGMB}(t_{\rm ewco})}
\simeq  \frac{48\sqrt{10}}{(2\pi)^3M_P} \,  T_{\rm ewco}^4
\int_{T_{\rm ewco}}^{T_{\rm max}} \! \frac{{\rm d}T }{T^3}
\,\!\!
\frac{\left[g_{*s}(T_{\rm ewco})\right]^{4/3}g_{*c}(T)}{\left[g_{*\rho}(T)\right]^{1/2}\left[g_{*s}(T)\right]^{7/3}}\,
\int_0^\infty {\rm d}k\,k^2\,
\hat\eta\left(T,\frac{k}{T} \right) 
 \;,
\label{eq:rho_GW_ewco}
\end{equation}
from which we can read off the spectrum of the GW energy density fraction per logarithmic wave number interval at the time 
of the electroweak crossover,  
\begin{eqnarray}
\Omega_{\rm CGMB}^{({\rm ewco})} (k_{\rm ewco}) &\equiv & 
\frac{1}{\rho(T_{\rm ewco})}\frac {{\rm d}\rho_{\rm CGMB}}{{\rm d}\ln k_{\rm ewco}} (T_{\rm ewco},k_{\rm ewco}) 
\nonumber \\ 
&\simeq & 
\frac{1440\sqrt{10}}{8\pi^5} \,  \frac{1}{M_P} \frac{1}{g_{*\rho}(T_{\rm ewco})} 
\frac{ k^3_{\rm ewco}}{T_{\rm ewco}^3} 
\int_{T_{\rm ewco}}^{T_{\rm max}} \! {\rm d}T 
\,
\times
\\ 
& &\times 
\,
\frac{\left[g_{*s}(T_{\rm ewco})\right]^{1/3}g_{*c}(T)}{\left[g_{*s}(T)\right]^{4/3}\left[g_{*\rho}(T)\right]^{1/2}}
\,
\hat\eta\left(T,\frac{k_{\rm ewco}}{T_{\rm ewco}}\,\left[\frac{g_{*s}(T)}{g_{*s}(T_{\rm ewco})}\right]^{1/3} \right)
 \;.
\nonumber
\end{eqnarray}
We have obtained this by taking into account that momenta redshift as 
\begin{equation}
 {k}(t) = {k}_{\rm ewco}\, \frac{a(t_{\rm ewco})}{ a(t)} = {k}_{\rm ewco}\, \left[\frac{g_{*s}(T)}{g_{*s}(T_{\rm ewco})}\right]^{1/3} 
\frac{T}{T_{\rm ewco}}
\end{equation}
and expressed the momentum space integral in \eqref{eq:rho_GW_ewco} in terms of $k_{\rm ewco}$.
The spectrum of the GW energy density fraction per logarithmic wave number interval at the present time is then 
\begin{eqnarray}
\Omega_{\rm CGMB} (k_{\rm ewco}) 
&=&
\frac{1}{2}\,
\left[\frac{g_{*s}({\rm fin})}{g_{*s}(T_{\rm ewco})}\right]^{4/3}
g_{*\rho}(T_{\rm ewco}) 
\ \Omega_\gamma\, \Omega_{\rm CGMB}^{({\rm ewco})} (k_{\rm ewco})
\\[1ex]
&\simeq & 
\frac{1440\sqrt{10}}{2\pi^2 (2\pi)^3M_P} \,   
\,\Omega_\gamma\,
\frac{\left[g_{*s}({\rm fin})\right]^{4/3}}{g_{*s}(T_{\rm ewco})}
\frac{ k^3_{\rm ewco}}{T_{\rm ewco}^3} 
\int_{T_{\rm ewco}}^{T_{\rm max}} \! {\rm d}T 
\,
\times
\nonumber \\ 
& &\times 
\,
\frac{g_{*c}(T)}{\left[g_{*s}(T)\right]^{4/3}\left[g_{*\rho}(T)\right]^{1/2}}
\,
\hat\eta\left(T,\frac{k_{\rm ewco}}{T_{\rm ewco}}\,\left[\frac{g_{*s}(T)}{g_{*s}(T_{\rm ewco})}\right]^{1/3} \right)
 \;,
\nonumber
\end{eqnarray}
where $g_{*s}({\rm fin}) = 3.931\pm 0.004$~\cite{Saikawa:2018rcs} is the number of effective degrees of freedom of the entropy density after neutrino decoupling\footnote{The quoted value is slightly larger than the simple expression $g_{*s}(T_0) =2+\frac{7}{8}\times 2\times 3\times \frac{4}{11}\simeq 3.909$ \cite{Kolb:1990vq} based on the approximation of instantaneous decoupling of the neutrinos before $e^+e^-$ annihilation.} 
and $\Omega_\gamma \equiv \rho_\gamma^{(0)}/\rho_c^{(0)}=2.4728(21)\times 10^{-5}/h^2$ the present fractional energy density of the CMB photons, with temperature $T_0 = 2.72548(57)$\,K and a present Hubble parameter $H_0=100\,h$\,km\,s$^{-1}$\,Mpc$^{-1}$.
The corresponding GW frequency today is given by 
\begin{equation}
f = \frac{1}{2\pi} \frac{a_{\rm ewco}}{a_0} k_{\rm ewco} = 
\frac{1}{2\pi} \left[\frac{g_{*s}({\rm fin})}{g_{*s}(T_{\rm ewco})}\right]^{1/3}\,
\left(\frac{T_0}{T_{\rm ewco}}\right)
k_{\rm ewco}\,.
\end{equation}
Therefore, 
\begin{equation}
\label{eq:kewco_Tewco}
\frac{ k_{\rm ewco}}{T_{\rm ewco}}
= 2\pi \left[\frac{g_{*s}(T_{\rm ewco})}{g_{*s}({\rm fin})}\right]^{1/3}
\frac{f}{T_0}
\,,
\end{equation}
and 
\begin{eqnarray}
\label{eq:Omega_CGMB_full_expression}
\Omega_{\rm CGMB} (f) 
&\simeq & 
\frac{1440\sqrt{10}}{2\pi^2 M_P} \,   
\,\Omega_\gamma\,
\left[g_{*s}({\rm fin})\right]^{1/3}
\frac{ f^3}{T_0^3} 
\times
 \\ 
& &\times 
\int_{T_{\rm ewco}}^{T_{\rm max}} \! {\rm d}T 
\,\,
\frac{g_{*c}(T)}{\left[g_{*s}(T)\right]^{4/3}\left[g_{*\rho}(T)\right]^{1/2}}
\,
\hat\eta\left(
T,2\pi \,
\left[\frac{g_{*s}(T)}{g_{*s}({\rm fin})}\right]^{1/3}\,
\frac{f}{T_0}
\right)
 \;.
\nonumber
\end{eqnarray}

For order of magnitude estimates and an analytic understanding of the result, one may exploit the fact that  the temperature dependence of the effective degrees of freedom of the energy density, entropy density, and 
heat capacity approximately agree amongst each other above the electroweak crossover, 
$g_{*\rho}(T)\approx g_{*s}(T)  \approx  g_{*c}(T)$, for $T\gtrsim T_{\rm ewco}$, see Appendix~\ref{app:effect_degr_freedom}. Therefore, 
Eq.~\eqref{eq:Omega_CGMB_full_expression} can be approximated by 
\begin{eqnarray}
\label{eq:Omega_CGMB_simpified_expression}
\Omega_{\rm CGMB}  (f)
&\simeq & 
\frac{1440\sqrt{10}}{2\pi^2 M_P} \,  
\,\Omega_\gamma\,
\left[g_{*s}({\rm fin})\right]^{1/3} 
\,
\frac{ f^3}{T_0^3} 
\times
\\ 
& &    \hspace{12ex} \times 
\int_{T_{\rm ewco}}^{T_{\rm max}} \! {\rm d}T 
\left[g_{*s}(T)\right]^{-5/6}\,
\hat\eta\left(
T,2\pi \,
\left[\frac{g_{*s}(T)}{g_{*s}({\rm fin})}\right]^{1/3}\,
\frac{f}{T_0}
\right)
 \;.
\nonumber
\end{eqnarray}
Furthermore, assuming that $g_{*s}$ and $\hat \eta$ are almost independent of temperature above the electroweak cross-over,  up to possible steps when new BSM degrees of freedom get relativistic, and exploiting the fact that the $T$ dependence of $\hat\eta(T,\hat k)$ is only logarithmic (from the running of the coupling constants with temperature), one may obtain an approximate  analytic expression by ignoring the temperature dependence of the integrand. Doing so the
energy density of the CGMB per logarithmic frequency interval, valid for $T_{\rm max} \gg T_{\rm ewco}$, can be approximated as: 
\begin{eqnarray}
\label{eq:Omega_CGMB_analytic_expression}
{{h^2}}\,\Omega_{\rm CGMB}  (f)
&\approx  &
\frac{1440\sqrt{10}}{2\pi^2} \,  
\,{{h^2}}\,\Omega_\gamma\,
\frac{\left[g_{*s}({\rm fin})\right]^{1/3}}
{\left[g_{*s}(T_{\rm max})\right]^{5/6}}\,
\frac{ f^3}{T_0^3}\, 
\frac{T_{\rm max}}{M_P} \,
\hat\eta\left(
T_{\rm max},2\pi \,
\left[\frac{g_{*s}(T_{\rm max})}{g_{*s}({\rm fin})}\right]^{1/3}\,
\frac{f}{T_0}
\right)
\\
&= & 
4.03\times 10^{-12} 
\, 
\left[\frac{T_{\rm max}}{M_P}\right]
\left[\frac{g_{*s}(T_{\rm max})}{106.75}\right]^{-5/6}
\left[ \frac{f}{\rm GHz}\right]^3
\hat\eta\left(
T_{\rm max},2\pi \,
\left[\frac{g_{*s}(T_{\rm max})}{g_{*s}({\rm fin})}\right]^{1/3}\,
\frac{f}{T_0}
\right)
 \;.
\nonumber
\end{eqnarray}
Its overall magnitude scales approximately linearly with the maximum temperature of the hot big bang.  
Therefore, it can play the role of a hot big bang thermometer.

Moreover, a measurement of the peak frequency of $\Omega_{\rm CGMB}$ provides a measurement of the relativistic degrees 
of freedom at $T_{\rm max}$. 
In fact, the peak frequency is, according to Eq.~\eqref{eq:Omega_CGMB_analytic_expression}, approximately determined by the peak of 
$\hat k^3 \hat \eta (T_{\rm max}, \hat k)$, which in turn can be estimated from its leading-log behaviour, 
\begin{equation}
\hat k^3 \hat\eta_{\rm LL} (T_{\rm max}, \hat k) =   
\frac{\hat k^4}{{\rm e}^{\hat k}-1} f(T_{\rm max})\,,
\end{equation} 
cf. Eq.~\eqref{eq:etaLL}. Its peak occurs at  
\begin{equation}
\label{eq:hatkmax}
\hat k_{\rm peak}^{\Omega_{\rm CGMB}} \approx  4 + W(-4/{\rm e}^4) \simeq 3.92 \,,
\end{equation}
where $W$ is the Lambert W function, leading to 
\begin{eqnarray}
\label{req:peak_freq_cgmb}
f_{\rm peak}^{\Omega_{\rm CGMB}} 
\approx  \frac{3.92}{2\pi}     \left[\frac{g_{*s}({\rm fin})}{g_{*s}(T_{\rm max})}\right]^{1/3}   T_0 
\simeq   74\,{\rm GHz} 
\,  
\left[\frac{g_{*s}(T_{\rm max})}{106.75}\right]^{-1/3}
\,.
\end{eqnarray}
Up to the factor $\left[\frac{g_{*s}({\rm fin})}{g_{*s}(T_{\rm max})}\right]^{1/3}$, it coincides with the 
peak frequency, 
\begin{equation}
f_{\rm peak}^{\Omega_{\rm CMB}} 
\simeq  \frac{3.92}{2\pi} \    T_0 \simeq 223\, {\rm GHz} 
\,,
\end{equation}
of the present energy fraction of the CMB per logarithmic frequency, 
\begin{equation}
\label{eq:CMB}
\Omega_{\rm CMB} (f)
=
\frac{16\,\pi^2 }{3 H_0^2 M_P^2} \frac{f^4}{{\rm e}^{2\pi f/T_0}-1}
\,.
\end{equation}
In fact, around the peak and to leading-log accuracy, the spectral form of the CGMB resembles the one of the CMB with an effective 
temperature 
\begin{equation}
T_{\rm CGMB} \simeq  \left[\frac{g_{*s}({\rm fin})}{g_{*s}(T_{\rm max})}\right]^{1/3}   T_0 \simeq 0.91\,{\rm K} \,  
\left[\frac{g_{*s}(T_{\rm max})}{106.75}\right]^{-1/3}
\,.
\end{equation}

Direct detection bounds or projected sensitivities are often expressed in terms of a characteristic dimensionless GW amplitude $h_c(f)$, 
which is related to the cosmic energy density fraction of stochastic GWs per logarithmic frequency interval 
as (we use the conventions of Ref.~\cite{Caprini:2018mtu})
\begin{equation}
\Omega_{\rm GW}(f) = \frac{2 \pi^2}{3 H_0^2} \, f^2 \, h_c^2(f)  \, ,
\label{eq:Omega_hc}
\end{equation}
that is, numerically, 
\begin{equation}
h_c(f) = 1.26 \times 10^{-18} \, \left[\frac{\mathrm{Hz}}{f}\right] \, \sqrt{h^2\,\Omega^{(0)}_{\rm GW}(f)} \, .
\label{eq:hc}
\end{equation}

For the CGMB, its overall magnitude scales approximately linearly with the square root of the maximum temperature of the hot big bang.  
Moreover, its peak frequency is approximately determined by the peak of 
$\hat k \,\hat \eta (T_{\rm max}, \hat k)$, which in turn can be estimated from its leading-log behaviour, 
\begin{equation}
\label{eq:LL_h_c_spectrum}
\hat k\, \hat\eta_{\rm LL} (T_{\rm max}, \hat k) =   
\frac{\hat k^2}{{\rm e}^{\hat k}-1} f(T_{\rm max})\,,
\end{equation} 
leading to  
\begin{equation}
\label{eq:hatkmax_h_c}
\hat k_{\rm peak}^{h_c^{\rm CGMB}} \approx 2 + W(-2/{\rm e}^2) \simeq 1.59 \,.
\end{equation}
Therefore, $h_c^{h_c^{\rm CGMB}}(f)$ is expected to peak approximately around 
\begin{eqnarray}
\label{req:peak_freq_cgmb_h_c}
f_{\rm peak}^{h_c^{\rm CGMB}} \approx  \frac{2 + W(-2/{\rm e}^2) }{4 + W(-4/{\rm e}^4) }\,f_{\rm peak}^{\Omega_{\rm CGMB}} 
\simeq \frac{1.59}{3.92}\,f_{\rm peak}^{\Omega_{\rm CGMB}} 
\simeq   30\,{\rm GHz} \ 
\left[\frac{g_{*s}(T_{\rm max})}{106.75}\right]^{-1/3}
\,.
\end{eqnarray}

More refined expressions for the frequencies $f^{\Omega_{\rm CGMB}}_{\rm peak}$ and $f^{h_c^{\rm CGMB}}_{\rm peak}$ for which $\Omega_{\rm CGMB}$ and $h_c^{\rm CGMB}$ become maximal in an arbitrary theory, taking into account the complete leading-order expression for $\hat k^3\,\hat\eta (T_{\rm max},\hat k)$  and $\hat k\,\hat\eta (T_{\rm max},\hat k)$, can be obtained as follows.
First, one can determine numerical approximations for  each loop function appearing in Eq.~\eqref{eq:etafull}, multiplied by $\hat k^3$ or $\hat k$, around its critical point, using a Taylor expansion of order 2 with coefficients determined numerically. While the functions $\hat\eta_{\rm HTL}$, $\hat\eta_{sf}$ have maxima, the loop functions $\hat\eta_{gg},$ $\hat\eta_{sg}$, $\hat\eta_{fg}$ have minima, and all the critical points have similar values of $\hat k$. As the latter three functions give only relatively small corrections to   $\hat\eta_{\rm HTL}$, it turns out that the location of the maxima of $\hat k^3\,\hat\eta$ and $\hat k\,\hat\eta$ are close to all the maxima and minima associated with the different loop functions, and hence doing Taylor expansions around the critical points allows for accurate determinations of the peak frequencies.

Using the notation of Sec.~\ref{sec:prod_gen_plasma}, we first define shorthands for the coefficients of the loop functions appearing in Eq.~\eqref{eq:etafull} and evaluated at $T=T_{\rm max}$:
\begin{align}\begin{aligned}
 c_{gg}=&\,\frac{1}{2}\sum_{n=1}^{{\cal N}_g}g_n(T_{\rm max})^2 N_n\,T_{n,\rm Ad},  &
 c_{sg}=&\,\sum_{n=1}^{{\cal N}_g}g_n(T_{\rm max})^2 N_n\,\sum_{\hat{\i}}   T_{n,\hat{\i}}, \\
 c_{fg}=&\,\frac{1}{2}\sum_{n=1}^{{\cal N}_g}g_n(T_{\rm max})^2 N_n\,\sum_{\hat \alpha}   T_{n,\hat{\alpha}}, &
 c_{sf}=&\,\frac{1}{4}\sum_{i\alpha\beta}|y^i_{\alpha\beta}(T_{\rm max})|^2. \\
\end{aligned}\end{align}
From this one may estimate the values of $\hat{k}^\Omega_{\rm peak}$, $\hat{k}^{h_c}_{\rm peak}$ as:
\begin{align}\label{eq:peakformula}
 \hat{k}^{\Omega/h_c}_{\rm peak}(T_{\rm max})\approx
\frac{
\sum\limits_{n=1}^{{\cal N}_g} N_n \hat{m}^2_n(T_{\rm max}) \,f^{\Omega/h_c}_{\rm HTL}(\hat m^2_{n}(T_{\rm max}))\, \hat k^{\Omega/h_c}_{\rm HTL}(\hat m^2_{n}(T_{\rm max}))+\sum\limits_{(xy)} c_{xy} \,f^{\Omega/h_c}_{xy} \hat k^{\Omega/h_c}_{xy}}{\sum\limits_{n=1}^{{\cal N}_g} N_n \hat{m}^2_n(T_{\rm max}) \,f^{\Omega/h_c}_{\rm HTL}(\hat m^2_{n}(T_{\rm max}))+\sum\limits_{(xy)} c_{xy} \,f^{\Omega/h_c}_{xy}  }.
\end{align}
Above, we have noted the explicit dependence on $T_{\rm max}$.
The quantities $f^{\Omega/h_c}_{\rm HTL}(\hat m^2_{n})$ and $\hat k^{\Omega/h_c}_{\rm HTL}(\hat m^2_{n})$ correspond to the second derivatives at the extrema and the location of the latter, respectively, for the functions 
\begin{align}
 \label{eq:HTLfunctions}
&\frac{\hat k^4}{16\pi(e^{\hat{k}}-1)}\,\log\left(1+4\frac{\hat{k}^2}{\hat m^2_n}\right), & &\frac{\hat k^2}{16\pi(e^{\hat{k}}-1)}\,\log\left(1+4\frac{\hat{k}^2}{\hat m^2_n}\right).
\end{align}
The aforementioned quantities depend on $\hat m^2_n$, and we estimated the dependency with a numerical fit.
The sum in $(xy)$ goes over the pairs $(gg)$, $(sg)$, $(fg)$, $(sf)$. The quantities $f^{\Omega/h_c}_{xy}$ are numbers related to the second derivatives of the functions $\hat{k}^3\,\eta_{xy}(\hat k)$ and  $\hat{k}\,\eta_{xy}(\hat k)$ at their respective extrema, while the $\hat  k^{\Omega/h_c}_{xy}$ are the locations of these extrema. 
We have computed the above numbers and functions as:
\begin{align}\label{eq:peakformula2}\begin{aligned}
   f^{\Omega}_{\rm HTL}(\hat m^2_n)= &\, 0.188097\,+{0.0689783}\,{(\hat m^2_n)^{-1/5}}    -{0.34907}\,{(\hat m^2_n)^{-1/10}},\\
   k^{\Omega}_{\rm HTL}(\hat m^2_n)= &\, 4.08639+0.266716 \,(\hat m^2_n)^{1/3}+0.0451801\, (\hat m^2_n)^{1/6},\\
      f^{h_c}_{\rm HTL}(\hat m^2_n)= &\, 0.0428322\, +{0.0103571}\,{(\hat m^2_n)^{-1/5}}-{0.0680904}\,{(\hat m^2_n)^{-1/10}},\\
   k^{h_c}_{\rm HTL}(\hat m^2_n)= &\, 1.77979+0.400398\, (\hat m^2_n)^{1/3}+0.181815\, (\hat m^2_n)^{1/6}.
\end{aligned}\end{align}
\begin{align}\label{eq:peakformula3}\begin{aligned}
 f^\Omega_{gg} = &\,0.0398501, & f^\Omega_{s,g} = &\,0.00996252, & f^\Omega_{f,g} = &\,0.00949888, & f^\Omega_{s,f} = &\,-0.00476284, \\
  k^\Omega_{gg} = &\,4.42592, & k^\Omega_{s,g} = &\,4.42592, & k^\Omega_{f,g} = &\,4.6548, & k^\Omega_{s,f} = &\,4.07387,\\
   f^{h_c}_{gg} = &\,0.00599662, & f^{h_c}_{s,g} = &\,0.00149915, & f^{h_c}_{f,g} = &\,0.00114337, & f^{h_c}_{s,f} = &\,-0.000980399, \\
  k^{h_c}_{gg} = &\,2.3873, & k^{h_c}_{s,g} = &\,2.3873, & k^{h_c}_{f,g} = &\,2.74138, & k^{h_c}_{s,f} = &\,2.11078,
\end{aligned}\end{align}

To obtain the peak frequencies, as follows from Eq.~\eqref{eq:Omega_CGMB_analytic_expression} one simply has to use:
\begin{align}\label{eq:peakformulaf}\begin{aligned}
 f_{\rm peak}^{\Omega_{\rm CGMB}}(T_{\rm max})=\frac{\hat k_{\rm peak}^\Omega(T_{\rm max})}{2\pi}\,\left(\frac{g_{*s}(\rm fin)}{g_{* s}(T_{\rm max})}\right)^{1/3}\,T_0\approx79.8\,{\rm GHz}\,\left[\frac{\hat k^\Omega_{\rm peak}(T_{\rm max})}{4.22}\right]\,\left[\frac{106.75}{g_{*s}(T_{\rm max})}\right]^{1/3},\\
  f_{\rm peak}^{h_c^{\rm CGMB}}(T_{\rm max})=\frac{\hat k_{\rm peak}^{h_c}(T_{\rm max})}{2\pi}\,\left(\frac{g_{*s}(\rm fin)}{g_{* s}(T_{\rm max})}\right)^{1/3}\,T_0\approx40.5\,{\rm GHz}\,\left[\frac{\hat k^{h_c}_{\rm peak}(T_{\rm max})}{2.14}\right]\,\left[\frac{106.75}{g_{*s}(T_{\rm max})}\right]^{1/3},
\end{aligned}\end{align}
where the values of $\hat k_{\rm peak}^\Omega(T_{\rm max})$, $\hat k_{\rm peak}^{h_c}(T_{\rm max})$ can be computed from Eqs.~\eqref{eq:peakformula}, \eqref{eq:peakformula2} and \eqref{eq:peakformula3}. In Eq.~\eqref{eq:peakformulaf} we normalized the values of $\hat k^\Omega_{\rm peak}(T_{\rm max})$ and $\hat k^{h_c}_{\rm peak}(T_{\rm max})$ by their values in the SM at $T_{\rm max}=10^{16}$ GeV. 

The 
above formulae, obtained by exploiting the approximation~\eqref{eq:Omega_CGMB_analytic_expression},  reproduce with an accuracy better than 1\% the peak 
frequencies of the different models considered (see later in Table~\ref{tab:peaks}) computed according to Eq.~\eqref{eq:Omega_CGMB_full_expression} from the integration over temperature of the full production rate. In the models to be analyzed in the following sections, we find that the differences in values of $\hat{k}^{\Omega/h_c}_{\rm peak}$ are smaller than the variation of $g_{*s}(T_{\rm max})$. For example, $\hat{k}^{\Omega}_{\rm peak}$ only changes by 5\% between the SM and the MSSM, while $\hat{k}^{h_c}_{\rm peak}$ changes by 15\%, and $g_{*s}(T_{\rm max})$ varies by 30\%. Thus the effect of $g_{*s}(T_{\rm max})$ dominates in Eq.~\eqref{eq:peakformulaf}, which predicts that peak frequencies decrease with growing values of $g_{*s}(T_{\rm max})$. The reduced variability of  $\hat{k}^{\Omega/h_c}_{\rm peak}$ can be understood from  \eqref{eq:peakformula} by noting that the model dependence enters through the values of the Debye masses $\hat m^2_n$ appearing in the HTL contributions, and through the coefficients $c_{xy}$. In general, the HTL contributions dominate both the numerator and denominator, so that there is an approximate cancellation of the leading model dependence through $\hat m^2_n$ in Eq.~\eqref{eq:peakformula}.

Similarly, we can also give approximate formulae for the values of  $\Omega_{\rm CGMB}$ and $h_c^{\rm CGMB}$ at their respective peak frequencies:
\begin{align}\label{eq:h2Omegaapprox}\begin{aligned}
 &h^2\Omega_{\rm CGMB}(f_{\rm peak}^{\Omega_{\rm CGMB}}(T_{\rm max}))\approx\,2.72\times10^{-8}\left(\frac{g_{*s}(T_{\rm max})}{106.75}\right)^{-11/6}\,\frac{T_{\rm max}}{M_P}\,\times\\
 &\times\left(\sum_n N_n \hat m^2_n\left(g^\Omega_{\rm HTL}(\hat m^2_n)\!+\!\frac{f^\Omega_{\rm HTL}(\hat m^2_n)}{2}(\hat{k}^{\Omega}_{\rm peak}\!-\!\hat{k}^{\Omega}_{\rm HTL}(\hat m^2_n))^2\right)+\sum_{xy}c_{xy}\left(g^\Omega_{xy}+\frac{f^\Omega_{xy}}{2}(\hat{k}^{\Omega}_{\rm peak}\!-\!\hat{k}^{\Omega}_{xy})^2\right)\right),\end{aligned}\end{align}
 \begin{align}\label{eq:hcapprox}\begin{aligned}
  &(h_c(f_{\rm peak}^{h_c^{\rm CGMB}}(T_{\rm max})))^2\approx\,1.21\times10^{-64}\left(\frac{g_{*s}(T_{\rm max})}{106.75}\right)^{-7/6}\,\frac{T_{\rm max}}{M_P}\,\times\\
 & \times\left(\sum_n N_n \hat m^2_n\left(g^{h_c}_{\rm HTL}(\hat m^2_n)\!+\!\frac{f^{h_c}_{\rm HTL}(\hat m^2_n)}{2}(\hat{k}^{h_c}_{\rm peak}\!-\!\hat{k}^{h_c}_{\rm HTL}(\hat m^2_n))^2\right)+\sum_{xy}c_{xy}\left(g^{h_c}_{xy}+\frac{f^{h_c}_{xy}}{2}(\hat{k}^{h_c}_{\rm peak}\!-\!\hat{k}^{h_c}_{xy})^2\right)\right).
\end{aligned}\end{align}
In the formulae above, $\hat m^2_n$ are meant to be evaluated at $T=T_{\rm max}$, and aside from the functions and constants of Eqs.~\eqref{eq:peakformula2} and \eqref{eq:peakformula3}, we introduced the quantities $g^{\Omega/h_c}_{\rm HTL}(\hat m^2_n)$ -- related to the peak values of the functions of Eq.~\eqref{eq:HTLfunctions} -- and  $g^{\Omega/h_c}_{xy}$, which correspond to the peak values of $\hat{k}^3\,\eta_{xy}(\hat k)$ and  $\hat{k}\,\eta_{xy}(\hat k)$. These functions and constants are given next:
\begin{align}\label{eq:peakformula0}\begin{aligned}
   g^{\Omega}_{\rm HTL}(\hat m^2_n)= &\,-0.821174\, -{0.314064}(\hat m^2_n)^{-1/5}+{1.53941}(\hat m^2_n)^{-1/10},\\
      g^{h_c}_{\rm HTL}(\hat m^2_n)= &\, -0.100919\, -{0.0264676}(\hat m^2_n)^{-1/5}+{0.163792}(\hat m^2_n)^{-1/10},\\
\end{aligned}\end{align}
\begin{align}\label{eq:peakformula02}\begin{aligned}
 g^\Omega_{gg} = &\,-0.174074, & g^\Omega_{sg} = &\,-0.0435185, & g^\Omega_{fg} = &\,-0.0416378, & g^\Omega_{sf} = &\,0.0189974, \\
   g^{h_c}_{gg} = &\,-0.0153403, & g^{h_c}_{sg} = &\,-0.00383507, & g^{h_c}_{fg} = &\,-0.00310212, & g^{h_c}_{sf} = &\,0.00205386.
\end{aligned}\end{align}

Again, we find that the approximate expressions~\eqref{eq:h2Omegaapprox} and \eqref{eq:hcapprox}  reproduce with an accuracy better than 3\%  (for $\Omega_{\rm CGMB}$) and 1\% (for $h_c^{\rm CGMB}$) the results for the different models (see later in Table~\ref{tab:peaks}) to be analyzed next. Aside from the model-dependence coming from  $g_{*s}(T_{\rm max})$, we note that, as opposed to the case of $\hat k^{\Omega/h_c}_{\rm peak}$, there is no approximate cancellation of the dependence on the values of $\hat m^2_n$. Thus, we anticipate more variability of the peak values of $h^2\Omega_{\rm CGMB}$ and $h_c^{\rm CGMB}$ across models for a fixed $T_{\rm max}$. Nevertheless, in weakly coupled extensions of the SM it is expected that the variations in $\hat m^2_n$ and $c_{xy}$ will be subleading with respect to changes in $g_{*s}(T_{\rm max})$. Under this assumption, the leading model dependence in the peak frequencies and in the peak values of $h^2\Omega_{\rm CGMB}$ and $h_c^{\rm CGMB}$ can be captured by $g_{*s}(T_{\rm max})$ alone, so that for a general model one has:
\begin{align}\label{eq:RelationPeaks}
 f^{\Omega/h_c}_{\rm peak}(T_{\rm max})\approx\left(\frac{g_{*s,\rm SM}(T_{\rm max})}{g_{*s}(T_{\rm max})}\right)^{1/3} f^{\Omega/h_c}_{\rm peak,SM}(T_{\rm max}),
\end{align}
(as follows from Eq.~\eqref{eq:peakformulaf} when one ignores changes in $\hat{k}^{\Omega/h_c}_{\rm peak}$), and
\begin{align}\label{eq:RelationOmegaHc}\begin{aligned}
 \Omega_{\rm CGMB}(f^{\Omega}_{\rm peak}(T_{\rm max}))\approx&\,\left(\frac{g_{*s,\rm SM}(T_{\rm max})}{g_{*s}(T_{\rm max})}\right)^{11/6} \Omega_{\rm CGMB,SM}(f^{\Omega}_{\rm peak,SM}(T_{\rm max})),\\
 h_c^{\rm CGMB}(f^{h_c}_{\rm peak}(T_{\rm max}))\approx&\,\left(\frac{g_{*s,\rm SM}(T_{\rm max})}{g_{*s}(T_{\rm max})}\right)^{7/12} h_c^{\rm CGMB,SM}(f^{h_c}_{\rm peak,SM}(T_{\rm max})),
\end{aligned}\end{align}
as implied by Eqs.~\eqref{eq:h2Omegaapprox}, \eqref{eq:hcapprox} when ignoring changes in $\hat m^2_n$, $c_{xy}$.

Throughout our explorations of minimal and non-minimal extensions of the SM in the next subsections, we find that Eq.~\eqref{eq:RelationPeaks} is satisfied with better than 15\% accuracy, while the relation for  $h_c^{\rm CGMB}$ in Eq.~\eqref{eq:RelationOmegaHc} works with better than 30\% accuracy. We find that by  simply comparing with the SM predictions, a measurement of both the peak frequency and the maximum value of $h_c^{\rm CGMB}$ could be used to estimate the hot big bang temperature $T_{\rm max}$ and the value of $g^{1/3}_{*s}(T_{\rm max})$ up to systematic effects below 15\% for $g^{1/3}_{*s}(T_{\rm max})$ and below 40\% for $T_{\rm max}$. These numbers correspond to the MSSM, while for example in SMASH the accuracies would be around 1\% for $g^{1/3}_{*s}(T_{\rm max})$ and 5\% for $T_{\rm max}$.

We also note that Eq.~\eqref{eq:RelationOmegaHc} predicts that, in weakly coupled theories, the SM is expected to give the highest power of thermally produced gravitational waves. Naively, in SM extensions one would expect  an increase in the instantaneous rate of production of gravitational waves, as the coefficients of the loop functions in Eq.~\eqref{eq:etafull} will receive additional contributions. However, the presence of additional degrees of freedom also implies that the waves produced at early times will suffer a larger redshifting, as the latter is proportional to $[g_{*s}(T_{\rm emission})/g_{*s}({\rm fin})]^{4/3}$ (note that $g_{*s}(T)^{-4/3}$ is indeed present in Eq.~\eqref{eq:Omega_CGMB_full_expression}). As long as the models are weakly coupled, the effect of redshifting will be dominant, and this is captured by the approximate relations in Eq.~\eqref{eq:RelationOmegaHc}.

To end this section, let us recall that the previous results are valid for $T_{\rm max}<M_P$, for which the gravitons never reach thermal equilibrium and one may ignore the backreaction from graviton annihilations and decays in the rates of Eq.~\eqref{evol1} and \eqref{eq:etafull}.
 For gravitons that were in thermal equilibrium for temperatures above the Planck scale, the prediction for the relative abundance of gravitational waves is that of a blackbody spectrum with an effective temperature obtained by redshifting the decoupling temperature $M_P$ by the expansion of the universe between the decoupling time and the present \cite{Kolb:1990vq}. This follows simply from noting that after decoupling, gravitons would stop interacting and start to propagate freely, with their momenta redshifting due to the expansion. As a consequence of this one has
 \begin{align}\label{eq:OmegaEqCGMB}
  \Omega_{\rm Eq. CGMB}(f)=&\,\frac{16\pi^2}{3M_P^2 H_0^2}\frac{ f^4}{e^{2\pi f/ T_{\rm grav}}-1}, &
T_{\rm grav}=&\, \frac{a(T=M_P)}{a(T=T_0)}\,M_P= \left(\frac{g_{*s}({\rm fin})}{g_{*s}(M_P)}\right)^{1/3} T_0.
 \end{align}
 For the equilibrated spectrum the peak frequencies are given by Eqs.~\eqref{req:peak_freq_cgmb} and \eqref{req:peak_freq_cgmb_h_c},
with $T_{\rm max}$ in these expressions replaced by the decoupling temperature $M_P$,  without any further model-dependence than the one coming from $g_{*s}$. The maxima of the spectra however scale differently with $g_{*s}$ than in Eqs.~\eqref{eq:RelationOmegaHc}. For the equilibrated spectrum one has
 \begin{align}\label{eq:OmegaHcMaxEq}\begin{aligned}
  h^2 \Omega_{\rm Eq. CGMB}(f^{\Omega_{\rm Eq. CGMB}}_{\rm peak})=&\,- \left(\frac{g_{*s}({\rm fin})}{g_{*s}(M_P)}\right)^{4/3}\frac{h^2 T_0^4}{3 \pi ^2 H_0^2 M_P^2} \,W(-{4}/{e^4}) \left(4+W(-{4}/{e^4})\right)^3\\
  =&\,2.23\times10^{-7}\left(\frac{106.75}{g_{*s}(M_P)}\right)^{4/3},\\
 h_c^{\rm Eq. CGMB}(f^{h_c^{\rm Eq. CGMB}}_{\rm peak})=&\,\frac{T_0}{\pi  M_P}\left(\frac{g_{*s}({\rm fin})}{g_{*s}(M_P)}\right)^{1/3}\sqrt{-2 W(-{2}/{e^2}) \left(2+W(-{2}/{e^2})\right)} \\
 =&\,3.49\times10^{-32}\,\left(\frac{g_{*s}({\rm fin})}{g_{*s}(M_P)}\right)^{1/3}.
 \end{aligned}\end{align}

\begin{figure}[t]
\begin{center}
\includegraphics[width=0.75\textwidth]{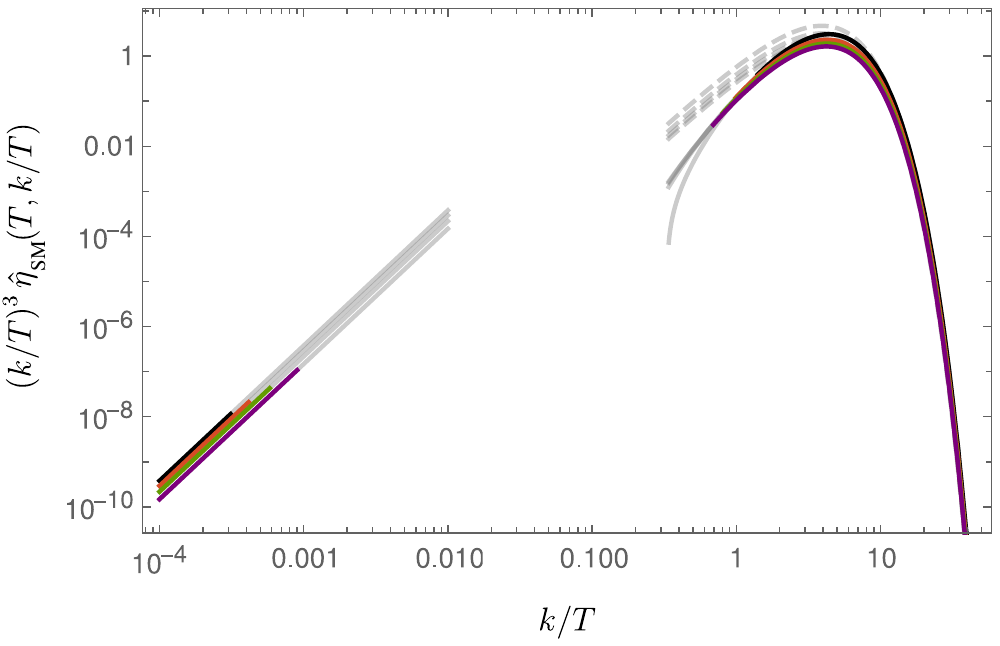}
\vspace{-4ex}
\end{center}
\caption{ Function $(k/T)^3\, \hat\eta_{\rm SM} (T,k/T)$ determining the background of stochastic gravitational waves produced in the thermal 
SM plasma, showing the hydrodynamic contributions (straighter solid lines for smallish $k/T$), the microscopic contributions at full leading order (curved solid lines for higher $k/T$), and their leading-log approximations (dashed lines). The lines are colored for the scales in which the calculations can be trusted, i.e. $k<\alpha^2_1\,T$ for the hydrodynamic contributions, and $k>m_{\rm 3}(T)$ for the microscopic ones. From top to bottom, the temperatures correspond to $T=10^3$ GeV (black), $T=10^8$ GeV (red), $T=10^{13}$ \,GeV (green) and $T=M_P$ (violet). The gauge and Yukawa couplings were evaluated at the renormalization $\bar \mu = 2\pi T$ using the respective 
two-loop renormalization group equations.}
\label{fig:leading_log_vs_complete_leading_order}
\end{figure}


\subsection{CGMB in the SM}
\label{sec:GWSM_SM}
In the SM we assign $n=1,2,3$ to the gauge groups U(1)$_Y$, SU(2)$_L$, SU(3). With the SM matter content one has
\begin{align}\begin{aligned}
N_{\rm species,SM}=&\,\frac{11}{2}, & N_{\rm leptons,SM} = &\, \frac{3}{2},\\
\hat{m}^2_{1,\rm SM}(T)=&\,\frac{11}{6} g_1(T)^2, & \hat{m}^2_{2,\rm SM}(T)=&\,\frac{11}{6} g_2(T)^2, & \hat{m}^2_{3,\rm SM}(T)=&\,2 g_3(T)^2.
\end{aligned}\end{align}
\begin{figure}[h]
\begin{center}
\includegraphics[width=0.75\textwidth]{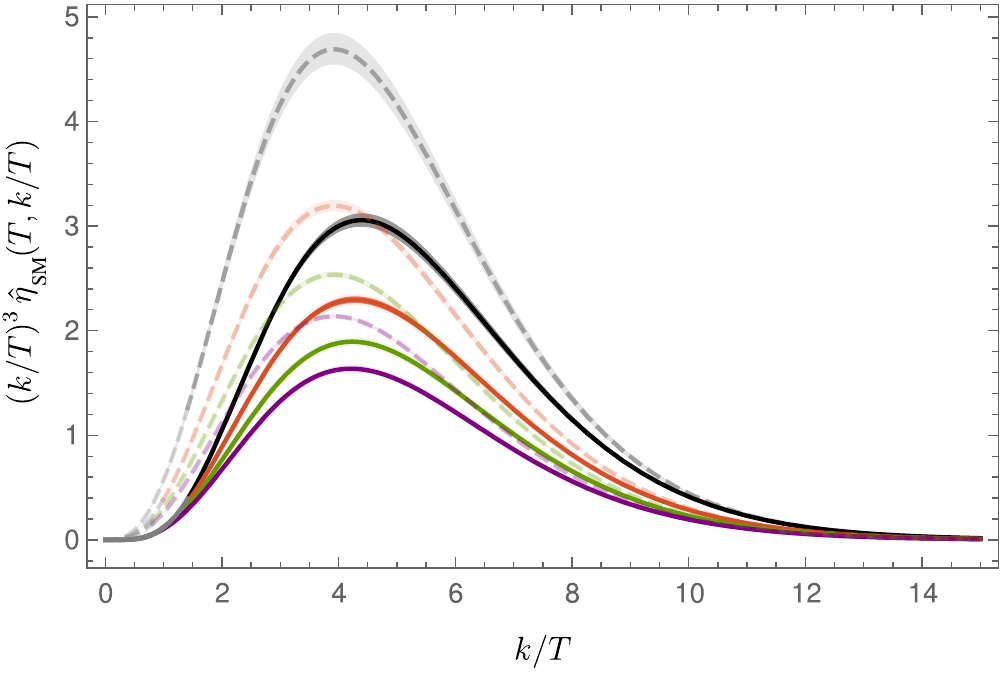}
\vspace{-4ex}
\end{center}
\caption{Microscopic contributions to $(k/T)^3\, \hat\eta_{\rm SM} (T,k/T)$ (solid lines) and their logarithmic approximations (dashed lines).  The lines are colored for the scales in which the calculations can be trusted, i.e.  $k>m_{\rm D3}(T)$. From top to bottom, the temperatures correspond to $T=10^3$ GeV (black), $T=10^8$ GeV (red), $T=10^{13}$ \,GeV (green) and $T=M_P$ (violet). The lines are surrounded by bands corresponding to the effect of varying the renormalization scale by a factor of 2 around the central scale $\bar \mu = 2\pi T$. The maxima of the leading logarithmic approximation occur at $k/T\simeq 3.92$, as anticipated in Eq.~\eqref{eq:hatkmax}. The maxima of the full leading order approximation are slightly shifted towards higher values.}
\label{fig:leading_log_vs_complete_leading_order_micro}
\end{figure}
%
With this one can fix $\bar\eta$ as well as the $\hat\eta_{\rm HTL}$ contribution of Eqs.~\eqref{eq:etafull} and \eqref{eq:etaHTL}. Computing as well the coefficients of the loop functions in terms of the representations and couplings in the SM leads to:
\begin{eqnarray}\label{eq:etaSM}
\hat\eta_{\rm SM}\left(T,\frac{k}{T} \equiv \hat{k}\right)  
\simeq  
 \left\{ 
  \begin{array}{ll}
    \displaystyle \frac{15.51}{g_1^4 \ln(5 / \hat{m}_{1,\rm SM})}, & \;\quad  \hat{k} \lesssim \alpha_1^2,\\
        \, \\
    \displaystyle\frac{}{}\hat{\eta}_{\rm HTL, SM}(T,\hat k)+(3g_2^2+12g_3^2)\eta_{gg}(\hat k ) &\\
   \displaystyle\frac{}{}
  +(g_1^2+3g_2^2)\eta_{sg}(\hat k )+(5g_1^2+9g_2^2+24g_3^2)\eta_{fg}(\hat{k})&  \;\quad \hat{k} \gtrsim {\rm max}\,\{\hat m_n\}.\\
  \displaystyle\frac{}{}+(3|y_t|^2+3|y_b|^2+|y_\tau|^2)\,\eta_{sf}(\hat k ),&
  \end{array}
  \right. 
\end{eqnarray}
In the previous equations we omitted for simplicity the logarithmic $T$-dependence of the couplings $g_i$, $y_i$ and the rescaled Debye masses $\hat{m}_i$. We have also ignored the Yukawa couplings of the lightest fermions.

A comparison between  the leading-log result of Eq.~\eqref{eq:etaLL} applied to the SM and the complete leading-order contribution in Eq.~\eqref{eq:etaSM} is shown in 
Figs. \ref{fig:leading_log_vs_complete_leading_order} and \ref{fig:leading_log_vs_complete_leading_order_micro}. For the calculations we used 2-loop RG equations for the SM couplings in the  $\overline{\rm MS}$ scheme \cite{Luo:2002ey}, supplemented with the $g_3$-dependent three-loop contribution to the running of $g_3$ \cite{Pickering:2001aq}, evaluated at a temperature-dependent renormalization scale. The couplings were fixed at low scales using values $m_t=172.9\, {\rm GeV},m_h=125.10\, {\rm GeV}$ for the physical top and Higgs masses. For the determination of $y_t$ from the top mass we used one-loop electroweak and three-loop QCD threshold corrections \cite{Hempfling:1994ar,Chetyrkin:1999qi,Melnikov:2000qh}, while for computing the Higgs couplings we used the full two-loop effective potential plus appropriate momentum corrections, as in Ref.~\cite{Degrassi:2012ry}. We made different choices of the renormalization scale, $\mu=\kappa 2\pi T$ with $\kappa=1/2,1,2$, in order to estimate theoretical uncertainties. 
Figs.~ \ref{fig:leading_log_vs_complete_leading_order} and \ref{fig:leading_log_vs_complete_leading_order_micro} illustrate how the leading-log result captures the leading-order result quantitatively
for $k/T\gtrsim 10$, whereas it overestimates the latter by a factor around two in the phenomenologically most interesting region $k/T\sim 4$. 
The peak positions of $\hat k^3\hat\eta_{\rm SM}(T,\hat k)$, on the other hand, are shifted slightly, by less then 10\,\%  from the generic value $\hat k_{\rm peak}^{\Omega_{\rm CGMB}}\simeq 3.92$, cf. Eq.~\eqref{eq:hatkmax}, estimated from the leading-log result. The true peaks correspond to values of $\hat k$ slightly about 4, of the order of 4.2 at high temperatures, cf. 
Fig.~\ref{fig:leading_log_vs_complete_leading_order_micro}. The locations of the peaks agree to better than $1$\,\% accuracy with Eqs.~\eqref{eq:peakformula}  and \eqref{eq:peakformulaf}, specialized to the SM. For the peaks of  $\hat k\hat\eta_{\rm SM}(T,\hat k)$ -- relevant for computing the peak frequency of the characteristic amplitude $h_c^{\rm CGMB}$ -- we find $\hat k^{h_c}_{\rm peak}\approx2.1$.
In Fig.~\ref{fig:leading_log_vs_complete_leading_order_micro} we show  with colored bands the variations from the change in renormalization scale; these are noticeably smaller in the full-leading order result and remain below  2\%.

\begin{figure}[h]
\begin{center}
\includegraphics[width=0.65\textwidth]{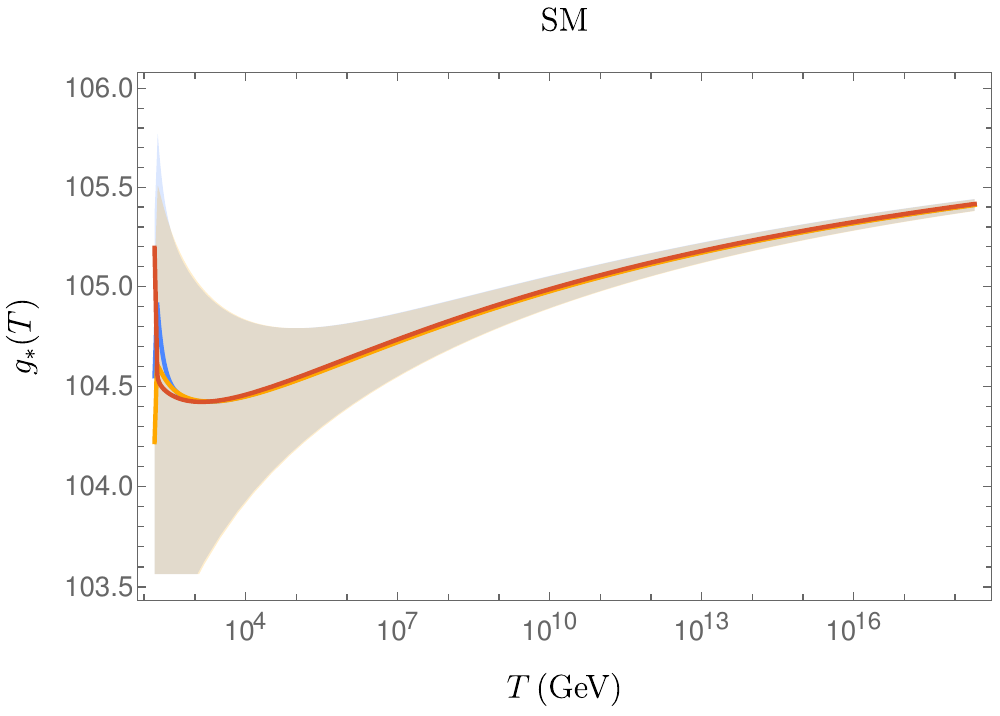}
\end{center}
\caption{$g^{\rm SM}_{*\rho}(T)$ (blue)  $g^{\rm SM}_{* s}(T)$ (orange) and  $g^{\rm SM}_{* c}(T)$ (red). The bands show the uncertainties coming from changing the RG scale within a factor of 2 and from shifting the unknown parameter $q_c$ of the three-loop QCD corrections to the pressure in Ref.~\cite{Kajantie:2002wa} between $-3000$ GeV and $3000 $ GeV. The calculation is expected to lose accuracy near the electroweak crossover around 160 GeV. For lower temperatures one should use the results of Refs.~\cite{Laine:2015kra,Saikawa:2018rcs}. 
}
\label{fig:gsSM}
\end{figure}

For the computation of the spectrum of thermally produced gravitational waves one has to use Eq.~\eqref{eq:Omega_CGMB_full_expression} and carry out the numerical integration. This requires knowledge of the functions $g_{*\rho}(T)$,  $g_{* s}(T)$ and  $g_{* c}(T)$. As reviewed in Appendix~\ref{app:effect_degr_freedom}, all these quantities can be derived from the thermal corrections to the effective potential --which correspond to minus the pressure of the thermal plasma-- and the use of thermodynamical relations. In our calculations we use the full one-loop contributions to the thermal potential, supplemented with three-loop QCD contributions \cite{Kajantie:2002wa}. As we consider gravitational wave production before the electroweak crossover, we can use perturbative results; for lower temperatures one requires more sophisticated techniques~\cite{Laine:2015kra,Saikawa:2018rcs}. The values used for $g^{\rm SM}_{*\rho}(T)$,  $g^{\rm SM}_{* s}(T)$ and  $g^{\rm SM}_{* c}(T)$ are shown in Fig.~\ref{fig:gsSM}.

%
\begin{figure}[t]
\begin{center}
\includegraphics[width=0.8\textwidth]{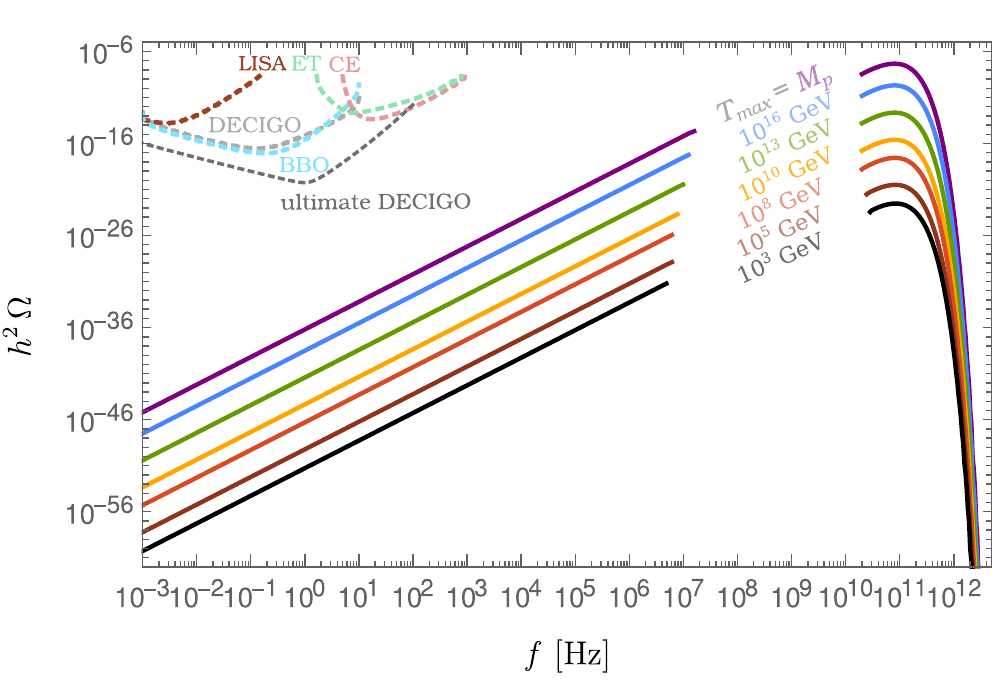}
\end{center}
\caption{Energy fraction of gravitational waves per logarithmic frequency interval from the primordial thermal plasma in the SM, cf. Eq. \eqref{eq:Omega_CGMB_full_expression}. 
From top to bottom, the solid lines correspond to the following maximal temperatures: $T_{\rm max}=M_P$ (violet), $10^{16}$ GeV (blue), $10^{13}$ GeV (green), $10^{10}$ GeV (orange),  $10^{8}$ GeV (red),  $10^{5}$ GeV (dark red),  $10^{3}$ GeV (black). The dashed lines give the projected sensitivities for planned gravitational wave experiments. 
}
\label{fig:h2Omega}
\end{figure}
%

The resulting spectrum of thermally produced gravitational waves is shown in Fig.~\ref{fig:h2Omega} for different values of the maximum temperature, together with the predicted sensitivities of upcoming gravitational wave experiments like the Big Bang Observer (BBO)~\cite{BBO_proposal}, the  Cosmic Explorer (CE)~\cite{Evans:2016mbw}, the Deci-hertz Interferometer Gravitational Wave Observatory 
(DECIGO)~\cite{Seto:2001qf}, the Einstein Telescope (ET)~\cite{Punturo:2010zz}, and LISA~\cite{Audley:2017drz}. The sensitivity projections were  taken from \cite{Schmitz:Zenodo}; for ultimate DECIGO we use the curve in Ref.~\cite{Ringwald:2020vei} based on Ref.~\cite{Kuroyanagi:2014qza}.

\begin{figure}[t]
\begin{center}
\includegraphics[width=0.65\textwidth]{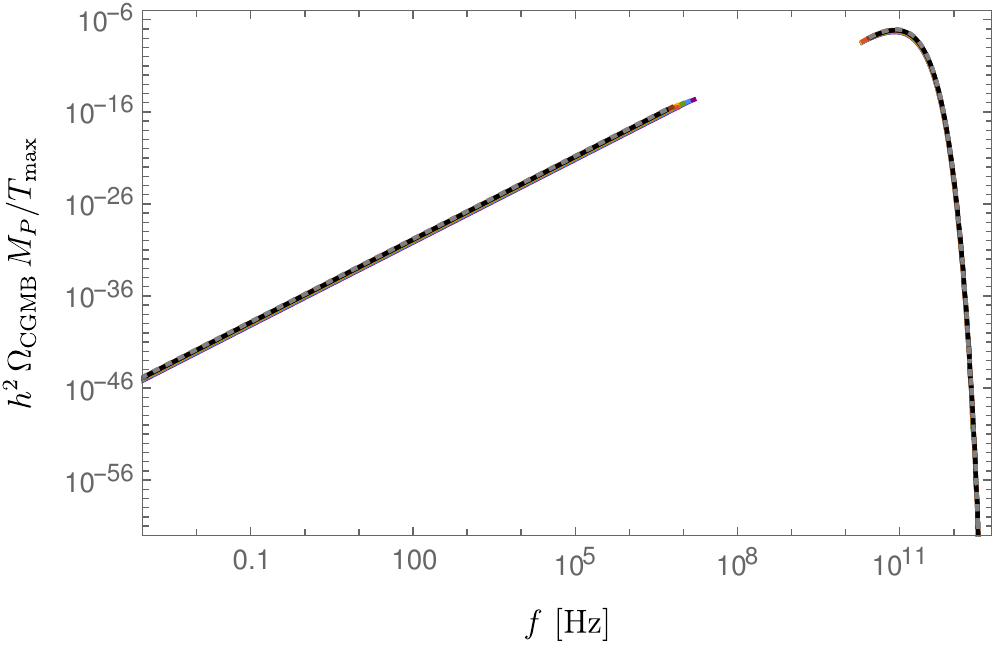}
\includegraphics[width=0.65\textwidth]{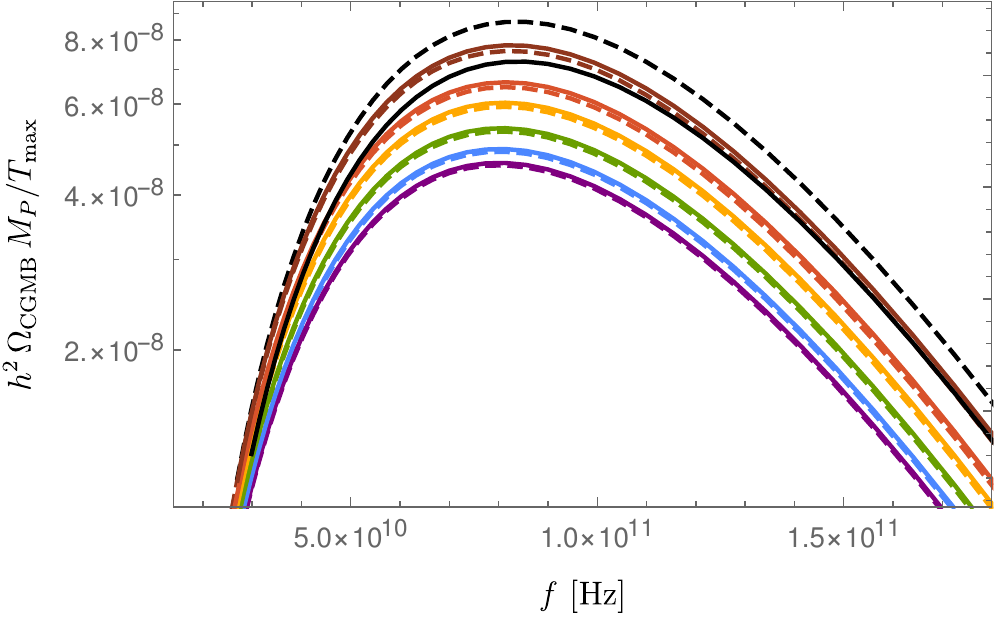}
\end{center}
\caption{Same spectra as in Fig. \ref{fig:h2Omega}, multiplied by $M_P / T_{\rm max}$. The dashed lines give the results obtained using the simplified formula~\eqref{eq:Omega_CGMB_analytic_expression}. 
}
\label{fig:h2Omega_rescaled}
\end{figure}

In Fig.~\ref{fig:h2Omega_rescaled} we show the spectra rescaled by $T_{\rm max}$ and compare these results with the analytic approximation \eqref{eq:Omega_CGMB_analytic_expression}, which predicts a value of $h^2\Omega_{\rm CGMB}/T_{\rm max}$ independent of $T_{\rm max}$ aside from variations in $g_{*s}(T_{\rm max})$.  The figure shows that the analytic prediction gives results with an accuracy better than 3\% near the peak for $T_{\rm max}\gtrsim10^5$ GeV.\footnote{In order to improve \eqref{eq:Omega_CGMB_analytic_expression} also for $T_{\rm max}< 10^5 $\,GeV, one has to into account the  additional negative 
contribution coming from the lower integration boundary $T_{\rm ewco}$ in \eqref{eq:Omega_CGMB_simpified_expression}. 
But for all practical purposes (that is for all ``observable" values of $T_{\rm max}$) the 
leading term coming from the upper integration boundary $T_{\rm max}$ in \eqref{eq:Omega_CGMB_simpified_expression} is 
dominant.}  Within this uncertainty, in accordance with the expectation from Eq.~\eqref{eq:Omega_CGMB_analytic_expression} the absolute value of $\Omega_{\rm CGMB}(f)$ scales approximately linearly with $T_{\rm max}/M_P$. Therefore, a measurement of it determines the maximum temperature of the hot big bang. 
The peaks in the spectra, for different $T_{\rm max}$, occur around $80$\,GHz, less than 10\,\% higher than the generic  estimate~\eqref{req:peak_freq_cgmb} 
based on the analytic approximation~\eqref{eq:Omega_CGMB_analytic_expression} and the leading-log result for $\hat\eta (T,\hat k)$, 
while they are reproduced with an accuracy of the order of 3\% or better (1\% or better for $h_c^{\rm CGMB}$) by the formulae~\eqref{eq:peakformulaf} and \eqref{eq:peakformula}. 

To end this section, let us note that the theoretical uncertainty of the above results for $h^2\Omega_{\rm CGMB}$ in the  SM  is of the order of 0.1\%. This has been estimated by considering the effect of varying the renormalization scale by a factor of 2, and by considering values  between -3000 GeV and 3000 GeV of the unknown  parameter $q_c$ appearing in the three-loop contributions to the QCD pressure of Ref.~\cite{Kajantie:2002wa}. Note that the final uncertainty is one order of magnitude lower than the maximal theoretical uncertainties found for $\hat\eta$; this  is  due to cancellations between the variations of $\hat\eta$ and the effective numbers of degrees of freedom.

\subsection{\boldmath CGMB in minimal BSM models explaining neutrino masses, DM, and the BAU}
\label{sec:GW_NUMSM_SMASH}

So far, our predictions were based on the assumption that the SM is valid up to the Planck scale, and the value of the temperature $T_{\rm max}$ was left unspecified.  However, there is a strong case for BSM physics. It is definitely required to explain neutrino masses and mixing, the origin of the non-baryonic DM, and the BAU. Therefore, we consider now two minimalistic extensions of the SM which solve also these problems. In addition to the latter issues, these models also accommodate realizations of the inflation mechanism, which can address the flatness and horizon problems associated with the observed lack of curvature and the striking homogeneity of the universe. As such, they have all the necessary ingredients to explain the cosmic history of the universe from inflation until the present time; in particular, this means that they give rise to concrete predictions of $T_{\rm max}$ as a result of the post-inflationary reheating dynamics. In turn, this gives refined predictions for the spectrum of gravitational waves originated from the thermal plasma.

The $\nu$MSM~\cite{Asaka:2005an,Asaka:2005pn} extends the SM by three right-handed SM singlet neutrinos,  
which have a GeV scale Majorana neutrino mass and mix with the three active left-handed neutrinos via Yukawa interactions with the SM Higgs field. 
This model may be valid up to the Planck scale. 
Neutrino masses and mixing are generated by the type-I seesaw mechanism~\cite{Minkowski:1977sc,GellMann:1980vs,Yanagida:1979as,Mohapatra:1979ia}. 
DM is comprised by a keV-scale neutrino mass eigenstate, and the BAU is produced 
by a low-scale leptogenesis mechanism involving neutrino oscillations~\cite{Akhmedov:1998qx}. Chaotic inflation can be provided by the Higgs field when allowing for a non-minimal gravitational coupling $S\supset -\int d^4x \sqrt{-g}\,\xi_H\,H^\dagger H\,R$ \cite{Bezrukov:2007ep,Bezrukov:2008ut}.  The CMB observations require a large, nonperturbative value of $\xi_H\sim 4\times 10^{4}\sqrt{\lambda_H}$, with $\lambda_H$ the Higgs self-quartic. Since the latter is determined by the Higgs mass and VEV,  $m_h\simeq 125$\,GeV,  $v\simeq 246$\,GeV, as $\lambda_H\simeq  m_h^2/2v^2$, one has 
$\lambda_H\simeq 0.13$ at low scales, leading to very large values of $\xi_H$. For critical scenarios in which the top mass allows a small $\lambda_H$ at the high scales relevant for inflation, one may get  $\xi_H\sim O(10)$  \cite{Hamada:2014iga,Bezrukov:2014bra}. Values of $\xi_H\gtrsim1$ have been connected with a lack of unitarity \cite{Barbon:2009ya,Burgess:2009ea}, yet arguments against this have been given e.g. in \cite{Bezrukov:2010jz,Gorbunov:2018llf,Ema:2019fdd}. In any case, one may have to pay the prize of uncertain predictions due to unknown nonperturbative corrections to the tree level results.  Nevertheless, ignoring this caveat, the tensor-to-scalar ratio 
$r={\mathcal P}_T/{\mathcal P}_S$ and the maximum temperature of the universe in the $\nu$MSM after reheating from Higgs inflation have been determined as~\cite{Bezrukov:2008ut}
\begin{equation}\label{eq:nuMSMranges}
r\simeq 0.0034, \hspace{6ex} 3.4\times 10^{13}\,{\rm GeV} \lesssim T_{\rm max}^{\nu{\rm MSM}} \lesssim 9.3\times 10^{13}\,{\rm GeV}
\,\left( \frac{\lambda_H}{0.13}\right)^{1/4}\,.
\end{equation}
These temperatures are much below the absolute upper bound following from the CMB constraint on the tensor-to-scalar ratio, $r<0.058$,\footnote{This corresponds to the Planck 2018 results including constraints from BICEP and baryon acoustic oscillations \cite{Akrami:2018odb}.} and the unphysical assumption of instantaneous and maximally efficient reheating to a radiation dominated universe,
\begin{equation}
\label{eq:reheating_constraint_Tmax_main_nuMSM}
T_{\rm max}^{\nu{\rm MSM}}
<   6.6\times 10^{15}\,{\rm GeV}
\,,
\end{equation}
cf. Appendix~\ref{app:upper_bound_on_T_max_ic}. 
The thermal plasma of the $\nu$MSM differs from the thermal plasma 
of the SM only slightly -- from the subleading effects of the Yukawas of the singlet neutrinos, that contribute to the $\eta_{sf}$ term in Eq.~\eqref{eq:etafull} -- and therefore the rate of production can be approximated with that in the SM, Eq.~\eqref{eq:etaSM}. In regards to the calculation of the present day spectrum using Eq.~\eqref{eq:Omega_CGMB_full_expression}, one has to use the values for $g_{*\rho}$, $g_{*s}$, $g_{*c}$ appropriate for the $\nu$MSM. We assume that the singlet neutrinos remain in thermal equilibrium above the electroweak crossover, so that the values of the effective degrees of freedom can be obtained from those of the SM by adding 3 units.

An alternative minimal extension of the SM explaining the origin of DM and the BAU is SMASH~\cite{Ballesteros:2016euj,Ballesteros:2016xej}.  
A SM singlet complex scalar field $\sigma$, which features a spontaneously broken global $U(1)_{\rm PQ}$ Peccei-Quinn (PQ) 
symmetry~\cite{Peccei:1977hh}, and a 
vector-like coloured Dirac fermion $Q$ are added to the field content 
of the $\nu$MSM. Exploiting the PQ mechanism, this model solves the strong CP problem. DM is 
comprised by the axion~\cite{Preskill:1982cy,Abbott:1982af,Dine:1982ah} -- the pseudo Nambu-Goldstone boson of the $U(1)_{\rm PQ}$ breaking~\cite{Weinberg:1977ma,Wilczek:1977pj} -- provided that the 
PQ breaking scale is in the range $1.3\times 10^9\lesssim v_\sigma/{\rm GeV}\lesssim 2.2\times 10^{11}$~\cite{Armengaud:2019uso}. The right-handed neutrinos get their Majorana masses also from spontaneous PQ symmetry breaking. 
The generation of the BAU proceeds via high-scale thermal leptogenesis~\cite{Fukugita:1986hr}. 
Finally, inflation can be accommodated for perturbative values of the non-minimal gravitational couplings~\cite{Ballesteros:2016euj,Ballesteros:2016xej}. Allowing for a non-minimal coupling $\xi_\sigma$ of the PQ field to the Ricci scalar, $S\supset -\int d^4x \sqrt{-g}\,\xi_\sigma\,\sigma^* \sigma\,R$, a mixture of the modulus of the complex PQ field,  $\rho=\sqrt{2}\,|\sigma|$, with
$h$, the neutral component of the SM Higgs doublet in the unitary gauge, is a viable inflaton candidate. Fitting the inflationary predictions to  the observed fluctuations in the CMB relates the size of the non--minimal coupling and the quartic coupling; for the latest Planck data \cite{Akrami:2018odb} this gives \cite{Ringwald:2020vei}
\begin{equation}
7\times 10^{-3} \lesssim \xi_\sigma \simeq 4\times 10^{4}\sqrt{\lambda_\sigma}\lesssim 1.
\end{equation}
The above window was obtained after ensuring a consistent post-inflationary history in which Planck's CMB pivot scale was matched to the appropriate mode during inflation.
The lower bound, $\xi_\sigma\gtrsim 7\times 10^{-3}$, arises from taking into account the upper limit on the 
tensor-to-scalar ratio, $r<0.058$, while the upper bound, $\xi_\sigma\lesssim 1$, arises from perturbativity and unitarity requirements\footnote{See however the above comments and references on the issue of unitarity for Higgs inflation.}.  
It corresponds to a lower limit on the tensor-to-scalar ratio, $r\gtrsim 4\times 10^{-3}$. As a consequence, the quartic coupling should be in the range  $7\times 10^{-13}\lesssim \lambda_\sigma\lesssim 5\times 10^{-10}$. 
The initial conditions for the standard hot big bang cosmology following inflation, non-perturbative preheating and perturbative reheating can be predicted from first principles in SMASH.  The maximum temperature of the thermalized SMASH plasma after reheating is obtained as~\cite{Ballesteros:2016xej}  
\begin{equation}\label{eq:SMASHranges}
8\times 10^9 \,{\rm GeV}\lesssim T^{\rm SMASH}_{\rm max}\lesssim 2\times 10^{10} \,{\rm GeV}
\,.
\end{equation}
Again, this is significantly below the upper bound on $T_{\rm max}$ following from the assumption of instant reheating and the CMB constraint $r<0.058$, which in this case gives
\begin{equation}
\label{eq:reheating_constraint_Tmax_main}
T_{\rm max}^{\rm SMASH}
<   6.4\times 10^{15}\,{\rm GeV}
\,,
\end{equation}

In order to calculate the CGMB in SMASH we can use the general expressions of Section~\ref{sec:prod_gen_plasma}. This requires knowing the BSM Yukawa couplings in SMASH. Stability in the $\sigma$ direction demands small couplings for the RH neutrinos \cite{Ballesteros:2016xej}, whose effect will be ignored as in the $\nu$MSM; this leaves the Yukawa couplings of the exotic vector quark. Assuming a small interaction with the down quarks (we consider the SMASH realization with  hypercharge $1/3$  for $Q$, in which such mixing allows the $Q$s to decay before nucleosynthesis), one has
\begin{align}
 {\cal L}\supset y_Q\, \sigma \bar Q P_L Q +c.c..
\end{align}
With the matter content in SMASH one has 
\begin{align}\begin{aligned}
N_{\rm species,SMASH}=&\,\frac{35}{6}, & N_{\rm leptons,SMASH} = &\, \frac{3}{2},\\
\hat{m}^2_{1,\rm SMASH}(T)=&\,\frac{35}{18}\, g_1(T)^2, & \hat{m}^2_{2,\rm  SMASH}(T)=&\,\frac{11}{6}\, g_2(T)^2, & \hat{m}^2_{3,\rm  SMASH}(T)=&\,\frac{13}{6}\,g_3(T)^2.
\end{aligned}\end{align}
With this one can fix $\bar\eta$ as well as the $\hat\eta_{\rm HTL}$ contribution of Eqs.~\eqref{eq:etafull} and \eqref{eq:etaHTL}. Computing as well the coefficients of the loop functions in terms of the representations and couplings in the SM leads to:
\begin{eqnarray}\label{eq:etaSMASH}
\hat\eta_{\rm SMASH}\left(T,\frac{k}{T} \equiv \hat{k}\right)  
\simeq  
 \left\{ 
  \begin{array}{ll}
    \displaystyle \frac{14.68}{g_1^4 \ln(5 / \hat{m}_{1,\rm SMASH})}, & \;\quad  \hat{k} \lesssim \alpha_1^2,\\
        \, \\
    \displaystyle\frac{}{}\hat{\eta}_{\rm HTL, SMASH}(T,\hat k)+(3g_2^2+12g_3^2)\eta_{gg}(\hat k ) &\\
   \displaystyle\frac{}{}
  +(g_1^2+3g_2^2)\eta_{sg}(\hat k )+\left(\frac{16}{3}g_1^2+9g_2^2+28g_3^2\right)\eta_{fg}(\hat{k})&  \;\quad \hat{k} \gtrsim {\rm max}\,\{\hat m_n\}.\\
  \displaystyle\frac{}{}+\left(3|y_t|^2+3|y_b|^2+|y_\tau|^2+\frac{3}{2}|y_Q|^2\right)\,\eta_{sf}(\hat k ),&
  \end{array}
  \right. 
\end{eqnarray}

For the computation of $g_{*\rho}$, $g_{*s}$, $g_{*c}$ in SMASH we proceed as in Ref.~\cite{Ringwald:2020vei}. In order to reliably follow the change in degrees of freedom across the PQ phase transition, one has to use an improved daisy resummation of thermal effects compatible with thermal decoupling. At low temperatures, the SMASH theory is matched to the SM plus the real part of $\sigma$ and the nearly massless axion; we include again three-loop QCD corrections plus corrections from the loss of chemical equilibrium of the axion due to its feeble interactions, which imply that the axion population has a different effective temperature than the rest of the plasma. For details, see Ref.~\cite{Ringwald:2020vei}; a summary is given in Appendix~\ref{app:effect_degr_freedom}.  Figure~\ref{fig:gsSMASH} shows results for $g^{\rm SMASH}_{*\rho}(T)$, $g^{\rm SMASH}_{* s}(T)$ and  $g^{\rm SMASH}_{* c}(T)$ for two benchmark points with $r=0.0037$ and $r=0.048$, taken from Ref.~\cite{Ringwald:2020vei}.

\begin{figure}[t]
\begin{center}
\includegraphics[width=0.5\textwidth]{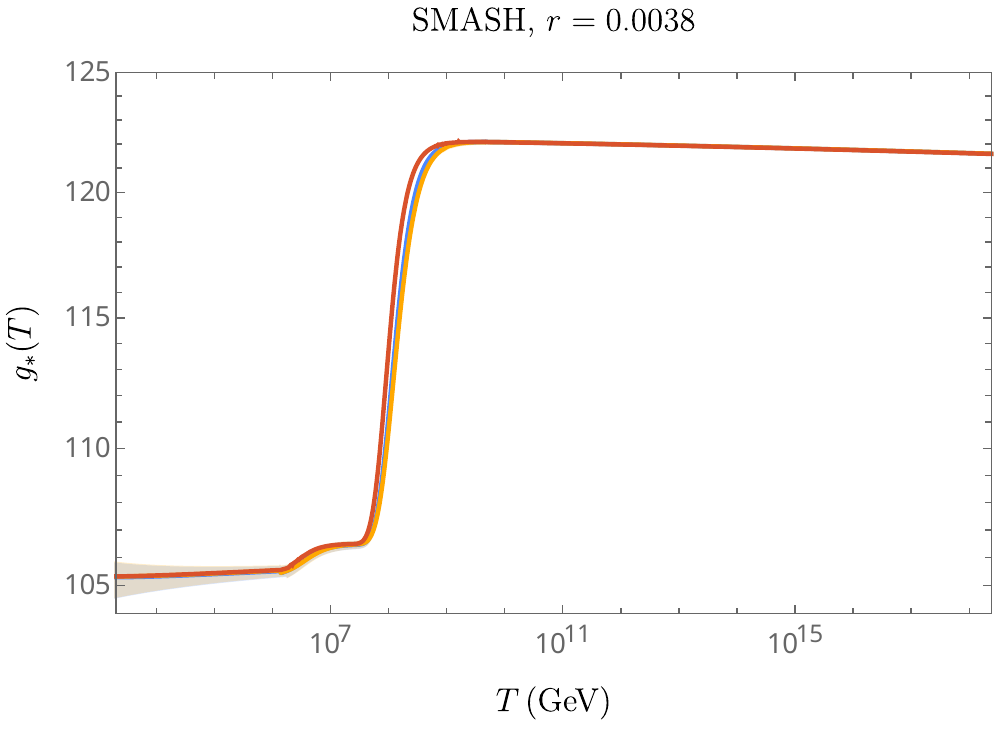}\includegraphics[width=0.5\textwidth]{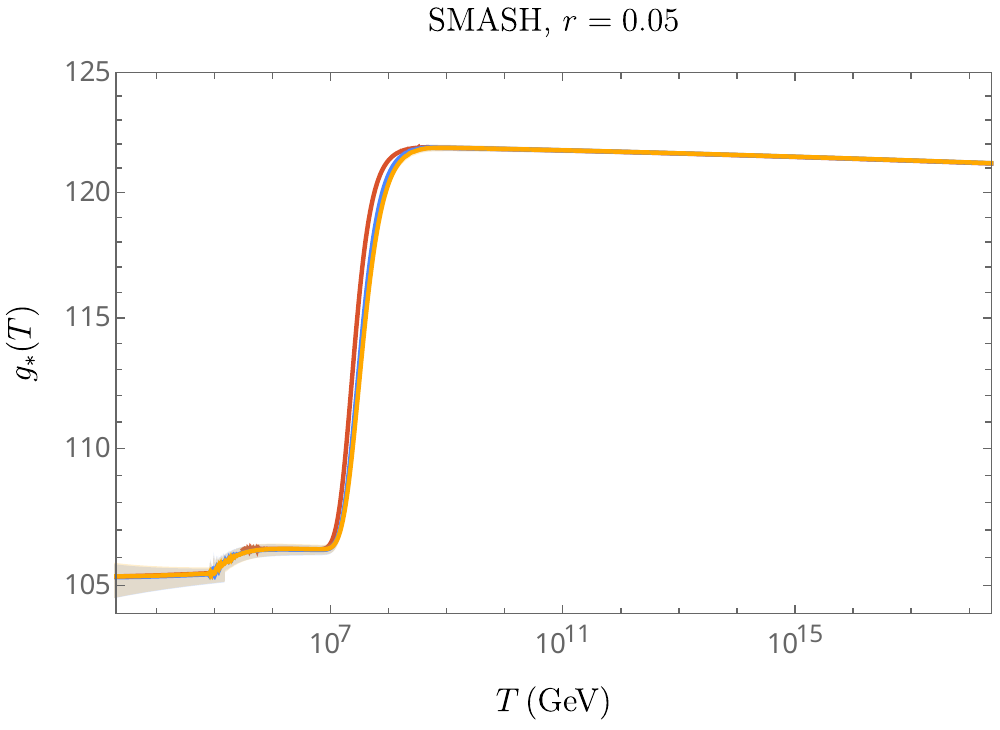}
\end{center}
\caption{$g^{\rm SMASH}_{*\rho}(T)$ (blue)  $g^{\rm SMASH}_{* s}(T)$ (orange) and  $g^{\rm SMASH}_{* c}(T)$ (red) for benchmark points with $r=0.0037$ (left) and $r=0.048$ (right). The bands show the uncertainties coming from changing the RG scale within a factor of 2 and from shifting the unknown parameter $q_c$ of the three-loop QCD corrections to the pressure in Ref.~\cite{Kajantie:2002wa} between $-3000$ GeV and $3000 $ GeV. The calculation is expected to lose accuracy near the electroweak crossover around 160 GeV.
}
\label{fig:gsSMASH}
\end{figure}

\begin{figure}[t]
\begin{center}
\includegraphics[width=0.95\textwidth]{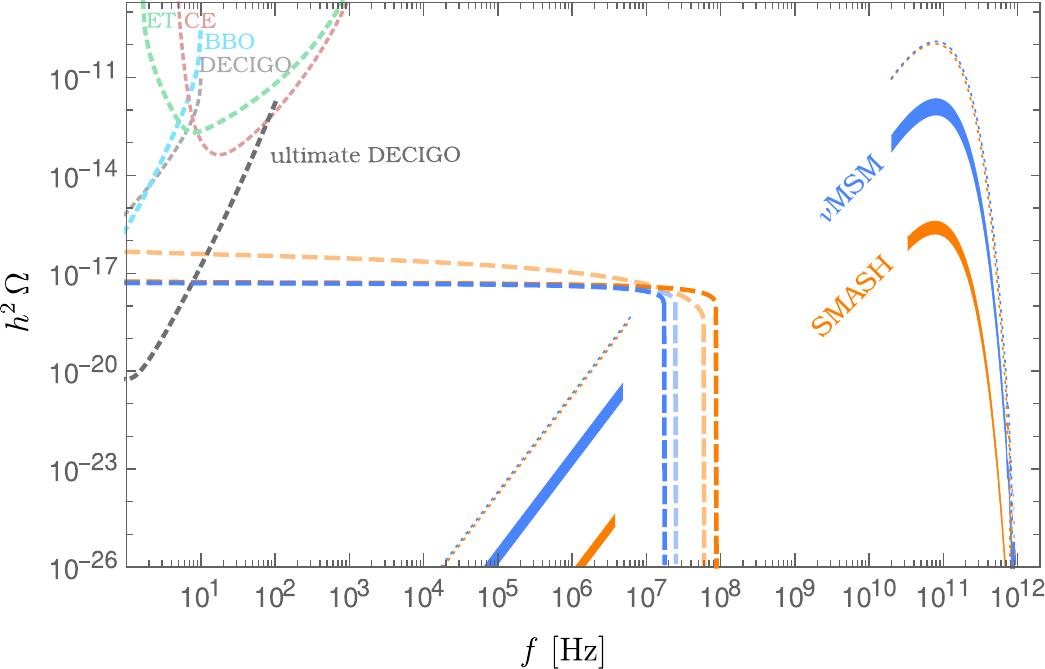}
\end{center}
\caption{Energy fraction of gravitational waves per logarithmic frequency interval  in the $\nu$MSM (blue) and in SMASH (orange). The solid bands correspond  to waves sourced from the primordial thermal plasma in the predicted range of reheating temperatures ($3.4\times 10^{13} \,{\rm GeV}\lesssim T^{\nu\rm MSM}_{\rm max}\lesssim 1.1\times 10^{14} \,{\rm GeV}$ for the $\nu$MSM and $8\times 10^9 \,{\rm GeV}\lesssim T^{\rm SMASH}_{\rm max}\lesssim 2\times 10^{10} \,{\rm GeV}$ for SMASH). The dotted lines give the predictions for the upper bound on the reheating temperature in Eq.~\eqref{eq:Tupmax} inferred from current CMB constraints. 
The dashed lines correspond to waves sourced by inflationary perturbations in the range allowed in each model.
For the $\nu$MSM, the darker dashed line corresponds to $r=0.0034$ and $T^{\nu\rm MSM}_{\rm max}=3.4\times10^{13} \,{\rm GeV}$, while the fainter line corresponds to $r=0.0033, \,T^{ \nu\rm MSM}_{\rm max}=10^{14}\,{\rm GeV}$.  In SMASH, dark orange corresponds to  the minimal value of the tensor-to-scalar-ratio $r=0.0037$, and light orange to the maximum value $r=0.058$.  
}
\label{fig:h2Omega_smash_universal}
\end{figure}

With the previous results for $\hat{\eta}$, $g_{*\rho}$, $g_{*s}$ and $g_{*c}$, and taking into account the ranges of $T_{\rm max}$ in Eqs.~\eqref{eq:nuMSMranges} and \eqref{eq:SMASHranges},  one can use Eq.~\eqref{eq:Omega_CGMB_full_expression} to calculate the predictions for the CGMB in the $\nu$MSM and SMASH. As the results for $\hat\eta$ assume massless fields, in the SMASH case we use  Eq.~\eqref{eq:etaSMASH} at high temperatures, and the SM result of Eq.~\eqref{eq:etaSM} below the temperature at which the axion field decouples. The latter was estimated as in Ref.~\cite{Ringwald:2020vei} by finding the temperature near the critical temperature of the PQ phase transition at which the trace of the stress-energy momentum tensor has a local maximum. The results are shown in Fig.~\ref{fig:h2Omega_smash_universal},
together with the inflationary Cosmic Gravitational Wave Background (iCGWB) due to the tensor modes generated by quantum fluctuations during inflation. For frequencies $f \gtrsim 10^{-16}\,{\rm Hz}$, which re-entered the horizon during radiation domination, the iCGWB can be calculated as~\cite{Saikawa:2018rcs},
\begin{equation}
\label{eq:omega_iCGWB}
\Omega_{\rm iCGWB}(f)\simeq \frac{1}{24}\,\Omega_{\gamma}\left[\frac{g_{*\rho}(T_{\rm hc}(f))}{2}\right]\left[\frac{g_{*s}(T_{\rm hc}(f))}{g_{*s}(\rm fin)}\right]^{-4/3}\,{\cal P}_T(f) \,,
\hspace{3ex} {\rm for}\ f \gtrsim 10^{-16}\,{\rm Hz}\,.
\end{equation}
In the equation above ${\cal P}_T(f)$ is the power spectrum of gravitational waves generated during inflation expressed in terms of the present frequency,
\begin{align}
 {\cal P}_T(f)=\,\left.\frac{2 H_{\rm inf}^2}{\pi^2M_P^2}\right|_{a_{\rm inf}H_{\rm inf}=2\pi f},
\end{align}
where $a_{\rm inf}$, $H_{\rm inf}$ are the scale factor and Hubble constant during inflation, which we have computed assuming non-critical inflation for the $\nu$MSM and without resorting to the usual slow-roll approximation, but rather by numerically solving the equation of motion for the inflationary background as a function of the number of efolds \cite{Ballesteros:2014yva,Ballesteros:2016xej}. $T_{\rm hc}(f)$ in Eq.~\eqref{eq:omega_iCGWB} is the temperature at which the mode corresponding to the frequency $f$ re-entered the horizon during radiation domination. It can be obtained by solving \cite{Saikawa:2018rcs}
\begin{align}
 T_{\rm hc}(f)=10^8 {\rm GeV}\times\frac{f}{1.2\,\rm Hz}\times\left[\frac{g_{*s}(\rm fin)}{3.91}\right]^{-1/3}\times [g_{*\rho}(T_{\rm hc}(f))]^{-1/2}\times [g_{*s}(T_{\rm hc}(f)]^{1/3}.
\end{align}
The iCGWB has an upper cutoff corresponding to frequencies that never exited the horizon during inflation. We have approximated this by a sharp feature, yet for frequencies near this threshold our calculations beyond the slow-roll approximation already show a drop in the power spectrum, as can be seen in Fig.~\ref{fig:h2Omega_smash_universal}.\footnote{See Ref.~\cite{Ito:2020neq} for a detailed study of the spectrum in this region beyond the slow-roll approximation.} Note that, for both SMASH and the non-critical $\nu$MSM,  the iCGGWB does not overlap with the peak in the CGMB, so that both sources of gravitational waves -- inflationary perturbations and thermal processes -- become distinguishable if experiments reach the appropriate sensitivity.

In the case of SMASH, the theoretical uncertainty in the calculations of $h^2\Omega_{\rm CGMB}/T_{\rm max}$ is of the order of 2\%. As before, this quantity corresponds to the variations of the results under changes of the renormalization scale and the unknown  three-loop contributions to the QCD pressure.

In our calculations we find that at sufficiently high temperatures $\hat k^3\hat\eta_{\rm SMASH}$ peaks near 
$\hat{k}^\Omega_{\rm peak}\sim 4.2$, and within 1\% range of the SM results at the same temperatures. Similarly we get $\hat{k}^{h_c}_{\rm peak}\sim 2.1$-2.2, within $1.5\%$ of the corresponding SM results. This means that Eq.~\eqref{eq:RelationOmegaHc} is satisfied with better than $\sim1$\% accuracy.
On the other hand, we find that the relation for $h_c$ in  Eq.~\eqref{eq:RelationOmegaHc} is satisfied with better than 4\% accuracy.  Under the assumption of a SMASH plasma and given a measurement of the peak frequency and the maximum value of $h_c^{CGMB}$, then  Eqs.~\eqref{eq:RelationPeaks} and \eqref{eq:RelationOmegaHc} would allow to estimate $g^{1/3}_{*s}(T_{\rm max})$ and $T_{\rm max}$  within errors around $1\%$ and $5\%$, respectively.
Some details of the peak frequencies and amplitudes for several SMASH benchmark points are given in Table~\ref{tab:peaks}.

\subsection{\boldmath CGMB in the MSSM}
\label{sec:GW_MSSM}

The MSSM goes beyond the SM by adding a fermionic (scalar) superpartners for all the SM bosons (fermions). Additionally, the model contains an extra scalar Higgs doublet with its  corresponding fermionic partners.

Given the matter content of the MSSM, and assuming that the scalar partners of the right-handed leptons contribute to $N_{
\rm leptons}$ in analogous manner to the usual leptons (i.e. adding the contribution ${1}/{2}\sum_{\hat i: T_{n,\hat{i}}=0,n>1}T_{1,\hat i}$ to Eq.~\eqref{eq:Nspeciesleptons}) one obtains:
\begin{align}\begin{aligned}
N_{\rm species,MSSM}=&\,\frac{33}{2}, & N_{\rm leptons,MSSM} = &\, \frac{9}{2},\\
\hat{m}^2_{1,\rm MSSM}(T)=&\,\frac{11}{2} g_1(T)^2, & \hat{m}^2_{2,\rm MSSM}(T)=&\,\frac{9}{2} g_2(T)^2, & \hat{m}^2_{3,\rm MSSM}(T)=&\,\frac{9}{2}g_3(T)^2.
\end{aligned}\end{align}
Computing the coefficients of the loop functions in Eq.~\eqref{eq:etafull}  leads to:
\begin{eqnarray}\label{eq:etaMSSM}
\hat\eta_{\rm MSSM}\left(T,\frac{k}{T} \equiv \hat{k}\right)  
\simeq  
 \left\{ 
  \begin{array}{ll}
    \displaystyle \frac{16.24}{g_1^4 \ln(5 / \hat{m}_{1,\rm MSSM})}, & \;\quad  \hat{k} \lesssim \alpha_1^2,\\
        \, \\
    \displaystyle\frac{}{}\hat{\eta}_{\rm HTL, MSSM}(T,\hat k)+(3g_2^2+12g_3^2)\eta_{gg}(\hat k ) &\\
   \displaystyle\frac{}{}
  +(22g_1^2+42g_2^2+96g_3^2)\eta_{sg}(\hat k )\\
  \displaystyle\frac{}{}+\left(\frac{11}{2}g_1^2+\frac{27}{2}g_2^2+36g_3^2\right)\eta_{fg}(\hat{k})&  \;\quad \hat{k} \gtrsim {\rm max}\,\{\hat m_n\}.\\
  \displaystyle\frac{}{}+(48g_3^2+21g_2^2+11g_1^2\\
  \displaystyle\frac{}{}+9|y_t|^2+9|y_b|^2+3|y_\tau|^2)\,\eta_{sf}(\hat k ),&
  \end{array}
  \right. 
\end{eqnarray}
 Note how in the $sg$ and $fg$ contributions the coefficients in front  of the gauge couplings squared are larger than in the previous models, due to the extra matter fields charged under the gauge interactions. Additionally, one has gauge coupling contributions in the coefficient of the loop function $\eta_{sf}$, as a consequence of the fact that
supersymmetry implies a relation between the Yukawa couplings of the gauge superpartners and the usual gauge couplings. Analogously, the coefficients of the usual Yukawa couplings are larger than before because  supersymmetry relates the usual Yukawa couplings to those of additional interactions involving scalar superpartners.

For our estimates of gravitational wave spectra in the MSSM, we have used the naive value of the effective number of relativistic degrees of freedom,
\begin{align}
 g_{*\rho,\rm MSSM}(T)\approx g_{*s,\rm MSSM}(T)\approx g_{*c,\rm MSSM}(T)\approx 228.75.
\end{align}
The reason for this simplification is that we lack knowledge of the QCD corrections to the pressure coming from scalar superpartners. For our numerical estimates we consider a simple scenario in which the dimensionful parameters in the MSSM that are not present in the SM are assumed to lie around the 2 TeV scale. We further assume that the lightest neutral Higgs state is SM-like, which can be realized with a small neutral Higgs mixing angle $\alpha$ (we take $\cot\alpha=10$) and a heavy pseudoscalar Higgs, taken to have a mass of 2 TeV. Demanding the correct mass of the $Z$ boson in the vacuum implies that the ratio of vacuum expectation values for the Higgs doublets $H_u$ and $H_d$ is $\tan v_u/v_d=8.5$. Given the SM-like low-energy limit, we evolve the couplings with the two-loop SM RG up to a scale of 2 TeV. At this scale  we match the SM to the MSSM by applying appropriate one-loop threshold corrections, and for  higher scales we use the 2 loop MSSM RG equations for the gauge and Yukawa couplings \cite{Jones:1974pg,Jones:1983vk,West:1984dg}. For the calculation of the spectrum of gravitational waves
we use  Eq.~\eqref{eq:etaMSSM} at high temperatures, and the SM result of Eq.~\eqref{eq:etaSM} below 2 TeV.

We give MSSM results for the peak frequencies and amplitudes in some benchmark points in Table~\ref{tab:peaks}.  In the MSSM, the values for $\hat k^{\Omega}_{\rm peak,MSSM}(T_{\rm max})$ for high temperatures lie around 4.40, within 5\% of their SM counterparts. Analogously, one has $\hat k^{h_c}_{\rm peak,MSSM}(T_{\rm max})\approx2.1$-2.2, within  15\% of the SM values for the same temperatures. We thus find that Eq.~\eqref{eq:RelationPeaks} holds with an accuracy better than 15\%, while the relation for $h_c$ in Eq.~\eqref{eq:RelationOmegaHc} holds up to deviations that remain below  30\%. Asuming an MSSM plasma and a hypothetical measurement of the peak frequency and the maximum value of $h_c^{\rm CGMB}$, then we find that Eqs.~\eqref{eq:RelationPeaks} and \eqref{eq:RelationOmegaHc} would allow to estimate $g^{1/3}_{*s}(T_{\rm max})$ and $T_{\rm max}$  within deviations below 5\% and 40\%, respectively.

%% file: observational_constraints_on_CGMB.tex
\section{Observational constraints on the CGMB}
\label{sec:observational_constraints}
\setcounter{equation}{0}

\subsection{Dark radiation constraint on the CGMB}
\label{sec:dark_radiation_constraint}

The CGMB acts as an additional dark radiation field in the universe. 
Any observable capable of probing the expansion rate of the universe, and hence its energy density, has therefore the potential ability to constrain the CGMB energy density ${\rho_{\rm CGMB}}$ present in that moment.
BBN and the process of photon decoupling of the CMB yield a very precise measurement of $H$, when the universe had a temperature of $T_{\rm BBN} \sim 0.1$ MeV and $T_{\rm CMB} \sim 0.3$ eV, respectively. 
A constraint on the presence of `extra' radiation is usually expressed in terms of an extra effective number of neutrinos species, 
$\Delta N_{\nu}$, 
\begin{equation}
 	\Delta \rho_\mathrm{rad} (T)=  \frac{\pi^2}{30}\, \frac{7}{4}\, \Delta N_{\nu} \, T^4\,. 
 \end{equation} 
Since the energy density in the CGMB must satisfy 
$\rho_{\rm CGMB}(T) \leq \Delta \rho_\mathrm{rad}(T)$, 
one finds a constraint on the CGMB energy density redshifted to today in terms of the number of extra neutrino species,
\begin{eqnarray}
\label{ConsRhoBBN}
h^2 \, \int\limits_0^\infty \frac{{\rm d}f}{f} \,  \Omega_{\rm CGMB}(f)
= h^2\, \frac{\rho_{\rm CGMB}^{(0)}}{\rho_c^{(0)}} 
\leq  h^2\Omega_{\gamma}
\left(\frac{g_{*s}({\rm fin})}{g_{*s}(T)}\right)^{4/3} \frac{7}{8}\,\Delta N_{\nu}
\simeq 5.7\times 10^{-6} \,\Delta N_{\nu} 
\,,
\end{eqnarray}
where  $\rho_c^{(0)} = 3 H_0^2 M_P^2$, and we have used $g_{*s}({\rm fin})\simeq 3.931$ and $g_{*s}(T = \mathrm{MeV}) \approx 10.75$. This bound corresponds roughly to a direct bound on the CGMB energy fraction per logarithmic frequency interval, 
\begin{eqnarray}
\label{ConsRhoBBN_approx}
h^2\,   \Omega_{\rm CGMB}(f) \lesssim h^2\,   \Omega_{\rm CGMB}(f_{\rm peak}^{\Omega_{\rm CGMB}}) < 
  5.7\times 10^{-6} \,\Delta N_{\nu} 
\,,
\end{eqnarray}
because it has a large width of order the peak frequency itself. 

The latest BBN constraints on $\Delta N_\nu$ can be found in Ref.~\cite{Fields:2019pfx}. The $^4\mathrm{He}$ alone is not very constraining due to degeneracies with the baryon-to-photon ratio $\eta_B$, so that the best constraints come from combining BBN measurements of $^4\mathrm{He}$ and deuterium abundances with CMB results. In this case Ref.~\cite{Fields:2019pfx} finds
$\Delta N_\nu<0.3$ at 95\%  implying, from Eq.~(\ref{ConsRhoBBN}),  
$
h^2 \, \rho_{\rm CGMB}^{(0)} / \rho_c^{(0)} < 1.7 \times 10^{-6}$.

A similar bound is obtained from other inferences from CMB~\cite{Smith:2006nka,Sendra:2012wh,Pagano:2015hma}. In particular, the analysis of Ref.~\cite{Pagano:2015hma} uses Planck data, together with CMB lensing, baryon acoustic oscillations and also deuterium abundances, and finds a constraint that goes down to 
\begin{equation}
\label{eq:dark_radiation_constraint}
h^2 \, \rho_{\rm CGMB}^{(0)} / \rho_c^{(0)}  < 1.2 \times 10^{-6}.
\end{equation} 
Not surprisingly, this is comparable to what is obtained from the BBN analysis in Ref.~\cite{Fields:2019pfx}, which also uses CMB data to pin down the baryon to photon ratio $\eta_B$. However, Ref.~\cite{Pagano:2015hma} only analyses adiabatic initial conditions. From the results of Refs.~\cite{Smith:2006nka,Sendra:2012wh}, one can infer that there is a gain when imposing homogeneous initial conditions, due to the breaking of degeneracies with neutrino parameters~\cite{Caprini:2018mtu}. This has been confirmed by Ref.~\cite{Clarke:2020bil}, which under the hypothesis of GW with homogeneous initial conditions finds
\begin{equation}\label{eq:DRstringent}
h^2 \, \rho_{\rm CGMB}^{(0)} / \rho_c^{(0)}  
< 2.9 \times 10^{-7}\,.
\end{equation}

Finally, it should be noted that, in the CMB context, the bound in Eq.~\eqref{ConsRhoBBN} is often quoted in terms of $\Delta N_{\rm eff}$, the effective number of extra neutrino species present in the thermal bath after $e^+e^-$ annihilation. In this case, instead of normalising at $T= {\rm MeV}$, one can choose a temperature below $e^+e^-$ annihilation, leading to a bound equivalent to Eq.~\eqref{ConsRhoBBN},
\begin{equation}
	h^2\, {\rho_{\rm CGMB}^{(0)}}/{\rho_c^{(0)}}  \leq h^2\Omega_{\gamma} ({7}/{8})\,\left({4}/{11}\right)^{4/3}\Delta N_{\rm eff} \simeq 5.6 \times 10^{-6} \, 
\Delta N_{\rm eff}\,.
\end{equation}
The current theoretical uncertainty of $\Delta N_{\rm eff}$ is of the order of 
$10^{-3}$ ~\cite{Bennett:2019ewm,Escudero:2020dfa,Akita:2020szl,Froustey:2020mcq}. If experiments were to reach this level of precision, one would obtain an upper bound of $h^2 \, \rho_{\rm CGMB}^{(0)} / \rho_c^{(0)}< 5.6\times10^{-9}$. 
%
\begin{table}
\begin{center}\small
\begin{tabular}{l||c||c||c||c}
 $\,$ & SM & $\nu$MSM & SMASH & MSSM\\[5pt]
 \hline
 \hline 
 \rule{0pt}{17pt}
$T_{\rm max}$ [GeV] $<$ & $(1.2$-$5.1)\times10^{19}$ & $(1.3$-$5.4)\times10^{19}$& $(1.4$-$6.0(1))\times10^{19}$ & $(2.3$-$9.4)\times10^{19}$\\[5pt]
\hline
 \rule{0pt}{17pt}
$T^{\Delta N_{\rm eff}=0.001}_{\rm max}$ [GeV] $<$ & $2.3\times 10^{17}$ &$2.4\times 10^{17}$  &$2.7\times 10^{17}$ &$4.39\times 10^{17}$
\end{tabular}
\end{center}
\caption{\label{tab:Tmax}Values of the maximal temperature allowed by the dark radiation constraints of Eqs.~\eqref{eq:DRstringent}- \eqref{eq:dark_radiation_constraint}, as well as the bound assuming $\Delta N_{\rm eff}$ equal to the theoretical uncertainty of $10^{-3}$. In the case of SMASH the numbers between brackets reflect the change in the last significant digit coming from choosing benchmark scenarios with different values of the tensor-to-scalar ratio. The results for trans-Planckian temperatures are not physically meaningful, as for $T>M_P$ one expects early-time equilibration of gravitons and a spectrum as in Eq.~\eqref{eq:OmegaEqCGMB}. We list the trans-Planckian temperatures simply to illustrate the reach of current dark radiation bounds.}
\end{table}

\begin{table}[t]
\begin{center}\small
\begin{tabular}{c||c||c|c|c|c}
 & $T_{\rm max}$ [GeV] & $f^{\Omega_{\rm CGMB}}_{\rm peak}$ [GHz] & $f^{h^{\rm CGMB}_c}_{\rm peak}$ [GHz] & $ h^2\Omega_{\rm CGMB}(f^{\Omega_{\rm CGMB}}_{\rm peak})$ &  $ h^{\rm CGMB}_c(f^{h^{\rm CGMB}_c}_{\rm peak} )$\\[5pt]
 \hline
 \hline
 \rule{0pt}{17pt}
 SM & $>M_P$ &74.45 & 30.26 & 2.27$\times10^{-7}$ &  1.17$\times10^{-32}$\\
    & 2.3$\times10^{17}$ & 80.09 & 40.48 & 4.47$\times10^{-9}$ &  1.42$\times10^{-33}$\\
     & 6.6$\times10^{15}$ & 80.23 & 40.69 & 1.34$\times10^{-10}$ &  2.45$\times10^{-34}$
\\[5pt]
\hline
\rule{0pt}{17pt}
& $>M_P$ & 73.75 & 29.98 & 2.19$\times10^{-7}$ &  1.16$\times10^{-32}$\\
$\nu$MSM & 2.4$\times10^{17}$ & 79.34 & 40.10 & 4.43$\times10^{-9}$ &  1.43$\times10^{-33}$\\
  & 6.6$\times10^{15}$ & 79.48 & 40.32 & 1.27$\times10^{-10}$ &  2.41$\times10^{-34}$\\
 & (3.4-11)$\times10^{13}$ & 79.73-79.67 & 40.69-40.60 & (7.02-22.34)$\times10^{-13}$ &  (1.78-3.19)$\times10^{-35}$
\\[5pt]
\hline
\rule{0pt}{17pt}
 & $>M_P$ & 70.99 & 28.85 & 1.88$\times10^{-7}$ & 1.11$\times10^{-32}$\\
SMASH & 2.7$\times10^{17}$ & 76.72 & 38.98 & 4.40$\times10^{-9}$ &  1.47$\times10^{-33}$\\
(r=0.0037)  & 6.4$\times10^{15}$ & 76.83 & 39.18 & 1.09$\times10^{-10}$ &  2.30$\times10^{-34}$\\
 & (8-20)$\times10^{9}$ & 77.56-77.44 & 40.35-40.22 & (1.64-4.02)$\times10^{-16}$ &  (2.79-4.37)$\times10^{-37}$
\\[5pt]
\hline
\rule{0pt}{17pt}
  & $>M_P$ & 71.06 & 28.88 & 1.89$\times10^{-7}$ &  1.11$\times10^{-32}$\\
SMASH & 2.7$\times10^{17}$ & 76.81 & 39.04 & 4.45$\times10^{-9}$ &  1.48$\times10^{-33}$\\
(r=0.05)  & 6.4$\times10^{15}$ & 76.91 & 39.24 & 1.10$\times10^{-10}$ &  2.31$\times10^{-34}$\\
 & (8-20)$\times10^{9}$ & 77.57-77.49 & 40.39-40.28 & (1.65-4.06)$\times10^{-16}$ &  (2.79-4.39)$\times10^{-37}$\\[5pt]
 \hline
 \rule{0pt}{17pt}
   & $>M_P$ & 57.50 & 23.37 & 8.09$\times10^{-8}$ &  9.02$\times10^{-33}$\\
MSSM & 4.4$\times10^{17}$ & 64.75 & 36.29 & 4.60$\times10^{-9}$ &  1.72$\times10^{-33}$\\
  & 5.5$\times10^{15}$ & 64.87 & 36.48 & 5.76$\times10^{-10}$ &  1.92$\times10^{-34}$
\end{tabular}
\end{center}
\caption{\label{tab:peaks}Values of peak frequencies and peak power spectra for the gravitational waves produced from the thermal plasma, for different models and maximum temperatures related to dark radiation and inflationary bounds, or to direct estimates of the reheating temperature. For a given model, the upper row corresponds to temperatures above the Planck mass, for which gravitons are expected to reach thermal equilibrium at early times, leading to an energy fraction around the dark radiation bound of Eq.~\eqref{eq:DRstringent}. The temperature in the second row is the upper bound corresponding to a hypothetical constraint  $\Delta N_{\rm eff}=10^{-3}$. In the third row one has the temperature bound of Eq.~\eqref{eq:reheating_constraint_Tmax}, which applies under the assumption of slow-roll inflation. If present, the range of $T_{\rm max}$ in the fourth row corresponds to direct estimates of the reheating temperature.}
\end{table}

We have turned 
the above limits 
 into upper bounds on $T_{\rm max}$ for the SM, the $\nu$MSM, SMASH and the MSSM, cf. Table~\ref{tab:Tmax}. The observational limits of Eqs.~\eqref{eq:dark_radiation_constraint} and \eqref{eq:DRstringent}  give bounds of the order of $10^{19}$\,GeV. These correspond to temperatures above the Planck scale, for which the gravitons can be expected to enter thermal equilibrium and the calculations based on Eq.~\eqref{eq:etafull} cannot be applied. Thus the previous dark radiation bounds cannot reliably constrain $T_{\rm max}$. For trans-Planckian temperatures one has to use the equilibrium form~\eqref{eq:OmegaEqCGMB} for the CGMB spectrum; integrating over the frequency so as to obtain the total energy fraction gives
 \begin{align}\label{eq:equilibriumDR}
  h^2
\, \frac{\rho_{\rm Eq. CGMB}^{(0)}}{ \rho_c^{(0)}}
=\left(\frac{g_{*s}({\rm fin})}{g_{*s}(M_P)}\right)^{4/3}\frac{h^2\pi^2 T_0^4 }{45 H_0^2 M_P^2}=3.0\times10^{-7}\,\left(\frac{106.75}{g_{*s}(M_P)}\right)^{4/3}.
 \end{align}
Intriguingly, this just about saturates the current dark radiation bound obtained assuming homogeneous initial conditions, Eq.~\eqref{eq:DRstringent}. Note that, in the case of early time equilibration of gravitational waves, one expects in fact homogeneity, and thus the relevant dark radiation bound is indeed given by Eq.~\eqref{eq:DRstringent} instead of \eqref{eq:dark_radiation_constraint}. Thus the current dark radiation bound is just on top of the value that corresponds to the contribution from gravitational waves that were in equilibrium at early times. Taking the significant digits of the bound of Eq.~\eqref{eq:DRstringent} seriously, then the result of Eq.~\eqref{eq:equilibriumDR} would imply that current dark radiation bounds are compatible with a CGMB with early time equilibrium in an extension of the SM in which $g_{*s}$ is augmented by a few degrees of freedom ($\Delta g_{*s}>2.8$ taking the naive value $g_{*s,{\rm SM}}=106.75$, $\Delta g_{*s}>4$ when including additional radiative corrections as summarized in Appendix \ref{app:effect_degr_freedom} (see Fig.~\ref{fig:gsSM})).

The next generation of CMB experiments is expected to improve the sensitivity on $\Delta N_{\rm eff}$ by one order of magnitude. 
Correspondingly, the upper bound on $T_{\rm max}$ may decrease by a factor of ten and thus reach the reduced
    Planck scale in the next decade. If future experiments were  to reach the theoretical uncertainty $\Delta N_{\rm eff}\sim 10^{-3}$,  then one would probe $T_{\rm max}$ at scales of the order of $10^{17}$\,GeV, as was already emphasized in Ref.~\cite{Ghiglieri:2020mhm}. Note that $T_{\rm max}$ bounds increase for models with more degrees of freedom, as expected from the scaling of Eq.~\eqref{eq:RelationOmegaHc}. More details for the peak frequencies and values of $\Omega_{\rm CGMB}$, $h_c^{\rm CGMB}$ for the maximal temperatures that follow from the dark radiation bounds are  given in Table~\ref{tab:peaks}.

\subsection{CMB Rayleigh-Jeans tail constraint on the CGMB}
\label{sec:cmb_RJ_constraint}

\begin{figure}[t]
\begin{center}
\includegraphics[width=0.9\textwidth]{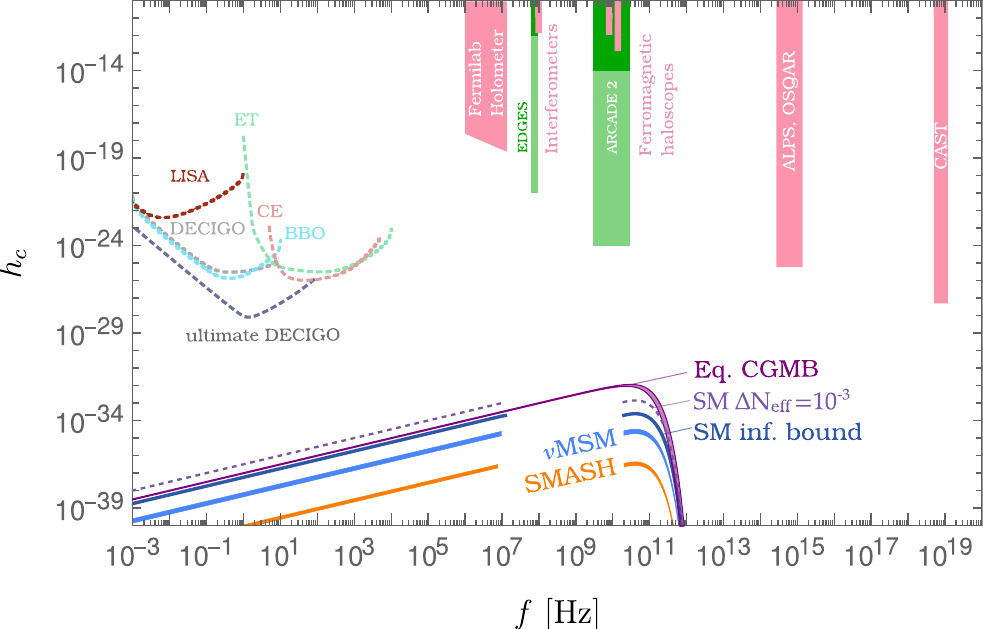}
\end{center}
\caption{The current upper bounds on the characteristic amplitude $h_c$ of a stochastic GW background from direct GW detection experiments (in red) and from 
the CMB Rayleigh-Jeans tail constraint (in green). Also shown are the projected sensitivities of planned laser interferometers and the predicted amplitudes of the CGMB, for the following five cases (from top to bottom at peak emission):
{\em i)} CGMB with early time equilibration,  in a band of models going from the SM to the MSSM, and corresponding to initial temperatures above the Planck mass and an approximate saturation of the bound of Eq.~\eqref{eq:DRstringent}, {\em ii)} SM plasma with $T_{\rm max}=2.3\times 10^{17}$\,GeV (corresponding to $\Delta N_{\rm eff}=10^{-3}$),
{\em iii)} SM plasma with $T_{\rm max}=6.6\times 10^{15}$\,GeV (upper limit consistent with slow-roll inflation, cf. \eqref{eq:reheating_constraint_Tmax}), 
{\em iv)} $\nu$MSM plasma with $3.4\times10^{13}\,{\rm GeV}\lesssim T_{\rm max}^{\nu\rm MSM}\lesssim 1.1\times10^{14}\,{\rm GeV}$, and {\em v)} SMASH plasma with $8\times10^9\,{\rm GeV}\lesssim T_{\rm max}^{\rm SMASH}\lesssim 2\times10^{10}$\,GeV (predicted by (pre-)heating in SMASH \cite{Ballesteros:2016euj,Ballesteros:2016xej}).}
\label{fig:hc_current_limits}
\end{figure}

In the presence of magnetic fields, GWs are converted into electromagnetic waves (EMWs) and vice versa. This is called the 
(inverse) Gertsenshtein effect~\cite{Gertsenshtein:1962,Boccaletti:1970,Zeldovich:1973,DeLogi:1977qe,Raffelt:1987im}. 
Recently, it has been shown that this conversion results in a  
distortion of the CMB, which can act therefore as a detector for MHz to GHz GWs generated before reionization~\cite{Domcke:2020yzq}. 
The measurements of the radio telescope EDGES have been turned into the bound
\begin{equation}
h_c^{\rm CGMB}  (f\approx 78\,{\rm MHz})<  10^{-12} \, (10^{-21})\,, 
\end{equation} 
for the  weakest (strongest) cosmic magnetic fields allowed by current astrophysical and cosmological constraints. Similarly, the observations of ARCADE 2 imply 
\begin{equation}
h_c^{\rm CGMB}  (3\,{\rm GHz}\lesssim f \lesssim 30\,{\rm GHz})<  10^{-14} \, (10^{-24})\,.
\end{equation} 
These upper bounds are displayed in Fig.~\ref{fig:hc_current_limits} as green exclusion regions. 
Future advances in radio astronomy and a better knowledge of 
cosmic magnetic fields are required in order that this method can get  competitive with the dark radiation constraint.  

%% file: laboratory_searches_for_CGMB.tex
\section{Laboratory searches for the CGMB}
\label{sec:laboratory_searches}
\setcounter{equation}{0}

In this section we will discuss current constraints on the CGMB from direct experimental GW searches in the laboratory and future possibilities to search for a stochastic GW background in the frequency range around the peak.

\subsection{\boldmath Current direct bounds from GW experiments}
\label{sec:current_constraints_on_h_c}

\begin{figure}[t]
\begin{center}
\includegraphics[width=0.85\textwidth]{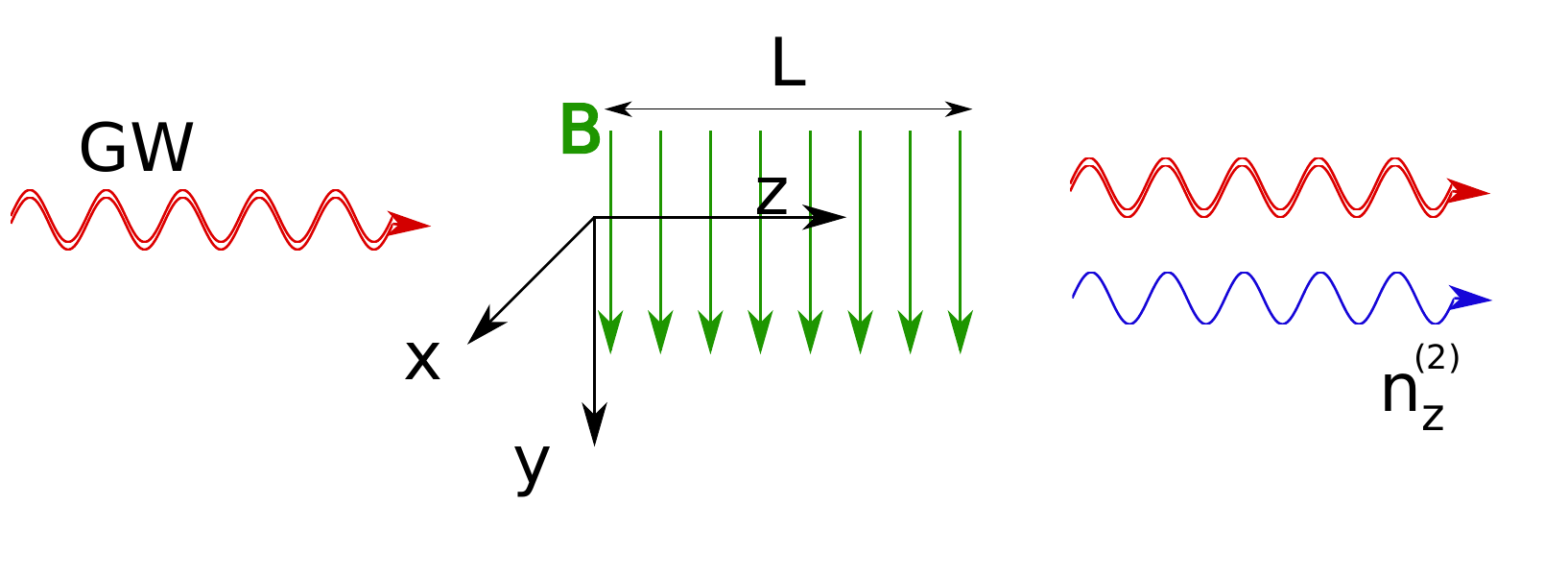}
\end{center}
\caption{Illustration of the inverse Gertsenshtein effect~\cite{Gertsenshtein:1962}. 
If a GW of frequency $f$ passes in vacuum through a transverse static magnetic field of strength $B$, an EMW is produced in the same direction and with the same frequency. 
Its EM power, at the terminal position of the magnetic field ($z=L$), is proportional to $f^2 h_c^2 (B L)^2$.}
\label{fig:inv_G}
\end{figure}

Current large-size ground-based laser-interferometric GW detectors, such as GEO, KAGRA, LIGO, and VIRGO \cite{Dooley:2015fpa,Somiya:2011np,Martynov:2016fzi,TheVirgo:2014hva}, are sensitive in the frequency range from about 10~Hz to 10~kHz. Their  technology is not necessarily ideal for studying very-high-frequency (VHF: 100~kHz$~-1$~THz) 
and ultra-high-frequency (UHF: above $1$~THz) 
GWs.
Several other small-size experiments have performed pioneering searches for stochastic GWs in the VHF and UHF range and 
put corresponding upper bounds which are confronted in Fig.~\ref{fig:hc_current_limits} to the CGMB prediction:
\begin{itemize}

\item A cavity/waveguide prototype experiment searched for polarization changes of electromagnetic waves, which are predicted to rotate under an incoming GW \cite{Cruise:2006zt}. It provided an upper limit,  
$h_c < 1.4\times10^{-10}$, on the characteristic amplitude of stochastic GWs at 100~MHz. 

\item 
Two laser interferometers with $0.75$~m long arms 
have been set-up as a so-called synchronous recycling interferometer~\cite{Nishizawa:2007tn}. They provided an upper limit, $h_c <1.4\times10^{-12}$, 
on the amplitude of a stochastic GW background at 100 MHz~\cite{Akutsu:2008qv}. 

\item 
The Fermilab Holometer,  consisting of separate, yet identical Michelson interferometers, with $39$~m long arms, has performed a  measurement at slightly lower frequencies. The upper limits, within 3$\sigma$, on the characteristic amplitude of stochastic GWs, 
are in the range $h_c < 25\times10^{-19}$ at 1~MHz down to a $h_c < 2.4\times10^{-19}$ at 13~MHz~\cite{Chou:2016hbb}.

\item Planar GWs induce resonant spin precession of electrons~\cite{Ito:2019wcb,Ito:2020wxi}. The same resonance is caused by coherent oscillation of hypothetical axion dark matter~\cite{Barbieri:1985cp}. Recently, searches for resonance fluorescence of magnons induced by axion dark matter have been performed and upper bounds on the axion-electron coupling constant have been 
obtained~\cite{Crescini:2018qrz,Flower:2018qgb}. These bounds can be translated to bounds on the amplitude of stochastic GWs: $h_c \lesssim 1.3\times10^{-13}$ at 14~GHz and $h_c \lesssim 1.1\times10^{-12}$ at 8.2~GHz~\cite{Ito:2019wcb,Ito:2020wxi}.

\item As mentioned earlier, in an external magnetic field, GWs partially convert into  EMWs~\cite{Gertsenshtein:1962,Boccaletti:1970,Zeldovich:1973,DeLogi:1977qe,Raffelt:1987im}, which can be processed with standard electromagnetic techniques and detected~\cite{Cruise:2012zz}, for example, by single-photon counting devices at a variety of wavelengths, cf. 
Fig.~\ref{fig:inv_G}. 
The authors of Ref.~\cite{Ejlli:2019bqj}  used data from existing facilities that have been constructed and operated with the aim of detecting axions or axion-like particles by their partial conversion into photons in magnetic fields: the light-shining-through-walls (LSW) experiments ALPS~\cite{Ehret:2009sq,Ehret:2010mh} and OSQAR~\cite{Ballou:2015cka,Pugnat:2013dha}, and the helioscope CAST~\cite{Zioutas:1998cc,GraciaGarza:2015sos}. They excluded GWs in the frequency bands from $\left(2.7 - 14\right)\times10^{14}$ Hz and $\left(5 - 12\right)\times10^{18}$~Hz down to a characteristic amplitude of $h_c < 6\times 10^{-26}$ and $h_c <  5\times 10^{-28}$, at 95\% confidence level, respectively. Using suitable EMW detectors sensitive to $h_c$ around its peak value at $\sim 40$\,GHz  one may 
exploit such axion experiments also for the search of the CGMB, as we will show in the next subsection. 

\end{itemize}
In summary: all the current upper bounds on the characteristic amplitude of stochastic GWs from direct experimental searches  are many orders of magnitude above 
the CGMB predictions.

\subsection{\boldmath Prospects of EM detection of the CGMB in the laboratory}
\label{sec:projected_constraints_on_h_c}

In this subsection, we will discuss the prospects of magnetic GW-EMW conversion experiments to  
probe  the CGMB\footnote{For a recent general review of detector concepts sensitive in the MHz to GHz range, see 
Ref.~\cite{1832786}.}. We will first concentrate on GW-EMW conversion in 
available static magnetic fields in vacuum with dedicated detectors appropriate for the tens of GHz range and 
then proceed to a proposal exploiting an additional VHF EM Gaussian beam in order to generate a conversion signal 
which is first order in $h_c$. 

\subsubsection{\boldmath Magnetic GW-EMW conversion in vacuum}
\label{sec:vacuum_magnetic_conversion}

In this subsection, we consider experiments exploiting the pure inverse Gertsenshtein effect~\cite{Gertsenshtein:1962}, cf. 
Fig.~\ref{fig:inv_G}. 
To this end, we assume that stochastic GWs of amplitude $h_c$ propagate through 
a transverse and 
constant magnetic $B$ in an evacuated tube of length $L$ and cross-section $A$ for a time $\Delta t$.
Then the  average power of the generated EMW, per logarithmic frequency interval, at the terminal 
position of the magnetic field ($z=L$ in Fig.~\ref{fig:inv_G}) 
is obtained as~\cite{Boccaletti:1970,DeLogi:1977qe,Cruise:2012zz,Ejlli:2019bqj}
\begin{equation}
 f\frac{{\rm d}P^{(2)}_{\rm EMW}}{{\rm d}{f}}
 \simeq 
\pi^2\, f^2 \, h_c^2 (f )\,B^2\, L^2 \,A
=  4.20\times 10^{-23} 
\,{\rm W}
\left[ \frac{f}{\rm 40\, GHz} \right]^2
\,\left[ \frac{h_c(f)}{10^{-21}}\right]^2
\,\left[ \frac{B}{\rm T} \right]^2
\,\left[ \frac{L}{\rm m} \right]^2\,\left[ \frac{A}{\rm m^2} \right]
\\
\,.
\label{eq:power_density_of_photons}
\end{equation}
The index ``2" denotes here the fact that the generated EMW power is second order in $h_c$. 
The associated expected average number of generated photons, per unit logarithmic frequency 
interval, is given by 
\begin{equation}
 f\frac{{\rm d}N^{(2)}_z}{{\rm d}{f}}
 \simeq 
\frac{\pi}{2}\, f  \,h_c^2 (f ) \,B^2\, L^2\,A\,\Delta t
= 1.59
\left[ \frac{f}{\rm 40\,GHz} \right]
\,\left[ \frac{h_c(f)}{10^{-21}}\right]^2
\,\left[ \frac{B}{\rm T} \right]^2
\,\left[ \frac{L}{\rm m} \right]^2
\,\left[ \frac{A}{\rm m^2} \right]
\,\left[ \frac{\Delta t}{{\rm s}}\right]
\,.
\label{eq:Flux_of_photons}
\end{equation}

These expressions are valid as long as the GWs and the generated EMWs are in phase coherence throughout their propagation in the 
magnetic field region. 
Under the assumption that the external B-field is surrounded by a circular beam tube of diameter $d$, coherent EMW generation is guaranteed if 
(see Appendix~\ref{app:3Deffects}): 
\begin{equation}
\label{eq:frequency_cutoff}
f\gg f_c \equiv \frac{c_{11}}{\pi^2} \frac{L}{d^2}\simeq 5.5\times 10^7\,{\rm Hz} 
\,\left[\frac{L}{\rm m}\right]
\,\left[\frac{\rm m}{d}\right]^2
,
\end{equation}
where $c_{11}=1.8$ and $d$ is the diameter of the beam tube. 
This effective lower frequency cut-off arises from the fact that 
the evacuated beam tube acts as an EM waveguide, in which the phase velocity of the EMW is higher than 
the phase velocity of light in vacuum, 
$v_{\rm EMW}=1/\sqrt{1- (f_c/f)^2}$

Around the peak frequency of the $h_c$ spectrum, $f_{\rm peak}^{h_c^{\rm CGMB}}\sim 30-40$\,GHz (see Table~\ref{tab:peaks}), 
one may either use heterodyne (HET) radio receivers or single photon detectors (SPDs) to search for an EM signal that was generated from magnetic conversion of the CGMB.

The sensitivity of the HET technique is limited by thermal noise in amplifiers and mixers (for an introduction, see Ref.~\cite{Kraus:1986}). 
In this context, it is useful to introduce an effective signal noise temperature $T_{\rm S}$ equal to the power of the 
generated EMW in a frequency bin $\Delta f$ around the peak frequency, 
\begin{eqnarray}
T_{\rm S} 
 =  
\frac{\Delta P^{\rm {(2)}}_{\rm EMW}}{\Delta f} 
\,.
\end{eqnarray}
Exploiting linear amplifiers with system noise temperature $T_{\rm sys}$,  
the  signal-to-noise ratio is determined then by~\cite{Kraus:1986}
\begin{equation}
\frac{\rm S}{\rm N} = \frac{T_{\rm S}}{T_{\rm sys}}\frac{1}{K_{\rm rec}}\sqrt{\Delta f\,\Delta t}\,,
\end{equation}
where $\Delta f$ is the pre-detection bandwidth of the receiver, $\Delta t$ is the measurement time, and 
$K_{\rm rec}$ is a receiver-system dependent dimensionless constant of order one\footnote{For example, $K_{\rm rec}=1$ for a total power receiver, 
$K_{\rm rec}=2$ for a Dicke receiver, see Ref.~\cite{Kraus:1986}.}
From this, we obtain the sensitivity of a magnetic GW-EMW conversion experiment with a heterodyne radiowave receiver  to the CGMB as 
\begin{eqnarray}
\nonumber
\left[ h^{\rm CGMB}_c \right]_{\rm sens}^{\rm HET} 
&\simeq &
9.65\times 10^{-21} 
\,\left[ \frac{{\rm S}/{\rm N}}{2} \right]^{1/2}
\,\left[ \frac{\Delta t}{\rm \pi\times 10^7\ s} \right]^{-1/4}
\,\left[ \frac{f}{\rm 40\ GHz} \right]^{-3/4}
\,\left[ \frac{\Delta f}{f} \right]^{-1/4}\,
\times
\\ 
\label{eq:GW_EMW_sensitivity_hc_heterodyne}
&& \hspace{10ex} \times 
  K_{\rm rec}^{1/2} 
\,\left[ \frac{T_{\rm sys}}{\rm 4\, K} \right]^{1/2}
\,\left[ \frac{B}{\rm T} \right]^{-1}
\,\left[ \frac{L}{\rm m} \right]^{-1}\,\left[ \frac{A}{\rm m^2} \right]^{-1/2}
\,.
\end{eqnarray}

\begin{table*}[t]
\,\hskip-.35cm\begin{tabular}{|l|c|c|c|c||c|c||c|c|}
\hline
             &    $B$~[T]         &  $L$ [m]  &    $d$~[m]   &  n$_{\rm tubes}$     &  $B L A^{1/2}$ & $f_c$ [Hz] & $[ h^{\rm CGMB}_c]^{\rm HET}_{\rm sens}$ &  
$[ h^{\rm CGMB}_c]^{\rm SPD}_{\rm sens}$ \\ \hline
ALPS IIc     &        5.3           &  211      &     0.05 & 1 & $49.6$\,Tm$^2$ & $4.6\times 10^{12}$  & --& -- \\ \hline
BabyIAXO         &      2.5            &  10      &   0.7 & 2 &  $21.9$\,Tm$^2$   & $1.1\times 10^{9}$  &  $4.41\times 10^{-22}$ & $3.52\times 10^{-25}$ \\ \hline
MADMAX         &      4.83            &  6      &   1.25 & 1 &  $32.1$\,Tm$^2$   & $1.9\times 10^{8}$  &  $3.01\times 10^{-22}$ & $2.40\times 10^{-25}$ \\ \hline
IAXO         &       2.5            &  20      &      0.7  	&	8	& $87.7$\,Tm$^2$    &  $2.2\times 10^{9}$ & $1.10\times 10^{-22}$ & $8.79\times 10^{-26}$\\ 
\hline  
\end{tabular}
\caption{Parameters of the magnetic field regions  of ALPs IIc~\cite{Bahre:2013ywa,Albrecht:2020ntd}, MADMAX~\cite{Brun:2019lyf,Calvelli:2020xxx}, 
BabyIAXO and IAXO~\cite{Ruz:2018omp}, 
used to estimate the minimum detectable GW amplitude through magnetic conversions of GWs (gravitons) to EMWs (photons) in vacuum: $B$ is the magnetic field magnitude, $L$ is the magnetic field length, $d$ is the diameter of the magnetized tube, and $B L A^{1/2}$, with $A=n_{\rm tubes}\pi d^2/4$, is the figure of merit for GW detection by magnetic conversion into EMWs. 
Also shown are the effective lower frequency cut-off  Eq.~\eqref{eq:frequency_cutoff} and the projected CGMB sensitivities 
around $f=40\ {\rm GHz}$, exploiting these magnetic field regions and the benchmark values in Eqs.~\eqref{eq:GW_EMW_sensitivity_hc_heterodyne} and \eqref{eq:GW_EMW_sensitivity_hc_SPD}.
ALPS IIc is not sensitive to the CGMB, because the lower frequency cut-off of its magnetic field region is around 5 THz. 
\label{tab:magnetic_conv_facilities}}
\end{table*}

\begin{figure}[t]
\begin{center}
\includegraphics[width=0.9\textwidth]{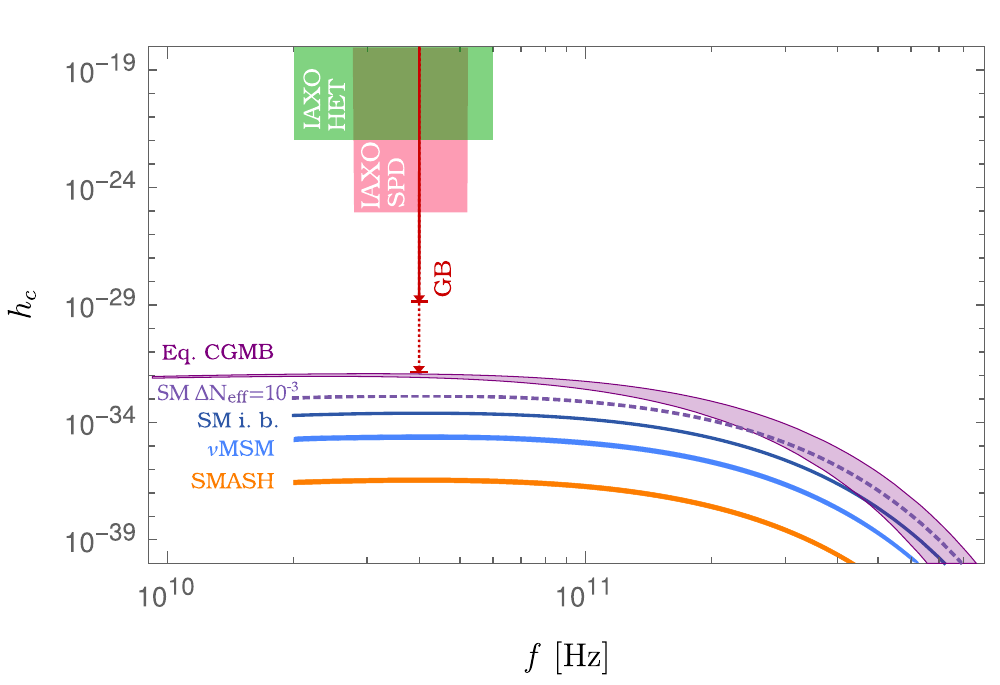}
\end{center}
\caption{Characteristic amplitude $h_c$ of the CGMB~\eqref{eq:hc} as in Fig.~\ref{fig:hc_current_limits}, but focusing on the frequency range near the peak and showing the predicted reach for experiments detecting GW-EMW conversions using heterodyne or single photon detectors in a static magnetic field, or resorting to an additional high-power 
40 GHz EM Gaussian beam. The dashed line shows an optimistic projection for the Gaussian beam technique accounting for future technological improvements, see main text.}
\label{fig:hc_future_limits}
\end{figure}

The figure of merit of  the magnetized region for conversion of GWs into EMWs is $B L A^{1/2}$, cf. Eq.~\eqref{eq:GW_EMW_sensitivity_hc_heterodyne}. This is shared also by 
LSW experiments exploiting optical cavities at the generation and regeneration side of the experiment and helioscopes searching for the magnetic conversion of axions into photons or vice versa. In Table \ref{tab:magnetic_conv_facilities} we show the parameters of 
the magnetic field region of the next generation of  axion experiments: the LSW experiment ALPS IIc \cite{Bahre:2013ywa,Albrecht:2020ntd}, the haloscope MADMAX~\cite{Brun:2019lyf,Calvelli:2020xxx}, and  the helioscopes BabyIAXO and IAXO \cite{Ruz:2018omp}. 
Unfortunately, the prospects to probe the CGMB exploiting these magnetic conversion facilities appear to be rather slim. For example, 
collecting the signal from all eight magnetized tubes of IAXO with a heterodyne radio receiver in a one year CGMB-EMW conversion experiment, the projected sensitivity 
given in Table~\ref{tab:magnetic_conv_facilities},  $[ h^{\rm CGMB}_c(f\approx 40\,{\rm GHz})]^{\rm HET}_{\rm sens}\approx 1.10\times 10^{-22}$,  is about ten orders  
of magnitude above the CGMB predictions with early time equilibration, corresponding to initial temperatures above the Planck mass and an approximate saturation of the bound of Eq.~\eqref{eq:DRstringent}, cf. Fig.~\ref{fig:hc_future_limits}.

%

The prospects are slightly better if progress  is made on single photon detection at photon energies around 
$\omega = 2\pi f=1.65\times 10^{-4} \,{\rm eV}[f/40\,{\rm  GHz}]$. The signal-to-noise ratio is given in this case 
by 
\begin{equation}
\frac{\rm S}{\rm N} = \frac{\Delta N_{\rm S\, counts}}{\sqrt{\Delta N_{\rm D\, counts}}}\,.
\end{equation}
Here, 
\begin{equation}
\label{eq:dark_counts}
\Delta N_{\rm S\, counts} \simeq \epsilon\, 
\frac{\pi}{2} \,h_c^2 (f ) \,B^2\, L^2\,A\,\Delta t\,\Delta \omega \,,
\end{equation}
denotes the number of signal counts in a time interval $\Delta t$ and an energy interval $\Delta \omega =2\pi \Delta f$
(cf. Eq.~\eqref{eq:Flux_of_photons}), with 
$\epsilon$ being the single photon detection efficiency and 
\begin{equation}
\Delta N_{\rm D\, counts} \simeq 
\Gamma_{\rm D}\,\Delta t 
\end{equation}
the number of dark counts, in terms of the  
 dark count rate $\Gamma_{\rm D}$. 
The sensitivity of a magnetic GW-EMW conversion experiment with an SPD detection system is then 
\begin{eqnarray}
\nonumber
\left[ h^{\rm CGMB}_c \right]_{\rm sens}^{\rm SPD} 
&\simeq & 
7.71 \times 10^{-24}
\,\left[ \frac{{\rm S}/{\rm N}}{2} \right]^{1/2}
\,\left[ \frac{\Delta t}{\rm \pi\times 10^7\ s} \right]^{-1/4}
\left(\frac{\Delta \omega}{\rm  10^{-4}\,eV}\right)^{-1/2}
\times 
\\ 
\label{eq:GW_EMW_sensitivity_hc_SPD}
&& \hspace{10ex} \times 
  \epsilon^{-1/2}\, 
\left[ \frac{\Gamma_{\rm D}}{10^{-3}\,\text{Hz}}\right]^{1/4}
\,\left[ \frac{B}{\rm T} \right]^{-1}
\,\left[ \frac{L}{\rm m} \right]^{-1}\,\left[ \frac{A}{\rm m^2} \right]^{-1/2}
\,.
\end{eqnarray}
If the experimental benchmark values 
chosen in Eq.~\eqref{eq:GW_EMW_sensitivity_hc_SPD} can be reached\footnote{In this context it is interesting to note that  
a quantum dot detector at 50 mK has achieved already a dark count rate of order mHZ 
in the photon energy range from $6.0$ to $7.1$\,meV~\cite{Komiyama:2000}.  
SPD with even lower dark count rates may be realized with Graphene-based Josephson junctions~\cite{Walsh:2017}. A research and development program
on dedicated SPD at sub-THz frequencies is also motivated by future axion experiments, such as the LSW experiment STAX~\cite{Capparelli:2015mxa} and the haloscope 
TOORAD~\cite{Marsh:2018dlj}.},   
the SPD sensitivity is about three orders of magnitude better than the HET sensitivity.
However, it is fair to say that, from today's perspective, vacuum magnetic GW-EMW conversion experiments will fail to beat 
the dark radiation constraint on $h_c^{\rm CGMB}$ by more than six orders of magnitude, cf. Table~\ref{tab:magnetic_conv_facilities} and Fig.~\ref{fig:hc_future_limits}.

\subsubsection{\boldmath Magnetic GW-EMW conversion in a VHF EM  Gaussian beam}
\label{sec:magnetic_conversion_in_GB}

The signal for magnetic GW-EMW conversion in vacuum, such as the generated EM power \eqref{eq:power_density_of_photons} or 
the number of generated photons \eqref{eq:Flux_of_photons},  is of second order in the tiny amplitude $h_c$ of the passing GWs. A number of modified schemes have been proposed which 
introduce in the magnetic conversion region certain powerful auxiliary EM fields oscillating at the frequency of the gravitational wave, 
such as plane EMWs~\cite{Li:2000du} or EM Gaussian beams (GBs)~\cite{Li:2003tv}, 
to obtain GW-induced EMWs which are first order in 
$h_c$. For $h_c\ll 1$, their signal strength  overwhelms the one from the second order EMWs induced by the inverse Gertsenshtein effect. However, this does not mean automatically that  the sensitivity of these 
modified magnetic conversion experiments is much larger than the one of the experiments based on the inverse Gertsenshtein effect in vacuum, because the 
powerful auxiliary EMWs tend to increase the noise floor and consequently to decrease the signal-to-noise ratio.

\begin{figure}[t]
\begin{center}
\includegraphics[width=0.9\textwidth]{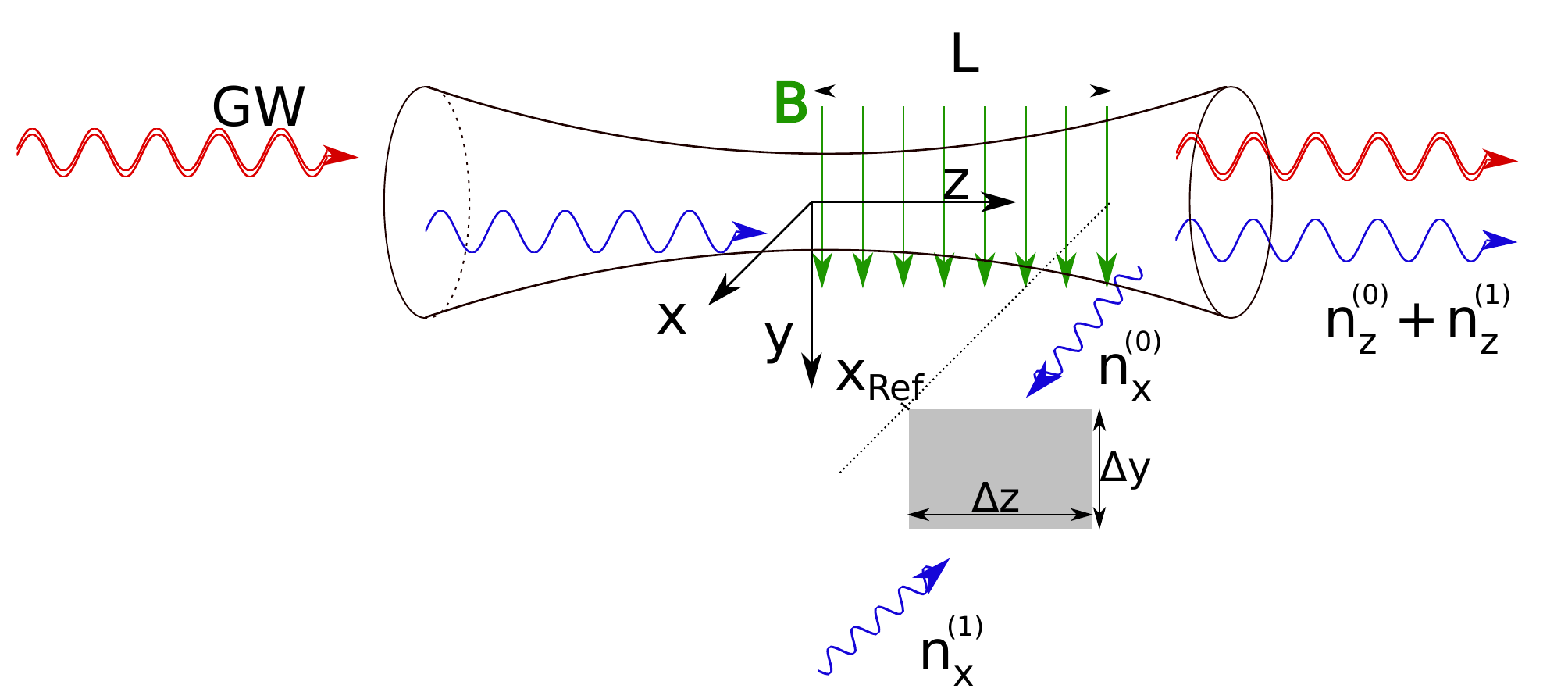}
\end{center}
\caption{Illustration of the proposal~\cite{Li:2003tv} to exploit an EM GB to induce a GW-EMW conversion signal which is first order in $h_c$.}
\label{fig:Li}
\end{figure}

The arguably most promising of these modified magnetic GW-EMW conversion detection proposals exploits a VHF EM GB 
to induce a first order signal in magnetic GW-EMW 
conversion~\cite{Li:2003tv,Li:2004df,Li:2006sx,Li:2008qr,Tong:2008rz,Stephenson:2009zz,Li:2009zzy,Li:2011zzl,Woods:2011xxx,Li:2013fna,Woods:2014aoa,Li:2014bma,Li:2015nti,Wang:2018esk}.
A continuous traveling wave EM GB with frequency $f_0$, propagating in the $z$-direction with linear polarization along the 
$x$-direction, passes through a  transverse static magnetic field,  cf. Fig.~\ref{fig:Li}. If a GW of  frequency $f=f_0$  propagates
along the $z$-direction, the resonant interaction of the GW with the EM fields of the GB and the static magnetic field will 
not only generate a longitudinal first order photon flux (denoted by $n_z^{(1)}$ in Fig.~\ref{fig:Li}),  which will be swamped by the background EM flux $n_z^{(0)}$ from the GB, 
but also a transverse first order photon flux (denoted by $n_x^{(1)}$ in Fig.~\ref{fig:Li}) in the direction perpendicular to the GB, which reads, for $z\geq L$, 
\begin{eqnarray}
f \frac{{\rm d} n_x^{(1)}}{{\rm d}{f}}\mid_{f_0} 
&\simeq &
\frac{1}{4}\, h_c(f_0)   B_y^{(0)} E_0 L   
\,\psi^{(1)}_x \left(\frac{w_0}{z_R} ,\frac{x}{w_0},\frac{y}{w_0},\frac{z}{z_R},\delta \right)
\\
\nonumber
&= &
\frac{3.15}{{\rm m}^2\,{\rm s}} 
\,\left[ \frac{h_c(f_0)}{10^{-36}} \right]
\left[ \frac{B_y^{(0)}}{\rm 1\,T}\right]
\left[ \frac{E_0}{\rm 5\times 10^5\,V/m}\right]
\left[ \frac{L}{\rm m}\right] 
\,\psi^{(1)}_x \left(\frac{w_0}{z_R} ,\frac{x}{w_0},\frac{y}{w_0},\frac{z}{z_R},\delta \right)\,.
\end{eqnarray}
Here $E_0$ is the amplitude of the electric field of the GB at the center of the beam at its waist, $w_0$ its waist radius, 
$z_R = \pi w_0^2 f_0=1.05\,{\rm m} [f_0/(40\,{\rm GHz})][w_0/(0.05\,{\rm m})]^2$ its Rayleigh range, $\delta$ the relative phase between the GW and the GB, 
and~\cite{Tong:2008rz}
\begin{eqnarray}
\psi^{(1)}_x \left(\frac{w_0}{z_R},x^\prime,y',z',\delta \right) &\simeq &
\frac{w_0}{z_R} 
\frac{\frac{y'}{z'}\exp\left( -\frac{x^{\prime 2}+y^{\prime 2}}{\left[ 1+z^{\prime 2}\right]}\right)}{\left[1+z^{\prime 2}\right]^{1/2}}  
\times 
\\
\nonumber 
&& \hspace{-24ex}
\left\{
\frac{1}{ \left[1+z^{\prime -2}\right]}
\cos \left( 
\frac{z^{\prime -1}(x^{\prime 2}+y^{\prime 2})}{ \left[1+z^{\prime -2}\right]}
 - \tan^{-1} z^\prime   + \delta \right)
-
\frac{z^\prime}{ \left[1+z^{\prime 2}\right]}
\sin\left( 
\frac{z^{\prime -1}(x^{\prime 2}+y^{\prime 2})}{ \left[1+z^{\prime -2}\right]}
 - \tan^{-1} z^\prime   + \delta \right)
\right\}
\,.
\end{eqnarray}

%
%

Depending on the overall sign of $\psi^{(1)}_x$, that is on $\delta$, this flux points either in the positive or negative $x$-direction. 
The idea is then to place at $x=\pm x_{\rm Ref}$ reflectors (e.g. fractal membranes) parallel to the $y$-$z$ plane, see Fig.~ \ref{fig:Li}, which could reflect and focus a portion of this flux to receivers and detectors placed  at positions $x=\pm x_{\rm Det}$ with $x_{\rm Det}>x_{\rm Ref}$  which are further away from the GB and therefore expected to suffer less from 
noise~\cite{Li:2004df}. 
The number of signal photons within a bandwidth $\Delta f_0$ around $f_0$ passing in a time interval $\Delta t$  
through a detector  surface element $\Delta S=\Delta y \Delta z$ 
in the $y$-$z$ plane, which extents from $y=0$ to $y=\Delta y$ in the $y$-direction and from $z=L$ to $z=L+\Delta z$ in the $z$-direction\footnote{We assume for simplicity that the receiver/detector surface is parallel to the 
reflector surface and has the same extensions.}, is then
\begin{eqnarray}
\label{eq:delta_N1_GB}
\Delta N_x^{(1)}(x_{\rm Det})
&\simeq &
\frac{\eta}{4}\, h_c(f_0 )  \, \left[ \frac{\Delta f_0}{f_0}\right] B_y^{(0)}E_0    L  
\,\Delta y \Delta z\,\Delta t\, 
 \mathcal F^{(1)}_x (x_{\rm Ref})
\\
\nonumber
&= &
3.15\,  \eta 
\,\left[ \frac{h_c(f_0)}{10^{-23}} \right]
\, \left[ \frac{\frac{\Delta f_0}{f_0}}{10^{-6}}\right]
\,\left[ \frac{B_y^{(0)}}{\rm 1\,T}\right]
\,\left[ \frac{E_0}{\rm 5\times 10^5\,V/m}\right] 
\,\left[ \frac{L}{\rm m}\right] 
\,\left[ \frac{\Delta y \Delta z}{\rm 0.01\,m^2}\right] 
\,\left[ \frac{\Delta t}{\rm s}\right]
\,\left[ \frac{\mathcal F^{(1)}_x (x_{\rm Ref})}{10^{-5}} \right]
\,,
\end{eqnarray}
where $0<\eta <1$ is the reflectivity of the reflector and  
\begin{eqnarray}
\mathcal F^{(1)}_x (x)
 &= &
\frac{1}{4\pi}
\frac{w_0 z_R}{\Delta y \Delta z}\,
\int\limits_0^{2\pi} {\rm d}\delta\, 
\left|
\int\limits_0^{\frac{\Delta y}{w_0}} {\rm d}y'\, 
\int\limits_{\frac{L}{z_R}}^{\frac{L+ \Delta z}{z_R}} {\rm d}z' 
\, \psi^{(1)}_x  \left(\frac{w_0}{z_R},x^\prime,y',z',\delta \right) \right| 
\,.
\end{eqnarray}

\begin{figure}[t]
   	\centering
    \includegraphics[width=0.48\textwidth]{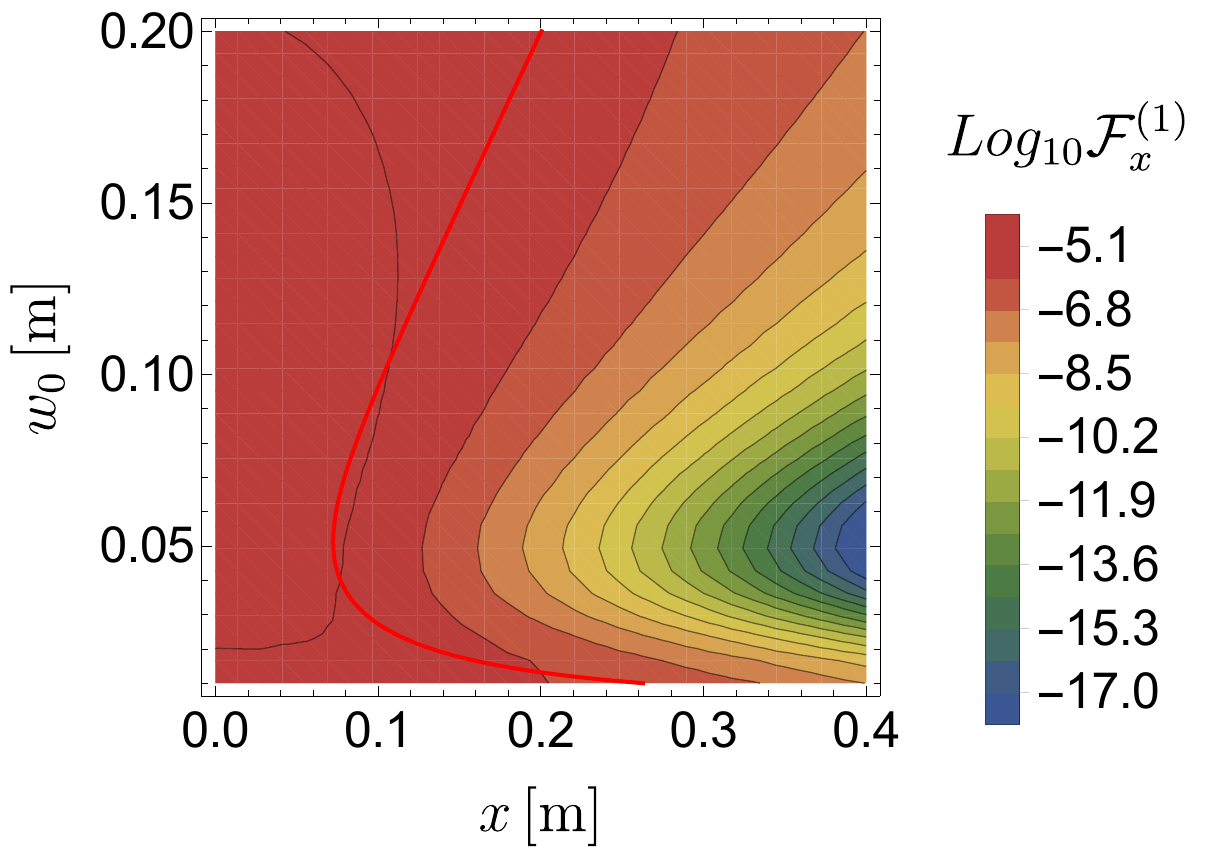}
    \includegraphics[width=0.48\textwidth]{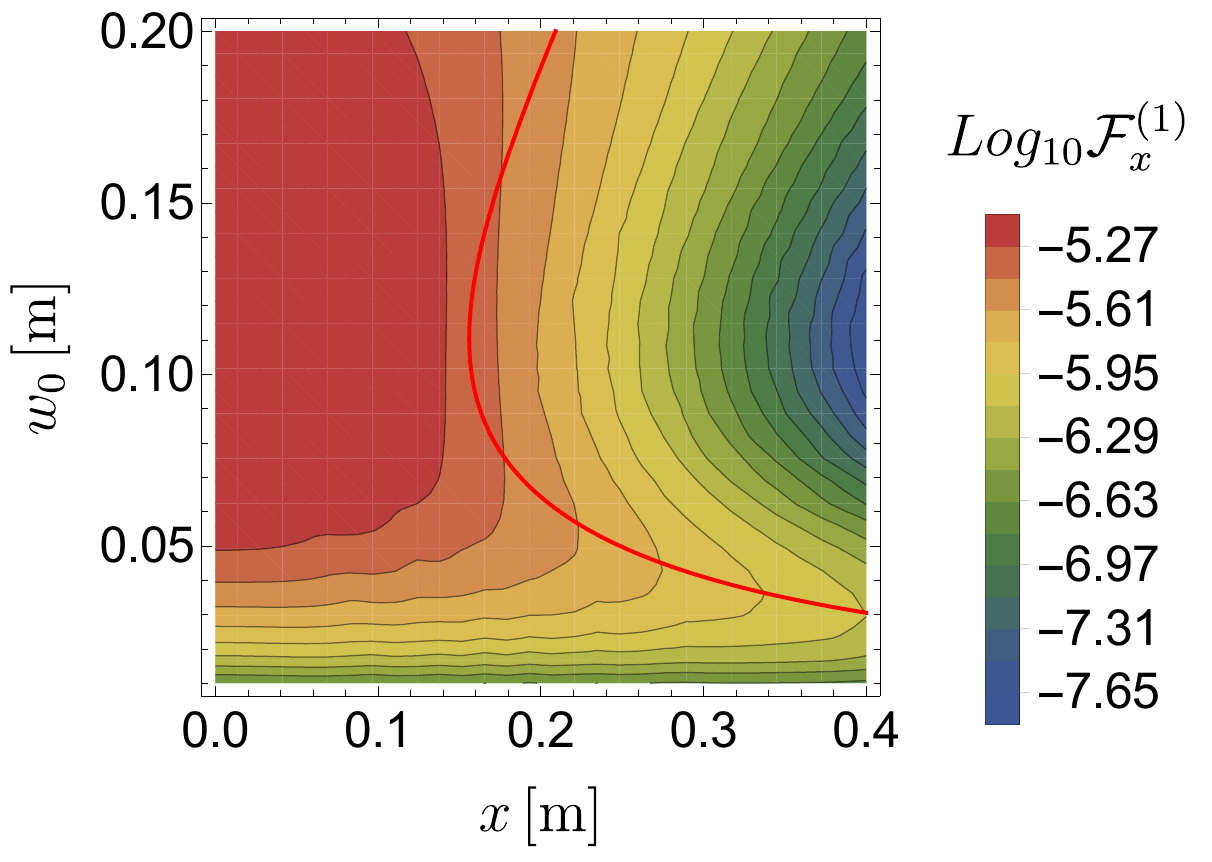}
    \includegraphics[width=0.48\textwidth]{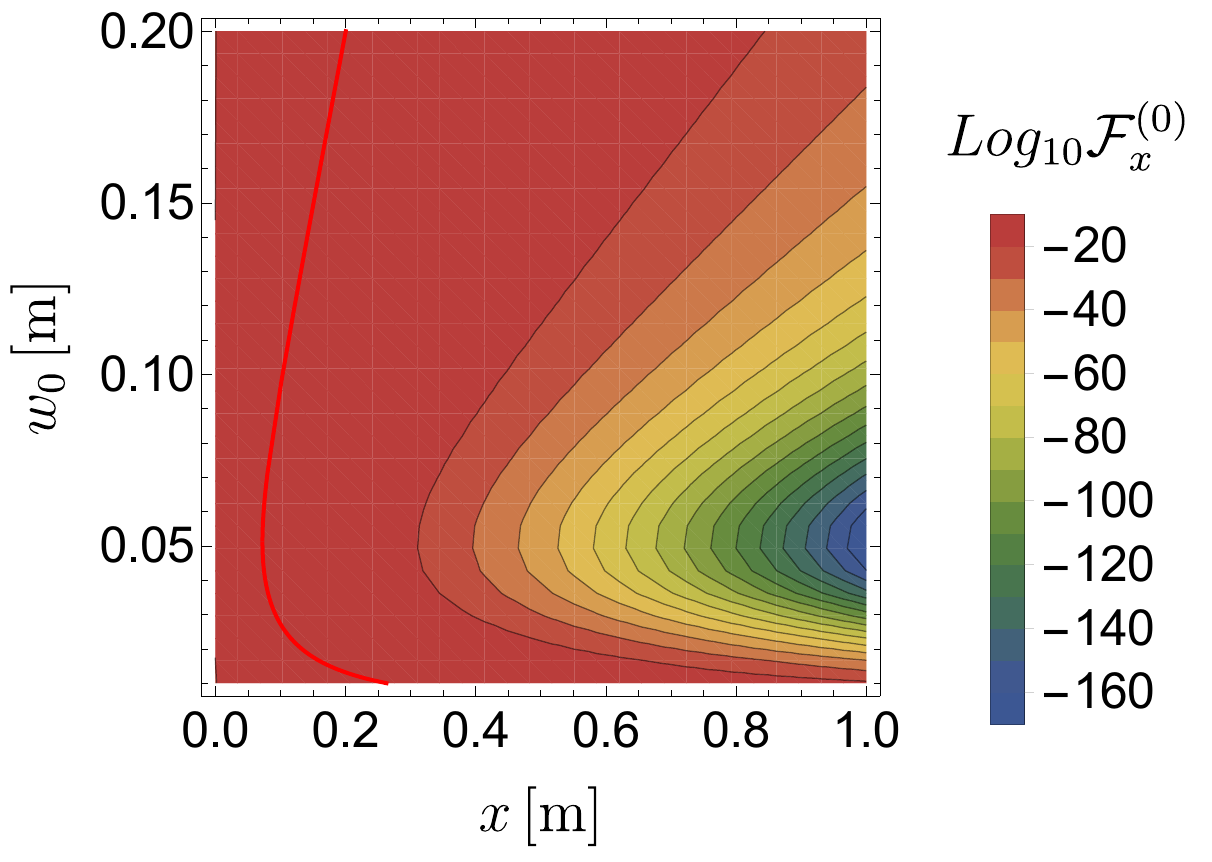}
    \includegraphics[width=0.48\textwidth]{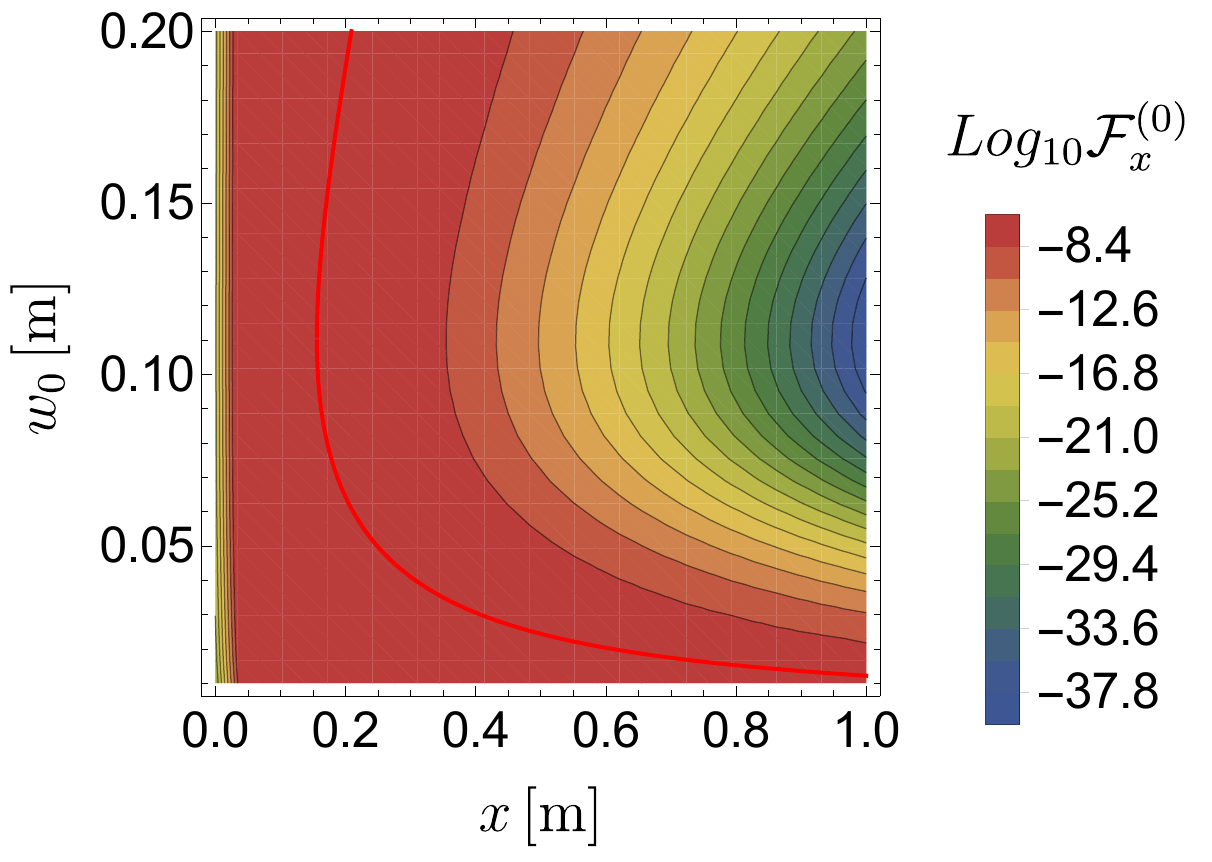}
	\caption{Contours of $\mathcal{F}_x^{(1)}$ (top panels) and $\mathcal{F}_x^{(0)}$ (bottom panels) 
in the $x$-$w_0$ plane, for 
$f_0=\SI{40}{\giga\hertz}$,  $\Delta y=\SI{0.1}{\metre}$, $\Delta z=\SI{0.1}{\metre}$, and $L=\SI{1}{\metre}$ (left panels)
and $L=\SI{5}{\metre}$ (right panels). The red lines indicate the distance in the $x$-direction for which   
$x = w(L+\Delta z) = w_0 \sqrt{1+[{(L+\Delta z)}/{z_R}]^2}$.} 
  	\label{fig:F1F0}
\end{figure}

Numerical results for $\mathcal F^{(1)}_x(x,w_0)$, for particular values of $f_0$, $\Delta y$, $\Delta z$, and $L$, are displayed in Fig.~\ref{fig:F1F0} (top panels).   
Right to the red lines, the Gaussian beam amplitude has dropped by a factor more than $1/e$. We assume that placing the reflector in this region will cause only minor disturbances of the signal photon flux. Based on this assumption we find a benchmark value of $10^{-5}$ for $\mathcal{F}_x^1$ which we have taken in 
Eq.~\eqref{eq:delta_N1_GB}.
As benchmarks for the amplitude $E_0$ and the relative bandwidth $\Delta f_0/f_0$ of the GB we have taken in 
Eq.~\eqref{eq:delta_N1_GB} values which can be achieved with a 
state-of-the-art free-running high-power (MW scale\footnote{The total power of a GB is given by $P_0=(\pi/4) w_0^2 E_0^2=1.30\,{\rm MW}\,[w_0/(0.05\,{\rm m})]^2[E_0/(5\times 10^5\,{\rm V/m})]^2$.}) gyrotron in this frequency range\footnote{It is interesting to note that a similar gyrotron has been proposed as the photon beam source of the axion LSW experiment STAX~\cite{Capparelli:2015mxa}. Therefore, in principle, one could extend STAX to a multi-purpose facility to search not only for axions, but also for GWs.}~\cite{Thumm:2020xxx,Jelonnek:2020xxx}.

The zeroth order flux in the $x$-direction 
propagates radially out from the GB’s axis, 
\begin{eqnarray}
n_x^{(0)} 
&=&
\frac{\alpha_p}{4\pi}
\frac{ E_0^2}{f_0} 
\,\psi^{(0)}_x \left(\frac{w_0}{z_R},x^\prime,y',z' \right)
\\ \nonumber
&=&
\frac{1.25\times 10^{31}}{{\rm m}^2\,{\rm s}} 
\,\alpha_p
\, \left[ \frac{f_0}{\rm 40\,GHz} \right]^{-1}
\,\left[ \frac{E_0}{\rm 5\times 10^5\,V/m}\right]^2
\,\psi^{(0)}_x\left(\frac{w_0}{z_R},x^\prime,y',z' \right)
\,,
\end{eqnarray}
where $\alpha_p\ll1$ is the ratio of the  $y$ to $x$ components of the GB electric field and~\cite{Tong:2008rz} 
\begin{equation}
\psi^{(0)}_x\left(\frac{w_0}{z_R},x',y',z'\right)=\frac{w_0}{z_R}\frac{\frac{x'}{z'}\exp\left(-2\frac{{x'}^{2}+{y'}^{2}}{\left[1+{z'}^{2}\right]}\right)}{\left[1+{z'}^{2}\right]\left[1+{z'}^{-2}\right]}
\,.
\end{equation}
The corresponding number of background photons passing in a time interval $\Delta t$ through the detector surface is then 
\begin{equation}
\label{eq:delta_N0_GB}
\Delta N_x^{(0)} (x_{\rm Det})
\simeq
1.25\times 10^{-11}
\,\alpha_p
\, \left[ \frac{f_0}{\rm 40\,GHz} \right]^{-1}
\,\left[ \frac{E_0}{\rm 5\times 10^5\,V/m}\right]^2
\,\left[ \frac{\Delta y \Delta z}{\rm 0.01\,m^2}\right] 
\,\left[ \frac{\Delta t}{\rm s}\right]
\, \left[ \frac{\mathcal F^{(0)}_x (x_{\rm Det})}{10^{-40}}\right] 
\,,
\end{equation}
with 
\begin{eqnarray}
\mathcal F^{(0)}_x(x)  
&= &
\frac{w_0z_R}{\Delta y\Delta z}\,\int\limits_0^{\frac{\Delta y}{w_0}}{\rm d}y'\,
\int\limits_{\frac{L}{z_R}}^{\frac{L+ \Delta z}{z_R}} {\rm d}z'\,\psi^{(0)}_x\left(\frac{w_0}{z_R},x^\prime,y',z'\right)  
\,.
\end{eqnarray}

Numerical results for $\mathcal F^{(0)}_x(x,w_0)$, for particular values of $f_0$, $\Delta y$, $\Delta z$, and $L$, are displayed in Fig.~\ref{fig:F1F0} (bottom panels).   
As its benchmark we have taken in \eqref{eq:delta_N0_GB} a value of $10^{-40}$ which is appropriate when the detectors are put at $x_{\rm Det}=0.45\,{\rm m}$ or 
$1\,{\rm m}$, for $L=1\,{\rm m}$ or $5\,{\rm m}$, respectively, cf. Fig.~\ref{fig:F1F0} (bottom panels). 
Therefore, this direct background from the GB is expected to be quite small if the receiver is placed sufficiently away from the beam. 
Moreover, it should occur simultaneously at the two detectors located at $x=\pm x_{\rm Det}$, while the signal ($n_x^{(1)}$ propagating towards $x=0$, see Fig.~\ref{fig:Li}), for fixed phase difference $\delta$, occurs only at one 
of the two detectors. Nevertheless, this consideration
neglects the possibility that the radiation from the GB is perturbed by the presence of the reflectors which in turn could disturb the signal photon flux. 
Furthermore the reflectors can be a noise source if the GB interacts with them. This kind of noise can be minimized by placing the reflectors 
at least right to the red lines in Fig.~\ref{fig:F1F0} where the GB amplitude has fallen off by $1/e$~\cite{Li:2009zzy}. 
However, the exact noise level, which is introduced by the interaction of the GB with the reflectors, has to be evaluated in a future study.  
If all these sources
can be dealt with and the apparatus can be designed in such a way that finally the dark count
rate~\eqref{eq:dark_counts} in SPD is the dominating background, then the sensitivity is given by
\begin{eqnarray}
\label{eq:GW_EMW_sensitivity_hc_GB_SPD_dark_background}
\left[ h^{\rm CGMB}_c \right]_{\rm sens}^{\rm GB} 
&\simeq & 
4.02\times 10^{-29} \,  \eta^{-1}
\,\left[ \frac{{\rm S}/{\rm N}}{2} \right]
\,\left[ \frac{\Delta t}{10^4\, s} \right]^{-1/2}
\, \left[ \frac{\frac{\Delta f_0}{f_0}}{10^{-6}}\right]^{-1}
\times 
\\ \nonumber
&&  \hspace{-10ex} \times 
\, \epsilon^{-1}
\, \left[ \frac{\Gamma_{\rm D}}{10^{-3}\,\text{Hz}}\right]^{1/2}
\,\left[ \frac{E_0}{\rm 5\times 10^5\,V/m}\right]^{-1}
\,\left[ \frac{B_y^{(0)}}{\rm 10\, T} \right]^{-1}
\,\left[ \frac{L}{\rm 5\,m} \right]^{-1}\,\left[ \frac{\Delta y \Delta z}{\rm 0.01\,m^2} \right]^{-1}
\, \left[\frac{\mathcal{F}_x^{(1)}(x_{\rm Ref})}{10^{-5}}\right]^{-1}
.
\end{eqnarray}
Here we have taken current state-of-the-art benchmarks for the various experimental parameters.

This is still three orders of magnitude above the CGMB predictions with early time equilibration, corresponding to $T_{\rm max}>M_P$ and an approximate saturation of the dark radiation bound~\eqref{eq:DRstringent}, cf. Fig.~\ref{fig:hc_future_limits}. 
However, the latter may be reached and eventually  surpassed 
by progress in the development of gyrotrons, SPD, and magnets. In fact, one may gain an order of magnitude in $h_c$ sensitivity by increasing the total power of the gyrotron by two 
orders of magnitude to $\sim 100$\,MW (and thus $E_0$ by one order of magnitude) and another order of magnitude in $h_c$ sensitivity by increasing the stable running time of the gyrotron by two orders of magnitude to $\Delta t\sim 10^6$\,s. The remaining order of magnitude one may gain by developing SPD with a dark count rate of order $10^{-5}$\,s.  Further improvements can 
be obtained by increasing the reflector size and by developing magnets with higher magnetic field and length. 
A sensitivity corresponding to $T_{\rm max}=M_P$ seems to be reachable in the not-so-distant future.
In case of the detection of a signal, one may explore the frequency region around the nominal frequency of the gyrotron in a range 
$\Delta f_0\sim 10^{-3} f_0 = 40\,{\rm MHz}\, [f_0/(40\,{\rm GHz})]$ by changing the acceleration voltage of the gyrotron.

%% file: discussion_outlook.tex
\section{Summary and outlook}
\label{sec:summary}
\setcounter{equation}{0}

Based on the pioneering work of 
Refs. ~\cite{Ghiglieri:2015nfa,Ghiglieri:2020mhm}, 
we have provided general formulae for the production rate of GWs from a primordial thermal plasma with sub-Planckian temperatures
in an arbitrary theory with gauge fields, real scalars and Weyl fermions (cf. Sect.~\ref{sec:prod_gen_plasma} and Appendix~\ref{app:complete_leading_order}) 
and derived general expressions for $\Omega_{\rm CGMB}(f)$, the current energy fraction of those GWs per logarithmic frequency interval 
(see Sect.~\ref{sec:current_stoch_backgd}). 
It is found to peak around $f_{\rm peak}^{\Omega_{\rm CGMB}}\simeq 80\,{\rm GHz}\,[106.75/g_{*s}(T_{\rm max})]^{1/3}$ (cf. Eqs.~\eqref{eq:peakformulaf},~\eqref{eq:RelationPeaks} and Table~\ref{tab:peaks})  -- hence we chose CGMB (for Cosmic Gravitational Microwave Background) as the acronym for this stochastic background. 
Its overall magnitude scales approximately linearly with the  
maximum temperature which the primordial plasma attained at the beginning of the standard hot big bang era (cf. 
Eq.~\eqref{eq:Omega_CGMB_analytic_expression}) while also depending on $g_{*s}(T_{\rm max})$. For weakly coupled theories, the peak emission satisfies the approximate scaling of Eq.~\eqref{eq:RelationOmegaHc}, (see also Eqs.~\eqref{eq:h2Omegaapprox} and \eqref{eq:hcapprox}) implying that for a given $T_{\rm max}$ the SM will typically maximize the CGMB with respect to its value in weakly coupled extensions. With the leading behaviour of the peak frequency and the magnitude of the CGMB being determined by $T_{\rm max}$ and $g_{*s}(T_{\rm max})$, 
the  CGMB can therefore act as a hot big bang thermometer  and, additionally, allow a measurement of the number of thermalized BSM degrees of freedom, $g_{*s}(T_{\rm max})-g_{*s,{\rm SM}}(T_{\rm max})$. 
As special cases, we have determined the CGMB spectrum for the cases of 
a SM (cf. Sect.~\ref{sec:GWSM_SM}),  a $\nu$MSM, a SMASH (cf. Sect.~\ref{sec:GW_NUMSM_SMASH}), 
as well as an MSSM (cf. Sect.~\ref{sec:GW_MSSM}) plasma. We confirmed that the leading model dependence is indeed captured by the effects of  $g_{*s}(T_{\rm max})$, so that within a broad class of weakly coupled SM extensions, a simple comparison of a hypothetical measurement of the CGMB peak with the SM prediction would allow to estimate  $g^{1/3}_{*s}(T_{\rm max})$  and $T_{\rm max}$ in a model-independent manner and with respective theoretical accuracies that should be better than the MSSM results of 15\% and 40\% (e.g. $\sim 1\%$ and $5\%$ in SMASH).

The previous features of the CGMB apply for $T_{\rm max}<M_P$, while for larger early-time temperatures one expects gravitons to thermalize and lead to the blackbody spectrum of Eq.~\eqref{eq:OmegaEqCGMB}, with peak frequencies and maxima scaling with $g_{*s}(M_P)$ as in Eqs.~\eqref{req:peak_freq_cgmb} and \eqref{eq:OmegaHcMaxEq}, and independent of the concrete value of $T_{\rm max}$ and of any additional model details. Here, a possible detection would allow a precise determination of $g_{*s}(T_{\rm max})$.

We have found that current dark radiation constraints from 
BBN and CMB on the total energy density fraction in GWs cannot yet probe the out-of-equilibrium gravitational wave emission with $T_{\rm max}<M_P$, as a naive application of the constraints implies a trans-Planckian  upper bound on 
$T_{\rm max}$ around $10^{19}$\,GeV, cf. Table~\ref{tab:Tmax}. Nevertheless we find the intriguing result that  the CGMB background with early time equilibration (i.e. with $T_{\rm max}>M_P$) just about saturates the dark radiation constraint of Eq.~\eqref{eq:DRstringent}, to be compared with the prediction of Eq.~\eqref{eq:equilibriumDR}. The former bound corresponds to homogeneous initial conditions for the gravitational waves, as appropriate under the assumption of thermal equilibrium at early times. Applying the bound of Eq.~\eqref{eq:DRstringent} and using the determination of $g_{*s,{\rm SM}}$ of Appendix \ref{app:effect_degr_freedom} (illustrated in Fig.~\ref{fig:gsSM}) would discard a CGMB with early time equilibration in models in which  $g_{* s}$ is augmented by less than $\sim4$ units. 

Further improvements on the dark radiation constraints would discard early time equilibration of the gravitational waves in a large class of models, and start constraining the out-of-equilibrium CGMB for sub-Planckian $T_{\rm max}$.
In case that future CMB constraints on dark radiation reach the theoretical uncertainty from the 
pure SM expectation ($\Delta N_{\rm eff} =0.001$), the upper bound of $T_{\rm max}$ can be improved to sub-Planckian values around $2\times 10^{17}$\,GeV, cf. Table~\ref{tab:Tmax}.  

Further progress should come from direct detection of GWs. However, all the current upper bounds from direct searches for stochastic GWs and also 
the projected sensitivities of planned GW detectors are at least nine orders of magnitude 
away from the prediction of the maximally allowed characteristic amplitude $h_c$ of the CGMB respecting the dark radiation constraint (corresponding to the CGMB with early time equilibration), cf. Fig. \ref{fig:hc_current_limits}. 

Conversion of GWs into EMWs in a static magnetic field has been identified as a promising search technique for stochastic GWs at
frequencies around $f_{\rm peak}^{h_c^{\rm CGMB}}\simeq 40\,{\rm GHz}\,[106.75/g_{*\rho}(T_{\rm max})]^{1/3}$, where the characteristic amplitude $h_c^{\rm CGMB}(f)$ of the CGMB attains its maximum, 
cf. Eq.~\eqref{eq:peakformulaf} and Table~\ref{tab:peaks}. 
We investigated the prospects of GW-EMW conversion in state-of-the-art superconducting magnets used in present and near future axion experiments and the detection of the generated EMWs/photons with dedicated detectors
appropriate around 40 GHz. The projected sensitivity of this technique, cf. Table~\ref{tab:magnetic_conv_facilities} and Fig.~\ref{fig:hc_future_limits}, turned out to fail to beat 
the dark radiation constraint on $h_c^{\rm CGMB}$ by about six orders of magnitude.
We then investigated the prospects of a proposal exploiting an additional EM Gaussian beam, delivered by a MW-scale 40 GHz gyrotron,  propagating along the magnetic conversion region in order to generate a transverse EMW conversion signal 
which is first order in $h_c$. 
Assuming state-of-the-art benchmarks for the gyrotron, the detector performance and the magnetic field strength and length,  
the projected sensitivity in $h_c$ is 
still three orders of magnitude above the maximum amplitude of the equilibrated CGMB (see Eq.~\eqref{eq:OmegaHcMaxEq}) which saturates the dark radiation bound, cf. Fig.~\ref{fig:hc_future_limits}.  
However, the latter may be reached by progress in the development of gyrotrons towards higher power and stable run time, single photon detection towards lower dark count rate, and superconducting magnets towards higher magnetic fields. 
The direct detection of the CGMB at a level corresponding to $T_{\rm max}\sim M_P$ by such a magnetic conversion experiment 
seems possible, although challenging.
In this connection, it should be emphasized that the search for the CGMB is truly a critical endeavour.
Any measurement of $T_{\rm max}$ above $6.6\times 10^{15}\,{\rm GeV}$ would be ground-breaking, since it would rule out 
standard inflation as a viable pre hot big bang scenario.

It would be very interesting to investigate the CGMB also in other BSM models with a complete and consistent cosmological 
history and thus giving a prediction of $T_{\rm max}$, such as for example the model in Ref.~\cite{Buchmuller:2012wn}. Furthermore, it seems
worthwhile to explore more deeply the possible synergies between axion and GW experiments which we have been touching upon in this 
paper. 

Currently, a community is forming which seriously considers the search for high-frequency gravitational waves~\cite{1832786}. Detecting the CGMB sets an ambitious, but rewarding goal for this enterprise.

%% file: acknowledgments.tex
\section*{Acknowledgments}
We acknowledge discussions with Walid Abdel Maksoud, Valerio Calvelli, Mike Cruise, Vladimir Fogel, John Jelonnek, Axel Lindner, Patrick Peter, J\"orn Schaffran, Manfred Thumm, Dieter Trines, and Yvette Welling. 
AR and JS acknowledge support by the Deutsche Forschungsgemeinschaft (DFG, German Research Foundation) under Germany's Excellence Strategy -- EXC 2121 ``Quantum Universe'' -- 390833306. CT acknowledges financial support by the DFG through SFB 1258 and the ORIGINS cluster of excellence.

%% file: appendix_complete_leading_order.tex
\section{\boldmath Loop functions for the rate of GW production from the primordial plasma at full leading order}
\label{app:complete_leading_order}
\setcounter{equation}{0}

Here we collect formula for the loop functions $\eta_{gg}(\hat k )$, $\eta_{sg}(\hat k )$, $\eta_{fg}(\hat k )$,  $\eta_{sf}(\hat k )$ appearing in Eq.~\eqref{eq:etafull}. We use the results of Ref.~\cite{Ghiglieri:2020mhm} with a simplified notation. One can define a set of six integrals ${\cal I}_{\pm}(\hat k), {\cal J}_{\pm}(\hat k), {\cal K}, {\cal L}$, in terms of which the above functions are expressed as:
\begin{align}\label{eq:etasIJKs}\begin{aligned}
  \eta_{gg}(\hat k )=&\,-4{\cal I}_{+}(\hat k)-4 {\cal J}_+(\hat k),\\
  \eta_{sg}(\hat k )=&\,-{\cal I}_+(\hat k)- {\cal J}_+(\hat k)=
\frac{1}{4} \eta_{gg}(\hat k ),\\
   \eta_{sf}(\hat k )=&\,4{\cal K}(\hat k)+2 {\cal L}(\hat k),\\
   \eta_{fg}(\hat k )=&\,4{\cal I}_{-}(\hat k)+4 {\cal J}_-(\hat k).
\end{aligned}\end{align}
The loop integrals  are given next:
\begin{align}\label{eq:IJKs}\begin{aligned} 
{\cal I}_{\pm}(\hat k)=&\,\frac{n_b(\hat k)}{4(4\pi)^3\hat{k}}\,\int_{-\infty}^{\hat{k}}dx\int_{|x|}^{2\hat{k}-x} dy\,\left\{(1+n_b(x)+n_b(\hat{k}-x))(y^2-x^2)\left(-\frac{2}{3}L^{\pm}_1\right.\right.\\
&\left.\left.+\frac{(y^2-3(x-2\hat{k})^2)(12 L^{\pm}_3+6y L^{\pm}_2+y^2L^{\pm}_1)}{6y^4}\right)+\frac{(1\pm3)\hat{k}^2\pi^2}{y^4}(y^2-x^2)\right\},\\
{\cal J}_{\pm}(\hat k)=&\,\frac{n_b(\hat k)}{4(4\pi)^3\hat{k}}\,\int_{\hat{k}}^{\infty}dx\int_{|2\hat{k}-x|}^{x} dy\,(n_b(x-\hat{k})-n_b(x))(y^2-x^2)\left(\frac{1}{3}(2M^{\pm}_1-y)\right.\\
&\left.-\frac{(y^2-3(x-2\hat{k})^2)}{6y^4}(12M^{\pm}_3+6yM^{\pm}_2+y^2M^{\pm}_1)\right),\\
{\cal K}(\hat k)=&\,-\frac{n_b(\hat k)}{4(4\pi)^3\hat{k}}\,\int_{-\infty}^{\hat{k}}dx\int_{|x|}^{2\hat{k}-x} dy\,(1+n_b(x)+n_b(\hat{k}-x))(y^2-x^2)L^{-}_1,\\
{\cal L}(\hat k)=&\,\frac{n_b(\hat k)}{4(4\pi)^3\hat{k}}\,\int_{\hat{k}}^{\infty}dx\int_{|2\hat{k}-x|}^{x} dy\,(n_b(x-\hat{k})-n_b(x))(y^2-x^2)\left(2M^{-}_1-y)\right..
\end{aligned}\end{align}
In the above expressions, one has 
\begin{align}
 n_b(x)=\frac{1}{e^x-1},
\end{align}
while the functions $L_i^{\pm}(x,y),M_i^{\pm}(x,y)$, with $i=1,2,3$ are defined below:
\begin{align}\begin{aligned}
L_1^{\pm}(x,y)=&\,\log\left(\frac{1 \mp e^{-\frac{1}{2}(x+y)}}{1 \mp e^{\frac{1}{2}(x-y)}}\right),\\
L_2^{\pm}(x,y)=&\,{\rm Li}_2(\pm e^{\frac{1}{2}(x-y)})-{\rm Li}_2(\pm e^{-\frac{1}{2}(x+y)}),\\
L_3^{\pm}(x,y)=&\,{\rm Li}_3(\pm e^{\frac{1}{2}(x-y)})-{\rm Li}_3(\pm e^{-\frac{1}{2}(x+y)}),\\
M_1^{\pm}(x,y)=&\,\log\left(\frac{1 \mp e^{-\frac{1}{2}(x-y)}}{1 \mp e^{-\frac{1}{2}(x+y)}}\right),\\
M_2^{\pm}(x,y)=&\,{\rm Li}_2(\pm e^{-\frac{1}{2}(x+y)})+{\rm Li}_2(\pm e^{-\frac{1}{2}(x-y)}),\\
M_3^{\pm}(x,y)=&\,{\rm Li}_3(\pm e^{-\frac{1}{2}(x+y)})-{\rm Li}_3(\pm e^{-\frac{1}{2}(x-y)}).
\end{aligned}\end{align}
As explained in the main text, the integrals above include a subtraction of infrared divergences, so that they remain finite. The subtraction corresponds to the last term in the integrand of ${\cal I}_\pm(\hat k)$.

%% file: appendix_eff_number_dog.tex
\section{\boldmath Effective number of degrees of freedom of the SM, the $\nu$MSM and the SMASH plasma}
\label{app:effect_degr_freedom}
\setcounter{equation}{0}

In the primordial plasma, the thermal contributions to the effective potential correspond to the free-energy density. Thermodynamic relations imply that the latter is equal to minus the plasma's pressure $p$:
\begin{align}
 p(\phi_i)=-\left[ V_{\rm eff}(\phi_i,T)- V_{\rm eff}(\phi_i,0)\right].
\end{align}
In the equation above, $\phi_i$ denote the scalar field backgrounds, and the subtraction of the second term above guarantees that the pressure is zero in the vacuum ($T=0$). We will assume that the system relaxes to the minimum of the effective potential, so that we will evaluate the backgrounds at the configurations $\phi_i=\bar\phi_i$ that extremize $V_{\rm eff}$:
\begin{align}
 \left.\frac{\partial V_{\rm eff}(\phi_i,T)}{\partial\phi_j}\right|_{\phi_i=\bar\phi_i(T)}=0.
\end{align}
The effective potential with its finite-temperature correction at one-loop plus higher-order QCD effects can be written as:
\begin{align}\label{eq:Veff}
 V_{\rm eff}(\phi_i,T)=V(\phi_i)+V^{\rm CW}(\phi_i,T)+V^T(\phi_i,T)+V^{\rm QCD}(T).
\end{align}
Above, $V$ is the tree-level potential, $V^{\rm CW}$ the usual one-loop Coleman-Weinberg contribution at zero temperature, $V^T$ the one-loop thermal correction to the potential, and $V^{\rm QCD}$ includes two and three-loop thermal corrections induced by QCD effects, which are known for zero field backgrounds. Note that we have included a temperature-dependence in the Coleman-Weinberg vacuum piece. The reason is that achieving accuracy in the finite-temperature effects near a phase transition requires a resummation of thermal contributions to the 2 point functions -- daisy resummation -- which can be implemented by replacing the effective masses in the vacuum propagators by temperature corrected masses; this explains the $T$ dependence in the Coleman-Weinberg piece. The daisy resummation is usually performed by keeping the leading terms of the corrections to the two point functions in a high-temperature expansion. However, as noted in Ref.~\cite{Ringwald:2020vei}, achieving accuracy \emph{across} a phase transition, i.e. for temperatures below the masses gained by particles during the transition, requires to go beyond the high-$T$ expansion. As we consider temperatures above the SM's electroweak crossover, for the SM and $\nu$MSM we use a traditional daisy resummation. For SMASH, for which the PQ phase transition lies much above the electroweak scale, we implement the improved resummation of Ref.~\cite{Ringwald:2020vei} for the corrections involving fields that acquire masses during the PQ transition.

In the Landau gauge and in the  $\overline{\rm MS}$ scheme, $V^{\rm CW}(\phi_i,T)$ and $V^T(\phi_i,T)$ are given by
\begin{align}\begin{aligned}
 V^{\rm CW}(\phi_i,T)=&\,\frac{1}{64\pi^2}\left[\sum_V m^4_V(\phi_i,T)\left(\log\frac{m^2_V(\phi_i,T)}{\mu^2}-\frac{5}{6}\right)+\sum_S m^4_S(\phi_i,T)\left(\log\frac{m^2_S(\phi_i,T)}{\mu^2}-\frac{3}{2}\right)
 \right.\\
 &\left.-\sum_F m^4_F(\phi_i,T)\left(\log\frac{m^2_F(\phi_i,T)}{\mu^2}-\frac{3}{2}\right)\right],\\
 V^T=&\,\frac{T^4}{2\pi^2}\left[\sum_B J_B\left(\frac{m^2_B(\phi_i,T)}{T^2}\right)-\sum_F J_F\left(\frac{m^2_F(\phi_i,T)}{T^2}\right)-\sum_GJ_B\left(0\right)\right].
\end{aligned}\end{align}
In the previous formulae, $V, S, F$ and $G$ correspond to massive gauge bosons, real scalars, Weyl fermions and ghosts. It is assumed that for gauge bosons one has to  sum over the three polarizations that propagate in the Landau gauge, while for fermions one has to sum over two helicities, and for ghosts one should count one degree of freedom per generator of each gauge group.  $m^2_{V/S/F}(\phi_i,T)$ are the field-dependent masses in the background of $\phi_i$, including the thermal corrections from the (improved) daisy resummation.  $\mu$ is the renormalization scale, while the thermal loop functions $J_B$ and $J_F$ are
 \begin{align}\label{eq:Js} \begin{aligned}
  J_B(x)=\int_0^\infty dy\, y^2\log\left[1-\exp(-\sqrt{x+y^2})\right],\\
   J_F(x)=\int_0^\infty dy\, y^2\log\left[1+\exp(-\sqrt{x+y^2})\right].
 \end{aligned}\end{align}
 The last missing piece in the effective potential in Eq.~\eqref{eq:Veff} corresponds to  higher loop QCD contributions, which have been computed in Ref.~\cite{Kajantie:2002wa} up to three-loop order in a theory with arbitrary massless QCD flavours. This allows to apply the QCD corrections for all the models of interest.

Once the pressure is calculated from the minimization of the effective potential that follows from the above equations, the energy density $\rho$, standard thermodynamic relations allow to compute the entropy density $s$ and the specific heat capacity $c=1/V(\partial U/\partial T)|_V$ as
\begin{align}
 \rho= &\,T\frac{\partial p}{\partial T}-p, & s=&\,\frac{\partial p}{\partial T}, & c=&\,T\,\frac{\partial^2 p}{\partial T^2}.
\end{align}

From the former quantities one defines the effective numbers of degrees of freedom $g_{*\rho}$, $g_{*s}$, $g_{*c}$ by writing
\begin{align}
 \rho = &\,\frac{\pi^2}{30}\,g_{*\rho}(T)T^4, & s = &\, \frac{2\pi^2}{45}\,g_{*s}(T) T^3,  & c = &\, \frac{2\pi^2}{15}\,g_{*c}(T) T^3
\end{align}

In our calculations of the effective potential, we use a renormalization scale proportional to the temperature, $\mu=\kappa 2\pi T$ with $\kappa\in\{1/2,1,2\}$. It should be noted that for the SM our perturbative calculations are not trustworthy for temperatures below the electroweak crossover, as the daisy resummation does not capture the decoupling of the degrees of freedom acquiring masses during the crossover. Furthermore, for lower temperatures the QCD interactions become nonperturbative, and other techniques are 
necessary~\cite{Laine:2015kra,Saikawa:2018rcs}. In this paper we neglect the gravitational wave production at temperatures below the electroweak crossover.

For SMASH we introduce an additional correction coming from the loss of chemical equilibrium of the axion, which implies a separate conservation of the entropy of the axion and the rest of the thermal bath, and thus  separate temperatures for both, to be denoted as $T_{\rm axion}$ and $T$. Labeling the results for $g^{\rm SMASH}_{*\rho}$,  $g^{\rm SMASH}_{*s}$,  $g^{\rm SMASH}_{*c}$ under the assumption of chemical equilibrium for the axion with the superfix ``eq'', the quantities accounting for decoupling can be written as \cite{Ringwald:2020vei}
\begin{align}\begin{aligned}
 g^{\rm SMASH}_{*\rho}=&\,g^{\rm SMASH, eq}_{*\rho}-1+\left(\frac{T_{\rm axion}}{T}\right)^{4},& g^{\rm SMASH}_{* s}= &\,g^{\rm SMASH,eq}_{* s}-1+\left(\frac{T_{\rm axion}}{T}\right)^3,\\
 g^{\rm SMASH}_{* c}= &\,g^{\rm SMASH, eq}_{* c}-1+\left(\frac{T_{\rm axion}}{T}\right)^3 .
\label{eq:gsrhosbathaxion}
\end{aligned}\end{align}

%% file: appendix_upper_bound_tmax_inflation.tex
\section{\boldmath Upper bound on $T_{\rm max}$ in inflationary cosmology}
\label{app:upper_bound_on_T_max_ic}
\setcounter{equation}{0}

The hypothesis of inflation postulates a period of accelerated expansion, $\ddot a > 0$, in the very early universe  \cite{Starobinsky:1980te,Guth:1980zm}, preceding the standard radiation-dominated era. It 
offers a physical model for the origin of the initial conditions of hot big bang cosmology. 
In slow-roll inflationary cosmology \cite{Linde:1981mu,Albrecht:1982wi},   
the energy scale at the end of inflation, $\rho_{\rm inf}$, can be inferred from the amplitude $A_S$ of scalar perturbations generated during inflation and the tensor-to-scalar ratio $r$  via 
\begin{equation}
\rho_{\rm inf} \equiv 3 H_{\rm inf}^2 M_P^2 \approx   \frac{3}{2}\, \pi^2\,r \,A_S M_P^4\,,
\end{equation}
where $H_{\rm inf}$ is the Hubble expansion rate during inflation. 
The upper limit on $r$ obtained from CMB observations of the BICEP2/Keck Array and Planck Collaborations 
provides an upper bound on $\rho_{\rm inf}$ \cite{Akrami:2018odb}, 
\begin{equation}
\label{eq:upp_bound_infl_scale}
\rho_{\rm inf}\, < \,  \left( 1.6\times 10^{16}\,{\rm GeV}\right)^4
\hspace{3ex} (95\%\ {\rm CL})\,.
\end{equation}
This may be turned into an upper bound on the maximum temperature of the post-inflationary universe by assuming 
instantaneous and thus 
maximally efficient reheating to a radiation dominated universe with energy density 
\begin{equation}\label{eq:Tupmax}
 \frac{\pi^2}{30} g_{*\rho}(T_{\rm max}^{\rm up}) T^{\rm up\,4}_{\rm max}  = \rho_{\rm inf} \,,
\end{equation}
which in combination with  \eqref{eq:upp_bound_infl_scale} leads to  
\begin{equation}
\label{eq:reheating_constraint_Tmax}
T^{\rm up}_{\rm max} = \left[\frac{\rho_{\rm inf}}{\frac{\pi^2}{30} \,g_{*\rho}(T_{\rm max}^{\rm up})} \right]^{1/4}
<   6.6\times 10^{15}\,{\rm GeV}
\left[ \frac{106.75}{g_{*\rho}(T_{\rm max}^{\rm up})} \right]^{1/4}
\,.
\end{equation}
In SMASH~\cite{Ballesteros:2016xej}, $g_{*\rho}(10^{15}\,{\rm GeV})\simeq 124.5$.

%% file: appendix_3d_eff_gw_emw.tex
\section{3D effects in GW-EMW conversion experiments}
\label{app:3Deffects}
\setcounter{equation}{0}

\subsection{Waveguide effect}\label{sec:waveguide_effects}
If the magnetic conversion volume is surrounded by a waveguide the phase velocity inside the waveguide will be different from the speed of light. We illustrate this for a circular waveguide which is infinite in $z$-direction and has radius $a$. For TE modes we have $E_z=0$ everywhere and the normal derivative of $B_z$ vanishes at the boundaries. After the application of all boundary conditions we obtain for the longitudinal $B$-field:
\begin{eqnarray}
B_z=F\sin\left(\nu(\phi-\phi_0)\right)J_\nu(k_c r)e^{-ik_z z}\,,
\end{eqnarray}
where $F,\phi_0$ are constants, $\nu$ is an integer, $r,\phi$ and $z$ are the three coordinates in the cylindrical coordinate system, $J_\nu$ is the Bessel function of the first kind and $k_c=\frac{p_{\nu n}}{a}$ is the transversal momentum, where $p_{\nu n}$ is the $n$th zero of $J_\nu'(x)$. The total wave vector is given by $k^2=k_c^2+k_z^2={\omega^2}$. From the longitudinal field $B_z$ we can derive all other electromagnetic fields in the cylindrical waveguide. The phase velocity is:
\begin{eqnarray}
v_p=\frac{\omega}{k_z}=\frac{\omega}{\sqrt{{\omega^2}{}-\left(\frac{p_{\nu n}}{a}\right)^2}}=\frac{1}{\sqrt{1-\left(\frac{p_{\nu n}}{a\omega}\right)^2}}.
\end{eqnarray}
If the radius of the waveguide is much larger than the wavelength we can expand the square root and get:
\begin{eqnarray}
v_p\approx 1+\frac{1}{2}\left(\frac{p_{\nu n}\lambda}{2\pi a}\right)^2
\,.
\label{eq:phase_vel}
\end{eqnarray}
Therefore the phase velocity is larger than the speed of light. 

For a coherent and thus efficient conversion of gravitons into photons inside a waveguide, they should be in phase all along the 
 waveguide. 
Therefore the phase difference between gravitons and photons has to be smaller than $\pi$ at the end of the waveguide:
\begin{eqnarray}
k_GL-k_z L\ll \pi,
\label{eq:phasediff}
\end{eqnarray}
where $k_G=\omega$ is the graviton dispersion, $k_z$ is the photon wave vector in $z$-direction and $L$ the length of the external $B$-field. Plugging equation~\eqref{eq:phase_vel} into equation~\eqref{eq:phasediff} yields the condition:
\begin{eqnarray}
L\ll \left(\frac{2\pi}{p_{\nu n}}\right)^2 \frac{a^2}{\lambda}.
\label{eq:phasediff2}
\end{eqnarray}
For the dominating $TE_{11}$ mode we find $p_{11}=1.8$ and therefore in this case the prefactor is $ \left(\frac{2\pi}{p_{\nu n}}\right)^2\approx 12$.

\subsection{Diffraction in open systems}

If no waveguide encloses the magnetized region we only have to take into account diffraction effects. For a magnetized region of length $L$ along the $z$-direction with a circular shape in the $xy$-plane we obtain the diffraction angel $\theta$ in the far field as:
\begin{eqnarray}
\tan\theta=1.22\frac{\lambda}{d},
\end{eqnarray}
where $d$ is the diameter of the magnetized region. For IAXO we get $\theta=0.6^{\circ}$ and therefore diffraction is negligible. In the case of ALPS IIc the diffraction angle is much larger and due to the huge length of ALPS IIc it cannot be neglected.